\documentclass[useAMS,usenatbib]{mn2e}

\usepackage{times}
\usepackage{latexsym}
\usepackage{graphicx}

\def\rd {{\rm d}}

\def\inch{"\,}
\def\Km{K$^{\prime}$}

\def\kms{km$\,$s$^{-1}$\,}

\def\htwo{$\mathrm{H}$\,{\sc ii}\,}
\def\s0{S$0$}
\def\ltsim{~\rlap{\lower -0.5ex\hbox{$<$}}{\lower 0.5ex\hbox{$\sim\,$}}}
\def\gtsim{~\rlap{\lower -0.5ex\hbox{$>$}}{\lower 0.5ex\hbox{$\sim\,$}}}
\def\deg{\hbox{$^\circ$}\,}
\def\degc{\hbox{$^\circ$}}
\def\eq{\!=\!}
\def\eg{e.g.,~}
\def\ie{i.e.,~}
\def\cf{cf.~}
\def\tab{Table~}
\def\fig{Fig.~}
\def\sec{Section~}
\def\apx{Appendix~}

\def\magsqarcsec{mag$/$\raisebox{-0.4ex}{\hbox{$\Box^{\prime\prime}$}\,}}

\def\hin{\hbox{$h_{\rm in}$}\,}
\def\hout{\hbox{$h_{\rm out}$}\,}
\def\hindhout{\hbox{$h_{\rm in}/h_{\rm out}$}\,}
\def\rbr{\hbox{$R_{\rm br}$}\,}

\def\munin{\hbox{$\mu_{0,\rm in}$}\,}
\def\munout{\hbox{$\mu_{0,\rm out}$}\,}
\def\mun{\hbox{$\mu_0$}\,}
\def\mubr{\hbox{$\mu_{\rm br}$}\,}
\def\z0{\hbox{$z_{0}$}\,}
\def\hz{\hbox{$h_{\rm z}$}\,}

\newcommand{\typet}{Type~II }
\newcommand{\typetc}{Type~II}
\newcommand{\typeiii}{Type~III }
\newcommand{\typeiiic}{Type~III}
\newcommand{\typeiiis}{Type~III-s }
\newcommand{\typeiiisc}{Type~III-s}

\newcommand{\typeiiidc}{Type~III-d}
\newcommand{\typeo}{Type~I }
\newcommand{\typeoc}{Type~I}
\newcommand{\typetpiii}{Type~II + III}

\title[On the 3 dimensional structure of edge-on disk galaxies]
{On the 3 dimensional structure of edge-on disk galaxies}

\author[M. Pohlen, S. Zaroubi, R.F. Peletier, \& R.~-J.~Dettmar]{Michael
Pohlen$^{1,2}$ \thanks{E-mail: pohlen@astro.rug.nl}, Saleem Zaroubi$^{1}$,
Reynier F. Peletier$^{1}$, \& Ralf-J\"urgen Dettmar$^{2}$\\ $^{1}$ Kapteyn
Astronomical Institute, University of Groningen, PO Box 800, NL-9700 AV
Groningen, The Netherlands \\ $^{2}$ Astronomisches Institut,
Ruhr-Universit\"at Bochum, Universit\"atsstrasse 150, D-44780 Bochum, Germany
\\ }
\begin{document}
%
%
\maketitle
\begin{abstract}
A simple algorithm is employed to deproject the two dimensional images
of a pilot sample of $12$ high-quality images of edge-on disk galaxies
and to study their intrinsic 3 dimensional stellar distribution.
We examine the radial profiles of the stars as a function of height above the
plane and report a general trend within our sample of an increasing
radial scalelength with height outside of the dustlane. This could be
explained by the widespread presence of a thick disk component in these
galaxies.
In addition, the 3-dimensional view allows the study of the vertical
distribution of the outer disk, beyond the break region, where we detect a
significant increase in scalelength with vertical distance from the major axis
for the truncated disks. This could be regarded as a weakening of the
``truncation'' with increasing distance from the plane.
Furthermore, we conclude that the recently revised classification of
the radial surface brightness profiles found for face-on galaxies is
indeed independent of geometry. In particular, we find at least
one example of each of the three main profile classes as defined in
complete samples of intermediate to face-on galaxies: not-truncated,
truncated and antitruncated. The position and surface brightness that
mark the break location in the radial light distribution are found to
be consistent with those of face-on galaxies.
\end{abstract}
\begin{keywords}
galaxies: photometry  -- 
galaxies: structure -- 
galaxies: fundamental parameters  -- 
galaxies: evolution  -- 
galaxies: formation 
galaxies: individual: ESO\,380-019, ESO\,404-018, FGC\,2339, 
IC\,4871, NGC\,522, NGC\,1596, NGC\,3390, NGC\,4179, NGC\,5290, 
NGC\,5290, NGC\,5981, UGC\,10459
\end{keywords}


\section{Introduction}
\label{introduction}
To study galaxy formation and evolution it is often helpful to imagine
that galaxies are made up of a small number of separate
components. Surface photometry of galaxies provides a very
simple way of parameterising these various components.
By modelling the surface-brightness distribution of a galaxy and thereby
parametrising the individual components, we obtain a common ground to measure,
compare, and sort large samples of galaxies.
Owing to their flat geometry the random orientation of a specific
galaxy disk provides us with different projected views of the,
presumably, common underlying 3 dimensional distribution. In
particular, one gets complementary views from observing face-on or
edge-on galaxies.
In the face-on view, one studies the radial distribution of the starlight
discussing, \eg non-axisymmetric features such as bars, rings, or spiral
arms. In the edge-on view on the other hand, one is able to study the vertical
distribution -- \eg disentangle thick disk, thin disk and stellar halos. Due
to a longer line-of-sight projection, the edge-on view allows in principle the
exploration of the radial profile to larger distances.

Early on, from face-on galaxy observations, several authors
\cite[e.g.][]{patterson1940, devaucouleurs1959} noticed that the radial light
distribution of stellar disks is well described by a simple exponential,
albeit the lack of a clear physical explanation for this fact.
Later on, \cite{free70} called these disks \typeo -- as opposed to some other
galaxies called \typet -- showing a break in their radial profile with a
downbending steeper outer region, or equivalently, a flatter inner region.
Using the edge-on view, \cite{vdk1979} discovered that the exponential decline
does not continue to the last measured point, but is {\it truncated} after
several radial scalelengths.
Afterwards, with the advent of CCD technology, several groups
\citep{barteldrees1994,pohlen2000b,florido2001,
degrijs2001,pohlen2001,kregel2002} studied larger samples of edge-on
galaxies to find out if and at what radius this truncation occurs.
More recently, \cite*{erwin2005}, \cite*{erwin2006}, and
\cite{pohlen2006} have analysed surface brightness profiles of face-on
to intermediate inclined galaxies and revised our view on disk truncations.
They identified three basic classes of surface brightness profiles
depending on the apparent break features or lack of them: 1) The well
known \typeo that has an exponential profile, with no break.  2)
\typet with a `downbending break'. 3) A completely new class,
called \typeiiic, also described by a broken exponential but with an
upbending profile beyond the break.
In this view the {\it classical truncations} described initially
in the pioneering work of \cite{vdk1979} and \cite{vdk1981} are in
fact part of a generalised Freeman \typet class \citep{free70}.
In this picture truncated galaxies show a surface brightness profile with a
shallow inner and a steeper outer exponential region separated at a relatively
well defined break radius.

These results are in agreement with those reported by \cite{pohlen2002a}, who
obtained deep photometry of three face-on galaxies and clearly demonstrated
that the sharp cut-offs\footnote{It is interesting to note that in this
concept the truncation radius is often equal to the last measured point and
therefore much further out compared to the break radius of a broken
exponential fit.} fitted to profiles of edge-on galaxies
\cite[e.g. by][]{vdk1981, barteldrees1994, pohlen2000b} are in fact not
complete but better described by a broken exponential.
Even before, \cite{degrijs2001} and \cite{pohlen2001} noted that for
edge-on galaxies the surface brightness distribution does not
disappear asymptotically (`vertically') into the background
noise. Interestingly, \cite{hunter2006a} find a similar broken
exponential shape for many objects in their large sample of irregular
galaxies.

The redefinition of the truncation as a point at which the two
exponential fits meet solves, among other things, the apparent
contradiction between simulations by \cite{holley2001} and the edge-on
observations of \cite{barteldrees1994} and \cite{pohlen2000a}. The
latter authors find that `truncations' are as `close' as twice the
radial scalelength from the centre in contradiction with the
simulations.  Furthermore, `early' truncations (\ie at $R\ltsim 2h$)
of edge-on galaxies have been always questioned by the `face-on
viewers', who had never observed a face-on galaxy with a sharp
truncation at a distance of twice the scalelength.
Despite the line-of-sight enhancement of the surface brightness level
in edge-on galaxies, star counts in nearby galaxies are able to
provide significantly deeper profiles with much higher S/N. Using this
completely independent method, \cite{ferguson2006} followed the
profile of the face-on galaxy M\,33 down to $\mu_{\rm lim}\!\sim\!30$
I-\magsqarcsec, which is currently unreachable with surface
photometry. Their result shows that the broken exponential description
is valid down to the last bin above the noise level and an actual {\it
edge} of the disk is not found, at least for this galaxy.
Nevertheless, there still seem to be some open issues for a complete
reconciliation of the renewed face-on with the `old' edge-on
view~\citep{dezeeuw2007} as to provide a consistent picture of the
three-dimensional structure, for example:
\begin{itemize} 
\item Is the broken exponential structure, now well observed for face-on
galaxies, quantitatively consistent with the truncated profiles observed for
edge-on galaxies?
\item Do we see a different kind of truncation in the deeper views
provided by the edge-on galaxies?
\item Can we identify the edge-on counterparts of the three main types of
surface brightness profiles (\typeoc, \typetc, and \typeiiic) observed for
face-on galaxies by \cite{erwin2006} and \cite{pohlen2006}?
\item Does the frequency of different profile classes found 
for the face-on galaxies match the edge-on results? 
\item Why did the former edge-on surveys, with large and deep samples, not
discover the new class of \typeiii profiles?
\end{itemize}
To answer these questions and disentangle the crucial problems with
the line-of-sight integration \cite[see e.g.][]{pohlen2004a} we
decided to deproject a sample of edge-on galaxies. This can only be
done using some geometric assumption. Here we choose the assumption of
axial symmetry. This will allow us to gain insight into the
three-dimensional galactic stellar disk structure by bringing together
the classical face-on and edge-on views.

The vertical structure may well be the key to understand the nature of the
different breaks in the surface brightness profiles, by setting additional
constraints on the various physical explanations put forward.
For example, some recent star formation models 
\cite[][]{elmegreen2006,li2006} are now able to produce the 
observed broken exponential structure, but, alternatively, 
\cite{debattista2006} also find downbending breaks using purely 
collisionless $N$-body simulations. 
Once the nature of the galactic stellar disk is determined it will be fully
justified to utilise these breaks for comparison of galaxies at various
redshifts as done by \cite{perez2004}, \cite{trujillo2005} or \cite{tamm2006},
and explain their appearance in cosmological simulations as shown by
\cite{governato2007}.
The remainder of this paper is organised as follows. 
\sec\ref{data} describes our sample selection and data. 
In \sec\ref{method} we introduce our method to deproject the 
edge-on galaxies. The results are given in \sec\ref{results}, 
we discuss them in \sec\ref{discussion} and conclude in 
\sec\ref{conclusion}. 
In \apx\ref{comments} we give detailed comments for 
all our galaxies and show in \apx\ref{figures} radial surface 
brightness profiles and isophote maps for all galaxies. 
\section{Data}
\label{data}
All optical images used in this paper were taken as part of a PhD
study on the radial structure of galactic stellar disks by
\cite{pohlen2001}.
Originally, the disk galaxies ($-2\!\le\!T\!\le\!7$) were chosen according to
the allocated observing time, the observatory (north/south), and the available
FOV ($D_{25}\!\gtsim\!2$\arcsec) to meet the following morphological selection
criteria verified by using images from the Digitized Sky Survey (DSS): 1) The
galaxies are edge-on with $i\gtsim 86$\degc. 2) Undisturbed with smooth
photometric features (see below). 3) They look similar to some prototypical
cases like NGC\,4565 or IC\,2531, this is done to make it possible to
consistently fit a simple disk model.

Galaxies with the following characteristics were rejected: 1) Apparent spiral
arms, indicating a lower inclination. 2) A significantly asymmetric or
disturbed disk, indicating strong interaction. 3) Two-sided or significantly
one-sided warped disks. 4) Galaxies apparently dominated by the light of their
bulge component (\ie there are almost no Sa/Sab galaxies, but several
lenticulars). 5) Galaxies that showed only a faint, patchy, not well defined
disk.
For more details of the original sample we refer the reader 
to \cite{pohlen2001} or \cite{pohlen2004b}.
The present sample of 11 galaxies is drawn from this base 
sample of 72 galaxies as a pilot sample covering a wide 
range of Hubble types and apparent vertical distributions 
(thin and thick) while trying to pick galaxies with very 
few disturbing fore- or background objects.  
We add an additional NIR image of one of the galaxies to 
observationally assess the influence of the dust.   
Global properties of the finally chosen galaxies are given in
Table~\ref{sample}. 
\begin{table*}
\begin{center}
{\footnotesize
\begin{tabular}{l  c c l  r@{.}l  r@{.}l  c  c r@{.}l c c}
\hline
\rule[+0.4cm]{0mm}{0.0cm}
Galaxy
&RA 
&DEC
&RC3
&\multicolumn{2}{c}{T}
&\multicolumn{2}{c}{Diam.}
&$v_{\sun}$
&$v_{\rm vir}$
&\multicolumn{2}{c}{D} 
&$v_{\rm rot}$
&$M^{\scriptscriptstyle 0}_{\rm B}$ \\[+0.1cm]
&\multicolumn{2}{c}{(J2000.0)}
&type 
&\multicolumn{2}{c}{}
&\multicolumn{2}{c}{[\ \arcmin\ ]}
&[\ \kms]
&[\ \kms]
&\multicolumn{2}{c}{[Mpc]} 
&[\kms]
&[mag]\\
\rule[-3mm]{0mm}{5mm}
{\scriptsize{\raisebox{-0.7ex}{\it (1)}}}
&{\scriptsize{\raisebox{-0.7ex}{\it (2)}}}
&{\scriptsize{\raisebox{-0.7ex}{\it (3)}}}
&{\scriptsize{\raisebox{-0.7ex}{\it (4)}}}
&\multicolumn{2}{c}{{\scriptsize{\raisebox{-0.7ex}{\it (5)}}}}
&\multicolumn{2}{c}{{\scriptsize{\raisebox{-0.7ex}{\it (6)}}}}
&{\scriptsize{\raisebox{-0.7ex}{\it (7)}}}
&{\scriptsize{\raisebox{-0.7ex}{\it (8)}}}
&\multicolumn{2}{c}{{\scriptsize{\raisebox{-0.7ex}{\it (9)}}}} 
&{\scriptsize{\raisebox{-0.7ex}{\it (10)}}} 
&{\scriptsize{\raisebox{-0.7ex}{\it (11)}}} \\[-0.2cm]
\hline\hline \\[-0.4cm]
NGC 522     &$01\,24\,45.9$ &$+09\,59\,40$&\texttt{.S..4$*$/} & 3&7   &2&6 &2727 &2729&37 &9& 168&$-$20.6\\
NGC\,1596   &$04\,27\,38.1$ &$-55\,01\,40$&\texttt{.LA..*/}   &$-$2&0 &4&1 &1508 &1227&17 &0&  98&$-$19.3\\
NGC\,3390   &$10\,48\,04.5$ &$-31\,32\,02$&\texttt{.S..3./}   & 3&0   &3&6 &2850 &2696&37 &4& 210&$-$21.4\\
NGC\,4179   &$12\,12\,52.6$ &$+01\,17\,47$&\texttt{.L..../}   &$-$2&0 &3&9 &1239 &1269&17 &6&\ldots&$-$19.6\\
ESO\,380-019&$12\,22\,02.1$ &$-35\,47\,32$&\texttt{.S..6$*$/} & 6&0   &3&6 &2937 &2793&38 &8& 240&$-$21.2\\
NGC\,5290   &$13\,45\,19.1$ &$+41\,42\,45$&\texttt{.S..4$*$/} & 3&7   &4&0 &2588 &2817&39 &1& 219&$-$20.8\\
NGC\,5981   &$15\,37\,53.4$ &$+59\,23\,29$&\texttt{.S..5\$/}  & 5&0   &2&5 &2528 &2813&39 &1& 251&$-$20.6\\
UGC\,10459  &$16\,35\,07.9$ &$+40\,59\,29$&\texttt{.S..6$*$.} & 6&0   &1&5 &9001 &9252&128&5& 218&$-$21.5\\
IC\,4871    &$19\,35\,42.4$ &$-57\,31\,06$&\texttt{.SXS7?/}   & 6&7   &3&1 &1927 &1726& 24&0& 115&$-$19.6\\
FGC\,2339   &$21\,44\,39.5$ &$-06\,41\,17$&\texttt{.S..6$*$/} & 6&0   &1&8 &3097 &3096& 43&0&  85&$-$19.5\\
ESO\,404-018&$22\,01\,10.2$ &$-32\,34\,44$&\texttt{.7?/}   & 7&0   &2&6 &2275 &2144& 29&8&  73&$-$19.1\\
\hline
\end{tabular}
}
\caption[]{The 11 galaxies of our pilot survey. \newline 
{\scriptsize{\it (1)}} Principal name, {\scriptsize{\it (2)}} right ascension,
{\scriptsize{\it (3)}} declination, {\scriptsize{\it(4)}} RC3 coded
Hubble-type, and the {\scriptsize{\it(5)}} Hubble parameter T \cite[all taken
from][]{rc3}, {\scriptsize{\it(6)}} diameter in arcminutes,
{\scriptsize{\it(7)}} heliocentric radial velocities, {\scriptsize{\it(10)}}
maximum deprojected rotation velocity of the gas, and the 
B-Band absolute magnitude {\scriptsize{\it(11)}} (taken from  LEDA). 
According to the heliocentric radial velocities corrected for the 
Local Group infall into the Virgo cluster {\scriptsize{\it(8)}} 
from LEDA, we estimated the {\scriptsize{\it (9)}} distances following 
the Hubble relation with the Hubble constant from the HST key project 
of $H_{0}\!=\!72$ km s$^{-1}$Mpc$^{-1}$ \cite[]{hst_h0}.
\label{sample}}
\end{center}
\end{table*}

%
The images were obtained in six different observing runs between 
1998 and 2004. Two runs ({\sl E1} and {\sl E2}) at the Danish 
1.54m telescope and one ({\sl E3}) at the 3.6m NTT telescope, both 
at the European Southern Observatory in Chile. 
One run ({\sl C1}) at the 1.23m telescope and one ({\sl C2}) 
at the 3.5m telescope both at Calar Alto in Spain. And finally 
one run ({\sl L1}) at the Hall 42\inch (1.06m) telescope of 
the Lowell Observatory at Anderson Mesa (USA).  
During the two runs with the 1.54m Danish telescope it was 
equipped with the DFOSC camera and the C1W7/CCD which is a 
2k\,x\,2k LORAL chip providing a field size of 
$\approx\!13\arcmin$ and a scale of 
$0.39\arcsec$pixel$^{-1}$. 
At the NTT we used EMMI, equipped with a 2k\,x\,2k Tektronix chip 
providing a field size of $\approx\!9\arcmin$ and a scale 
of $0.27\arcsec$pixel$^{-1}$. 
The run at the Calar Alto 1.23\,m telescope was done in 
service mode with the Site\#18b chip, a 2k\,x\,2k SITE CCD 
with 24\,$\mu$m pixel size, providing an unvignetted field 
of $\approx\!10\arcmin$ and a scale of $0.5\arcsec$pixel$^{-1}$.
At the 3.5m telescope we used the wide-field near infrared camera
OmegaPrime with a 1k\,x\,1k pixel HAWAII HgCdTe array by Rockwell,
providing a field of $\approx\!7\arcmin$ at a scale of 
$0.4\arcsec$pixel$^{-1}$.
During the run at Anderson Mesa the 42\inch telescope was 
equipped with the NSF CCD used with 2x2 binning providing 
a field size of $\approx\!4.9\arcmin$ and a scale of 
$\approx\!0.73\arcsec$pixel$^{-1}$.
For the optical imaging the available standard Bessel 
R and V filter at ESO, and Johnson R at Calar Alto and Anderson 
Mesa are used. The NIR image with OmegaPrime is in the \Km-band. 
Table \ref{observ} summarises the detailed observational 
parameters for each image. 

\begin{table*}
\begin{center}
\begin{tabular}{l c c c c c c }
\hline
\rule[+0.3cm]{0mm}{0.0cm}
Galaxy
&Filter 
&Date
&Site
&Exp.time
&Seeing
&Coadds \\
 &&[mmyy]&&[min]&[\arcsec]&\#\,x\,[s] \\[+0.1cm]
\rule[-3mm]{0mm}{5mm}{
\scriptsize{\raisebox{-0.7ex}{\it (1)}}}
&{\scriptsize{\raisebox{-0.7ex}{\it (2)}}}
&{\scriptsize{\raisebox{-0.7ex}{\it (3)}}}
&{\scriptsize{\raisebox{-0.7ex}{\it (4)}}}
&{\scriptsize{\raisebox{-0.7ex}{\it (5)}}}
&{\scriptsize{\raisebox{-0.7ex}{\it (6)}}}
&{\scriptsize{\raisebox{-0.7ex}{\it (7)}}} \\[-0.2cm]
\hline\hline \\[-0.4cm]
NGC\,522     & R & 1198 &{\sl L1}&60.0& 3.0  &  6x600       \\
NGC\,1596    & R & 0198 &{\sl E1}&45.0& 1.3  &8x300, 2x150  \\
NGC\,3390    & R & 0198 &{\sl E1}&51.7& 1.6  &3x600, 3x300, 2x200 \\
NGC\,4179    & V & 0599 &{\sl E2}&65.0& 1.7  &1x600, 1x480, 1x360, 1x300, 9x240 \\
ESO\,380-019 & V & 0599 &{\sl E2}&60.0& 1.6  &  6x600       \\
NGC\,5290    & R & 0699 &{\sl C1}&60.0& 2.0  &  3x600       \\ 
NGC\,5290    &\Km& 0204 &{\sl C2}&38.0& 1.2  & 38x60        \\ 
NGC\,5981    & R & 0699 &{\sl C1}&60.0& 1.7  &  6x600       \\ 
UGC\,10459   & R & 0699 &{\sl C1}&60.0& 1.6  &  6x600       \\
IC\,4871     & V & 0599 &{\sl E2}&60.0& 1.4  &  6x600       \\
FGC 2339     & R & 0700 &{\sl E3}&30.0& 0.9  &  3x600       \\
ESO\,404-018 & V & 0599 &{\sl E2}&60.0& 1.3  &  6x600       \\
\hline \\
\end{tabular}
\caption{Observing log for the individually combined images. \newline 
{\scriptsize{\it (1)}} Galaxy name, {\scriptsize{\it (2)}} filter, 
{\scriptsize{\it (3)}} observing date, {\scriptsize{\it (4)}} 
site, {\scriptsize{\it (5)}} the total coadded on-source exposure 
time, {\scriptsize{\it (6)}} seeing, as measured on
the final combined image, {\scriptsize{\it (7)}} the number of 
individual images with their individual exposure times.}
\end{center}
\label{observ}
\end{table*}
For more details about the data reduction (i.e. flatfielding, 
mosaicing, and sky subtraction) and photometric calibration we 
refer to \cite{pohlen2001,pohlen2004b}.
Details of the NIR data reduction can be found in 
\cite{olof}. 

\subsection{Preparation of the images} 
\label{prepa}
\begin{figure*}
\includegraphics[width=5.0cm,angle=270]{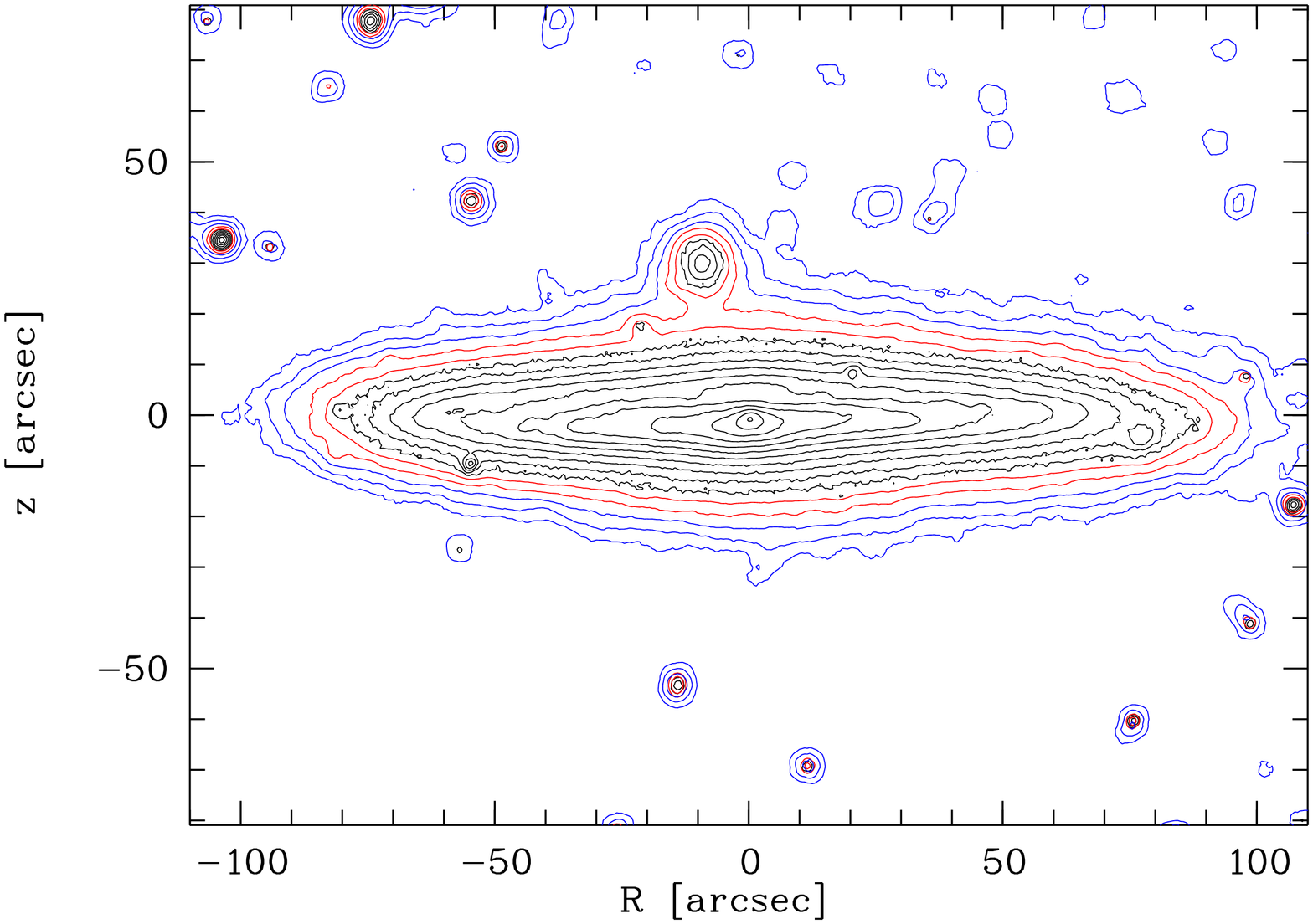}   
\includegraphics[width=5.0cm,angle=270]{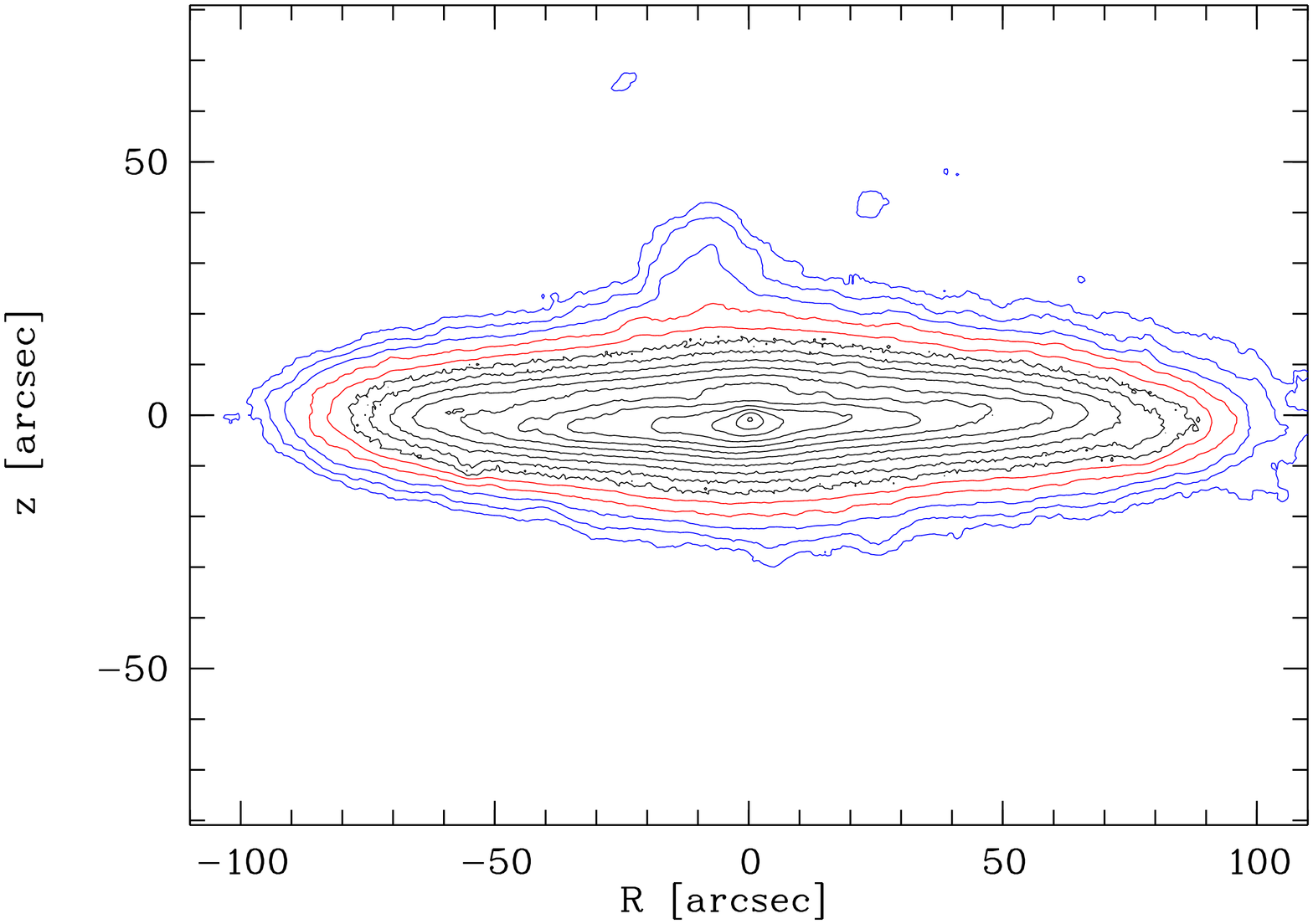}    
\includegraphics[width=5.0cm,angle=270]{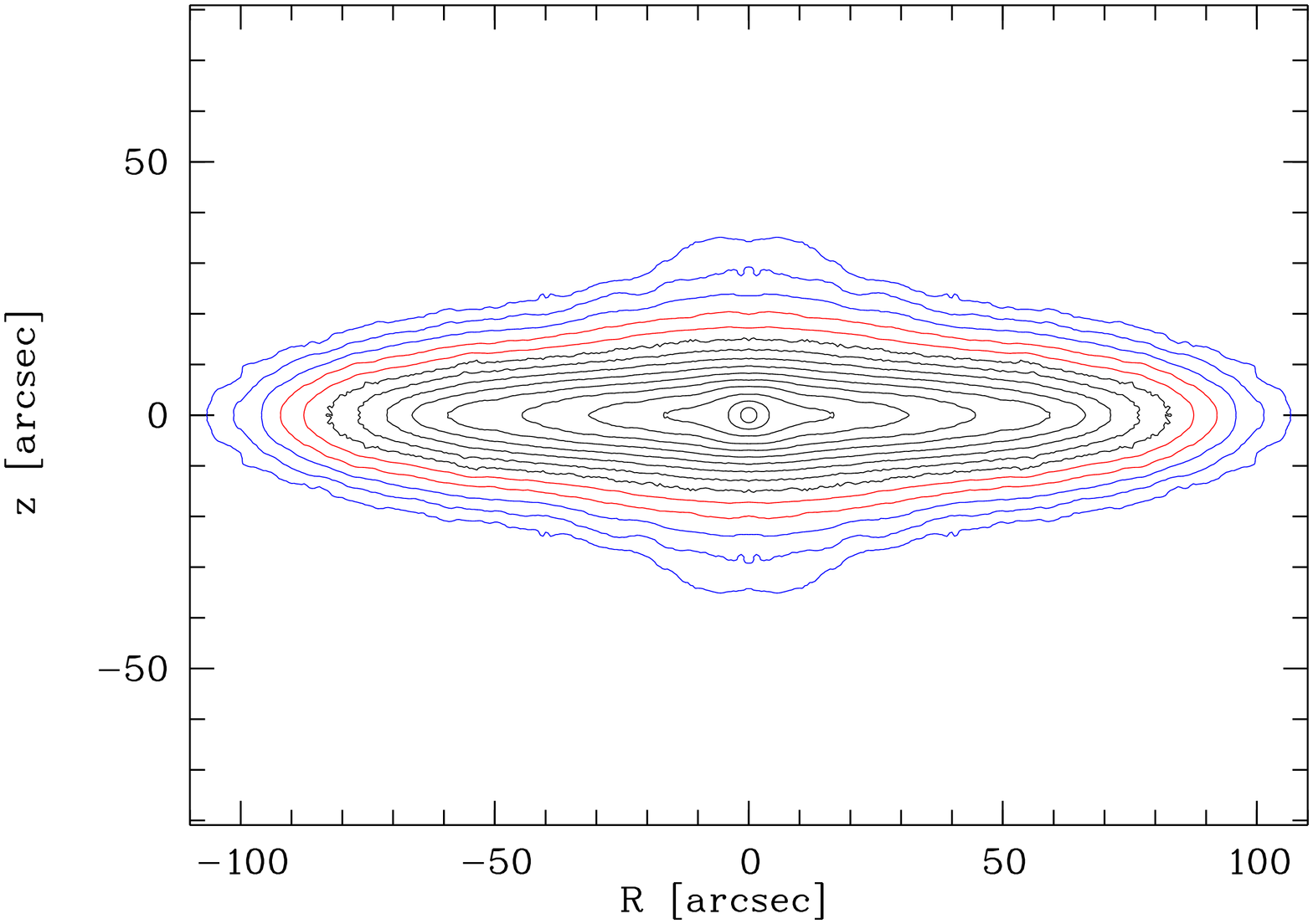}   
\includegraphics[width=5.0cm,angle=270]{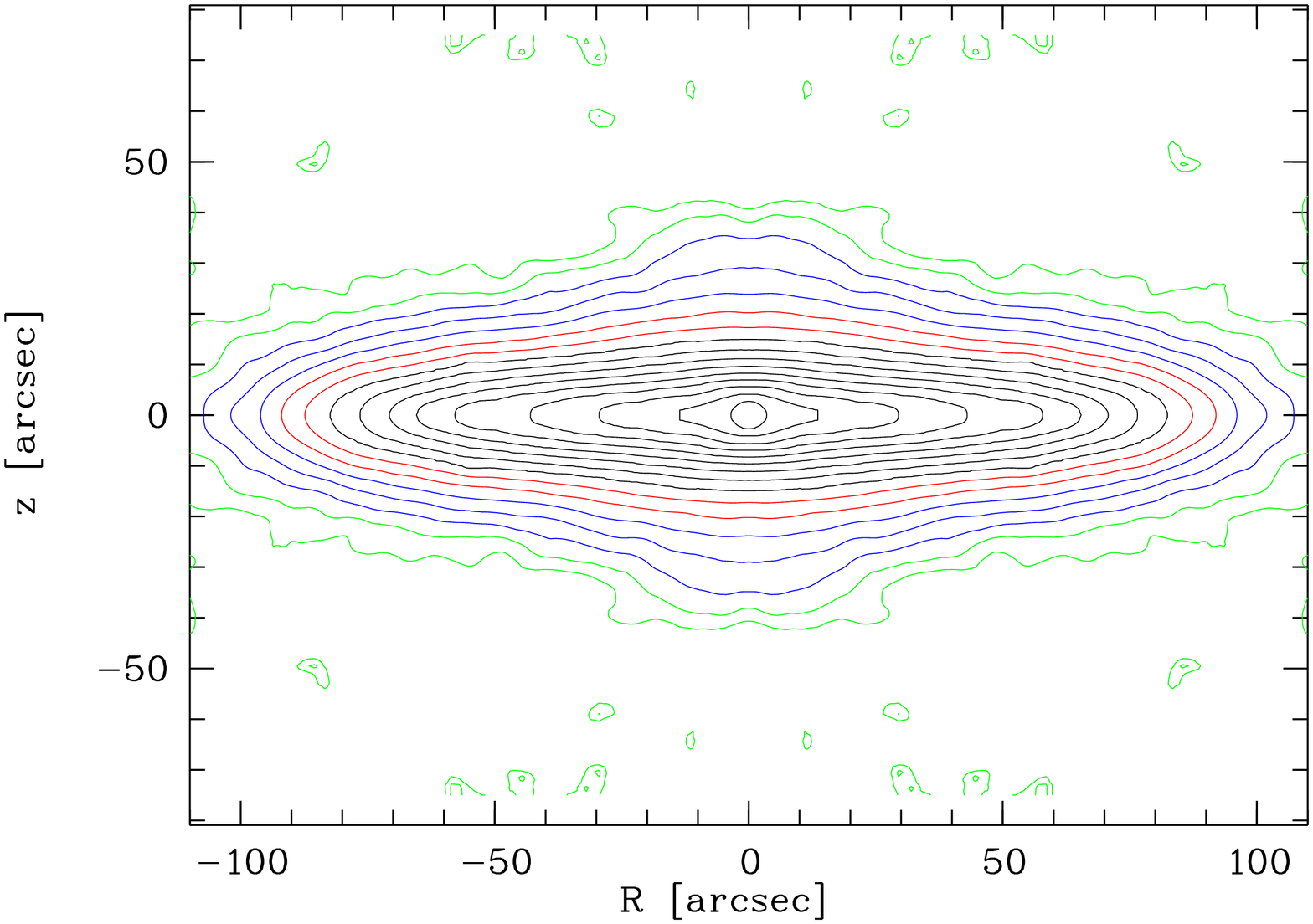} 
\caption{Preparation of the images. \newline Contour maps of NGC\,5981 showing
the four main steps in image preparation.  The {\it upper left panel} shows
the rotated, original image.  The {\it upper right panel} shows the image
after masking of fore-and background sources and interpolating across the
masks.  The {\it lower left panel} shows the galaxy after averaging the four
quadrants.  The {\it lower right panel} shows the galaxy after Gaussian
smoothing.  The equidistant contour levels plotted are in all cases the same,
except for the smoothed version where two additional outer contours are
added.}
\label{prepima}  
\end{figure*}

Before deprojecting our galaxy images we first have to prepare the images,
here we describe the procedure followed in this work (see
\fig\ref{prepima}).
As a first step we mask all the foreground stars and the background
galaxies. To do this semi-automatically, we use SExtractor \citep{bertin1996}
to create an object catalogue. Depending on the measured ellipticity and the
provided star/galaxy classifier we separated stars from galaxies, which in
turn get assigned circular masks with radii depending on their measured flux.
The output is a DS9 region file which is overlayed on the galaxy image and
visually inspected for each of the masked objects. For highly elliptical
background galaxies the mask is changed to an ellipse.
In addition we have to adapt the radii of the masks. 
Typically one opts for a conservative masking (\ie 
prefer to mask more than necessary and leave 
`larger holes') to make sure that all non-galaxy features 
are removed. 

However, since our deprojection method cannot deal with 
masked areas (actual holes) we have to interpolate the 
flux across our mask. By doing this, one is forced to chose 
the mask radius as small as possible in order to not 
introduce too much unwanted structure. 
This approach clearly leaves residuals on the masked, 
interpolated image which will be discussed later (\cf\sec\ref{asymmetries}). 
In a second step the galaxy image is rotated to align the galactic plane
parallel to the major axis using the smallest angle of rotation according to
their true position on the sky (chip). The rotation and centering is done by
blinking the adjacent sides and minimising the difference by eye. Depending on
the intrinsic symmetry this can be done down to an accuracy of $\pm0.5$\degc.
The third step is to cut the galaxy into 4-quadrants
and average them. This procedure increases the S/N 
and minimises the residuals from the interpolated, masked 
regions and gives a more regular averaged galaxy. 
The influence of the dustlane and intrinsic asymmetries 
will be discussed later (\cf\sec\ref{asymmetries}).    
The final step is smoothing the image. Since our deprojection 
code only handles one radial cut at a time, we choose the kernel 
size for the Gaussian smoothing to match the vertical increment between
successive deprojections in order to get fully independent 
cuts but using all available rows.  
%

\section{Method}
\label{method}

In this section we describe and test the method used to recover the 3D
structure of the available disk galaxies. We start with describing the
deprojection algorithm (\sec\ref{deprojmethod}) and discuss the noise
properties of the reconstructed 3D distribution, especially after the original
image has been smoothed with a Gaussian kernel (\sec\ref{noise}). To further
reduce the noise we applied an additional denoising (\sec\ref{denoise}).
In \sec\ref{shape} we discuss in detail our choice of fitting function for the
radial surface brightness distribution. Then we describe how the radial
profiles of the recovered distribution for a given $z$ are actually fitted
(\sec\ref{fitting}). In \sec\ref{tests} we apply the method and the fitting
procedure on a large sample of artificial galaxy images in order to test the
method's ability to recover the underlying 3D distribution. In this set of
artificial images the influence of the inclination of the galaxy and the
existence of dust on the recovery of the 3D distribution is tested.  Finally,
we estimate the effects of our image preparation on the resulting profiles
using real galaxies in \sec\ref{asymmetries}.  
\subsection{Deprojection} 
\label{deprojmethod}

The deprojection of axially symmetric quantities is a classical
problem in astronomy \cite[e.g.,][]{lucy74}. It is considered for
determining the 3D stellar orbit structure in elliptical galaxies
\cite[e.g.,][]{dehnen93,dehnen94,bdi90}, and in determining cluster 3D
structure from X-ray observations \citep{fabricant84} and more
generally from X-ray, Sunyaev-Zeldovich and cluster weak-lensing data
\citep{zaroubi1998,zaroubi2001}.  
In the case of edge-on galaxies, \cite{florido2001,florido2006} have applied 
an inversion method to obtain deprojected profiles before, but without
studying a dependence with the vertical distance from the plane. 
Here we apply a simple analytical method that employs the, so called,
Fourier Slice Theorem in order to recover the intrinsic 3D stellar
distribution from our sample of high quality 2D edge-on disk galaxy
images.
The disk galaxy images are deprojected as follows. We adopt the
convention that bold-face symbols denote 3D quantities (e.g., ${\bf k}
= (k_x, k_y, k_z)$).  Let the observer's coordinate system be defined
with the Cartesian axes $(x,y,z)$, with the $y$ axis aligned with the
line-of-sight.  The disk galaxy axis of symmetry lies along the $z$-axis which
is perpendicular to the line-of-sight.  Let $I(x,z)=\int\lambda({\bf x})\,dy$
denote a projected quantity (image) of the source function $\lambda$.
The 3D Fourier transform  of the source function is related to the
image by
\begin{eqnarray}
\tilde{\lambda}(k_x,0,k_z)   
  &  = &\int e^{[-i(k_x x + k_z z)]}
I(x,z) \, \rd x \, \rd z  \nonumber \\
  & =  & \tilde{I}(k_x,k_z).  \label{eqn:simagereln} 
\end{eqnarray}
Note that functions marked with \ $\,\tilde{}$ \ are in Fourier space.

This relation could be easily understood since the galaxy has a
rotational symmetry, \ie $\lambda({\bf r}) = \lambda(r,z)$ and its
Fourier transform is $\tilde{\lambda}({\bf k})=\tilde{\lambda}(k,k_z)$
where $k=\sqrt{k^2_x + k^2_y}$.  The deprojection is then obtained by
inverse Fourier transform of $\tilde{I}(k_x,k_z)$ and replacing $k_x$
with $k$, namely by using the following equation:
\begin{equation}
\lambda(r,z) = {\frac{1}{(2 \pi)^2}} \int e^{i k_z z}  
              \tilde{I}(k, k_z) J_0(k\, r) 
k\,dk\,dk_z. 
\label{eqn:fouriersliceeqn}
\end{equation}
Here $J_0$ is the zero order Bessel function.

Notice that for a specific radial slice, i.e., a cut at a given $z_0$
and parallel to the major axis, the integral~(\ref{eqn:fouriersliceeqn})
can be written simply as,
\begin{equation}
\lambda(r,z_0) = {\frac{1}{(2 \pi)}} \int  \hat{\cal{I}}
        (k,z_0) J_0(k\, r) 
k\,dk\, . 
\label{eqn:zsliceeqn}
\end{equation}
Where $\hat{\cal{I}} (k, z_0)$ is the Fourier transform over the image
x-direction. This way of writing the equation enables us to easily
deproject certain radial slices of the image, or collapsing the whole
z-direction of the image to recreate the equivalent of a face-on view
of the galaxy.

In principle, it is easy to obtain an equation similar
to~(\ref{eqn:fouriersliceeqn}) for the general case of a disk with
an arbitrary inclination angle. However, one can show that for such
cases there is a cone in k-space inside which there is no information
on the 3D structure in the image.  The general case is beyond the
scope of this paper and is not needed for the range of inclinations we
restrict our selves to i.e., to $i\gtsim 86$\deg (see
section~\ref{tests}). The interested reader is referred to
\cite{zaroubi1998,zaroubi2001}.

Furthermore, one can show that the Fourier Slice Theorem approach for
inverting edge-on galaxies is actually another way of writing the
inverse Abel transform \citep{binney87}. The advantage of the Fourier
space language here is that it avoids taking direct derivatives of
noisy data, hence, making the inversion more stable.

\begin{figure}
\hspace*{-0.4cm}
\includegraphics[width=9cm,angle=0]{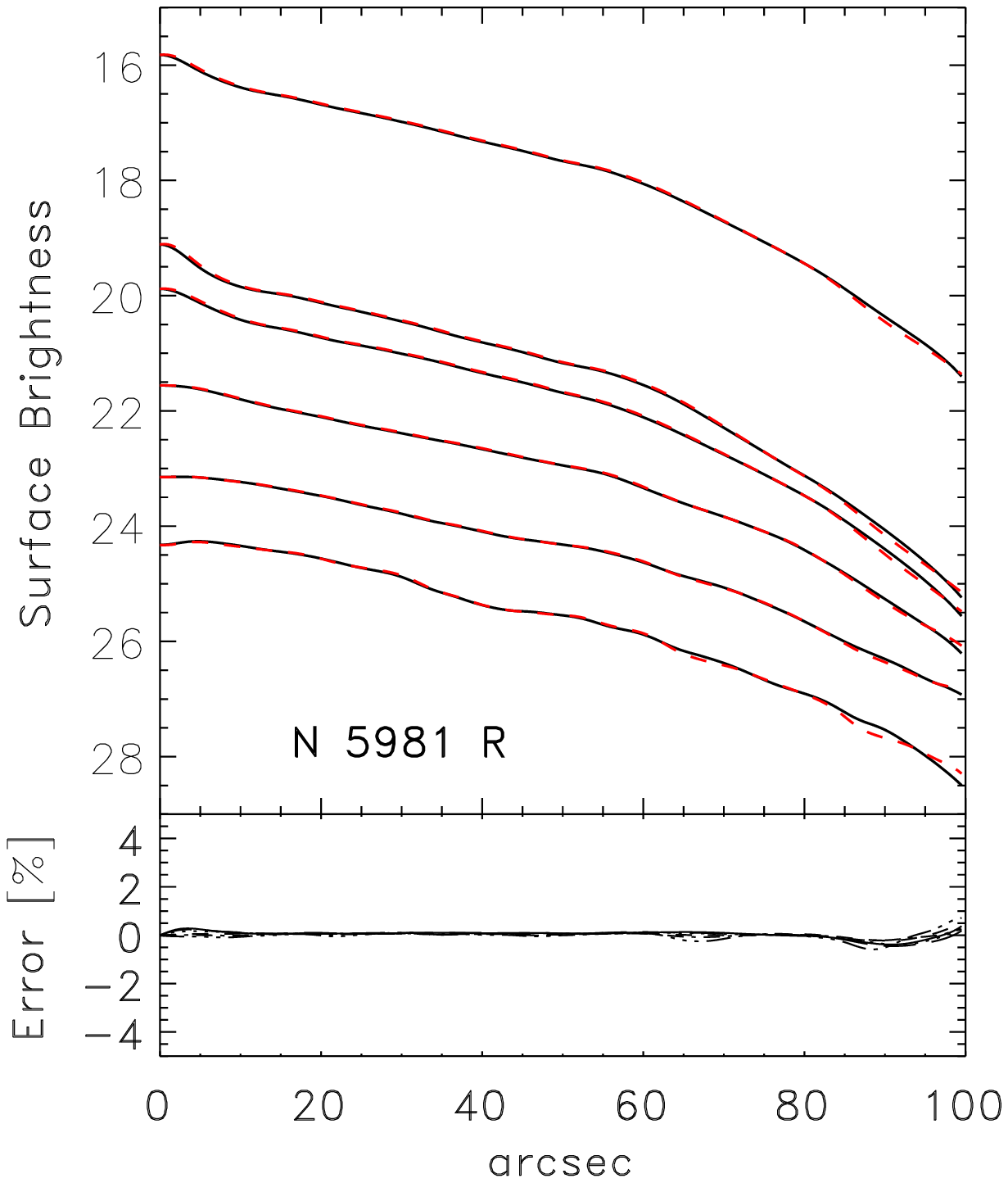}
\caption{Test of deprojection method for NGC\,5981. \newline In the
upper frame the radial profiles of the symmetrised image
{\it (dashed-lines)} and those of the deprojected-reprojected image
{\it (solid-lines)} are shown. The uppermost profile is the vertically
integrated profile, while each of the others represents a cut at
increasing vertical distance from the major axis. The lower frame
shows the percentage error one obtains from this procedure.
\label{deprep}}
\end{figure}

The applied Fast Fourier Transform (FFT) assumes periodic boundary
conditions which has to be satisfied by our images. Therefore, one has
to pad the image with at least twice the number of the image pixels
along each axis. However, since FFT of sharp transitions will give a
strong aliasing signal in the deprojected slice, we do not apply zero
padding. Rather we padded the image with white noise background with
the same rms as the noise in the observed image. The residual small
differences at the edge changing from the artificial white noise to
the actual noise in the background of the image is washed out by the
final wavelet denoising and Gaussian smoothing of the image
(see~\sec\ref{noise} and \ref{denoise}).
Further tests with more careful tapering of the image's edge with a
Fermi-Dirac function without adding noise showed that there is no
significant influence on the deprojected profiles.
As a zero-order test, we deproject the image of NGC\,5981 and
then reproject it back. The result is that we obtain profiles that are
identical to those of the original image (\cf\fig\ref{deprep}) down to
$< 0.5\%$ error level. Most of this error actually comes from the
reprojection -- not the deprojection -- procedure. The reason is simply
that the deprojection is a differential quantity that depends on the
intensity of the image at a given radius, whereas the reprojection is
an integration procedure that will depend on the intensity of galaxy
at every radius including radii beyond those explored in the image
(i.e., after the noise dominates over the signal.)
A more thorough testing of the method in which we compare the
deprojected 3D structure with the original one will be presented in
\sec\ref{tests}.
%
%
\subsection{Noise properties and smoothing} 
\label{noise}
To explore the noise properties, let us suppose that the image only
contains white noise, with rms $\sigma$, \ie $ \langle I (x_1, z_1)
I (x_2, z_2) \rangle = (2 \pi)^2 \delta(x_1 - x_2) \delta(z_1 - z_2)
\; \sigma^2$.  The noise correlation function in the deprojected image
is given by
\begin{equation}
 \langle \lambda(r_1, z_1) \lambda (r_2,z_2) \rangle \nonumber =
\sigma^2\frac{\delta(z_1 - z_2) }{ (2 \pi)} \int k^2 J_0(k r_1) J_0(k
r_2) dk.
\end{equation}
Not surprisingly, the high frequency components dominate and hence some
smoothing must be employed. The analytic nature of the correlation
matrix is an attractive feature of the method. 

Another nice feature of the method is that any isotropic smoothing
function applied to the image will depend only on the modulus of the
radial wavevector, $|{\bf k}|$, in the deprojection.  This can easily
be seen from the previous equations.  A simple and practical Gaussian
smoothing filter, with smoothing scale $R_s$, gives an auto-correlation of
the noise of the form 
\begin{eqnarray}{\cal{N}}(r) & \equiv & \langle \lambda(r, z)^2 \rangle \nonumber\\ & =& { \,
_2F_2\left(\frac{1}{2},\frac{3}{2};1,1;-\frac{r^2}{{R_s}^2}\right)}/\left({8\sqrt{\pi
}{R_s}^{5}}\right) \, ,
\end{eqnarray}
where $_2F_2$ is a hyper-geometric function. This has a particularly
simple asymptotic form,

\begin{equation}
\lim_{r \to \infty}{\cal{N}}(r) \,\propto r^{-1} \, .
\end{equation}
 
Since we are interested in the large scale features of the reconstruction a
Gaussian smoothing is applied on the image.  The choice of the FWHM of the
smoothing kernel is discussed in~\ref{prepa}.

Despite these simple analytic features of the noise properties, we
sample the noise properties of the smoothed 3D structure created by
the method with deprojecting noisy slices of the image (with a given z
well outside the galaxy). For each image we deproject some 10 randomly
selected lines, average their rms and use that as our noise
estimate. The profile of the reconstructed noise rms is shown in every
set of reconstructed profiles (see \fig\ref{dprofpfits}).

The noise is dominated not by pixel to pixel variations, but 
by any residual large scale asymmetries in the background 
as well as residuals from the masking (\cf\sec\ref{prepa}). 
The latter cause the somewhat bumpy structure of the 
resulting deprojected profiles (see \fig\ref{dprofpfits}).   

The seeing acts in effect as an additional Gaussian
smoothing of the image. This type of smoothing flattens the brightness
profile around the center but has very little influence on the
scalelength measurements.
%
%
\subsection{Wavelet Denoising}
\label{denoise} 
In order to reduce the apparent bumpy structures along the radial profiles we
apply an extra denoising algorithm based on wavelet expansion. The employment
of this algorithm is not necessary and we tested that it does not change the
quantitative results of the deprojection in any significant way, it however
reduces the apparent ``bumpiness'' in the radial profiles.
The denoising scheme adopted here follows the method proposed by
\cite{donoho1992} \cite[see also][]{donoho1994, zaroubi2000}
where the wavelet coefficients are reduce with a constant equal
to the rms of the white noise. This procedure is known as the soft
thresholding of wavelet coefficients. The wavelet functions used here
are the Daubechies' discrete wavelets \citep{daubechies1988} with 
12 parameters (DAUB12).
%
\subsection{Profile shape and fitting function}
\label{shape} 

The recent studies of profile shapes for face-on disk galaxies
\citep{erwin2005,erwin2006,pohlen2006} and irregular galaxies
\citep{hunter2006a} suggest that the {\it broken exponential} is a
good fitting function for the radial light distribution
(\cf\sec\ref{introduction}).
Technically, this results in a discontinuity between the inner and outer
slope, which is unphysical and of course not
implied. \cite{pohlen2006} find an extended {\it transition or break
region}, which is not as sharp as measured before \cite[\eg
by][]{pohlen2002a}. From their sample only a few clean and undisturbed
truncated cases (\eg NGC\,5300 or NGC\,7437) exhibit a very sharp
transition between the inner and outer disk region. Most often their
\typet breaks appear with a more gradual transition zone of a size up
to $\sim\!4$\,kpc.
Despite the fact that the exponential function in itself has no clear
physical explanation, the broken exponential is a reasonable fitting
function that describes the shape of the profiles in a simple and
uniform way.
The key point of this description is that the {\it outer disks} of
truncated galaxies -- regardless of any transition zone and the exact
shape of the outer profile -- makes up a significant fraction of the
physical size of the their stellar disks. For example, the ratio
between the inner and the outer disk is typically 2 to 1 in radius
\citep{pohlen2006}, therefore in terms of area the outer disk is even
slightly larger.
In some cases authors ignore the outer region altogether 
\cite[e.g.,][]{vdk1981,pohlen2000b} and use an exponential with a sharp {\it
cut-off} as their fitting function of choice.
However, this is not the case for edge-on galaxies in general as clearly
apparent from the examination of the radial profiles of, \eg
NGC\,5907 in \cite{vdk1981} or IC\,4393 in \cite{pohlen2000b}. We see
the same for several galaxies used in our study. For example, the
major axis profiles of ESO\,380-019, ESO\,404-018, NGC\, 5981,
UGC\,10459, and NGC\,522 (see \fig\ref{mapspprofs} in
\apx\ref{figures}) are obviously better described by a broken
exponential fit \citep{pohlen2001, pohlen2002b}.
Many further examples in the literature \cite[\eg
see][]{jensen,sasaki, naeslund,byun,pohlen2001,degrijs2001} show that
the broken exponential shape is indeed a very good fitting function
even for edge-on galaxies.  However, we should mention that some
authors \cite[e.g.][]{florido2001,florido2006}, still prefer the
description of a smooth, but complete truncation.

To summarise, since the profiles of face-on galaxies are well
parameterised by a broken exponential model, and those of edge-on
galaxies seem to be, it is useful to apply the same parametrisation
for the profiles of the edge-on galaxies studied here.
%
%
\subsection{Parameter fitting}
\label{fitting}

In order to quantify the behaviour of the recovered 3D stellar
distribution of the galaxies, we fit the deprojected radial profiles, for
a given $z$, with a simple broken exponential function of the form 

\[ \mu(R) = \left\{ \begin{array}{ll}
        \munin\;\, + 1.086 * R/\hin & \mbox{if $ R \le $\rbr};\\
        \munout + 1.086 * R/\hout & \mbox{if $R > $\rbr}.\end{array} \right. \]

In principle, the model includes 5 free parameters, the inner and
outer scalelengths $\hin$ and $\hout$, the inner and outer central
surface brightnesses, $\munin$ and $\munout$ and the location of the
break radius, $\rbr$, which marks the boundary between the inner and
outer regions.  However by using the downhill simplex method
\citep{nelder65} with initial starting values of only the scalelength
and central brightness parameters we avoid using the break radius as a
free parameter.  This is achieved by deciding \textit{a-priory} that
the inner slope is either shallower or steeper than the outer one.
This enables the break radius to be determined by the crossing point
of two exponential fits found by the algorithm.
We find that in all cases, except NGC\,1596, the inner profile is
shallower than the outer profile. In addition, when the ratio between
the two fitted radial scalelengths is about $1\pm 0.2$, we fit it
with a single exponential profile of the form $\mu(R) = \mun + 1.086 *
R/h$.

In order to avoid including the bulge and/or bar regions in the
fitting procedure, i.e., avoid a detailed bulge to disk decomposition,
we used the ``marking the disk'' method with fixed inner, $b_1$, and
outer, $b_2$, boundaries \cite[see e.g.][]{pohlen2006}.
The inner boundary  is chosen to exclude the region that is
obviously dominated by the bulge plus bar (shoulder) component.
The outer boundary is set at the point at which the surface brightness
profile is comparable to the noise level (see \sec\ref{noise}).
In the case of an extended bump clearly visible in the profile, caused
for example by a residual from the masking, the boundary is moved
further inside.
Sometimes a sharp rise or drop appears beyond the outer boundary
of the recovered profiles (see, \eg the face-on equivalent profiles of
IC\,4871 or NGC\,5290 R-band in \fig\ref{dprofpfits}).  This is caused
by the Fast Fourier transform (FFT) operation used in the method. The
FFT procedure requires periodic boundary conditions. To avoid
contamination of the profile at hand by these conditions we increase
the size of the image by a certain factor. However, this periodicity
still shows up in the noise dominated regions and sometimes close 
enough to, but never within, the region of interest.

The five starting values for the downhill simplex minimisation are
obtained by a single exponential fit to the inner or outer region with
an approximate break radius for each galaxy.
%
\subsection{Tests with artificial data}
\label{tests}
\subsubsection{Model grid}
To study the influence of the dust and possible non-$90\deg$
inclination on the scalelength measurements in our deprojected
profiles we used a grid of 96 mock images (400x400 pixels) kindly
provided by Simone Bianchi. Using a Monte Carlo radiative transfer
method \citep*{bianchi96}, he calculated single disk models at a range
of edge-on inclinations ($80\degc\!-\!90\degc$) with different input parameters
for the stellar and dust distribution.
A similar approach is adopted by \cite*{moellenhoff2006} to measure
the influence of dust and inclination for less inclined 
galaxies.  
\begin{figure*}
\includegraphics[width=18cm,angle=0]{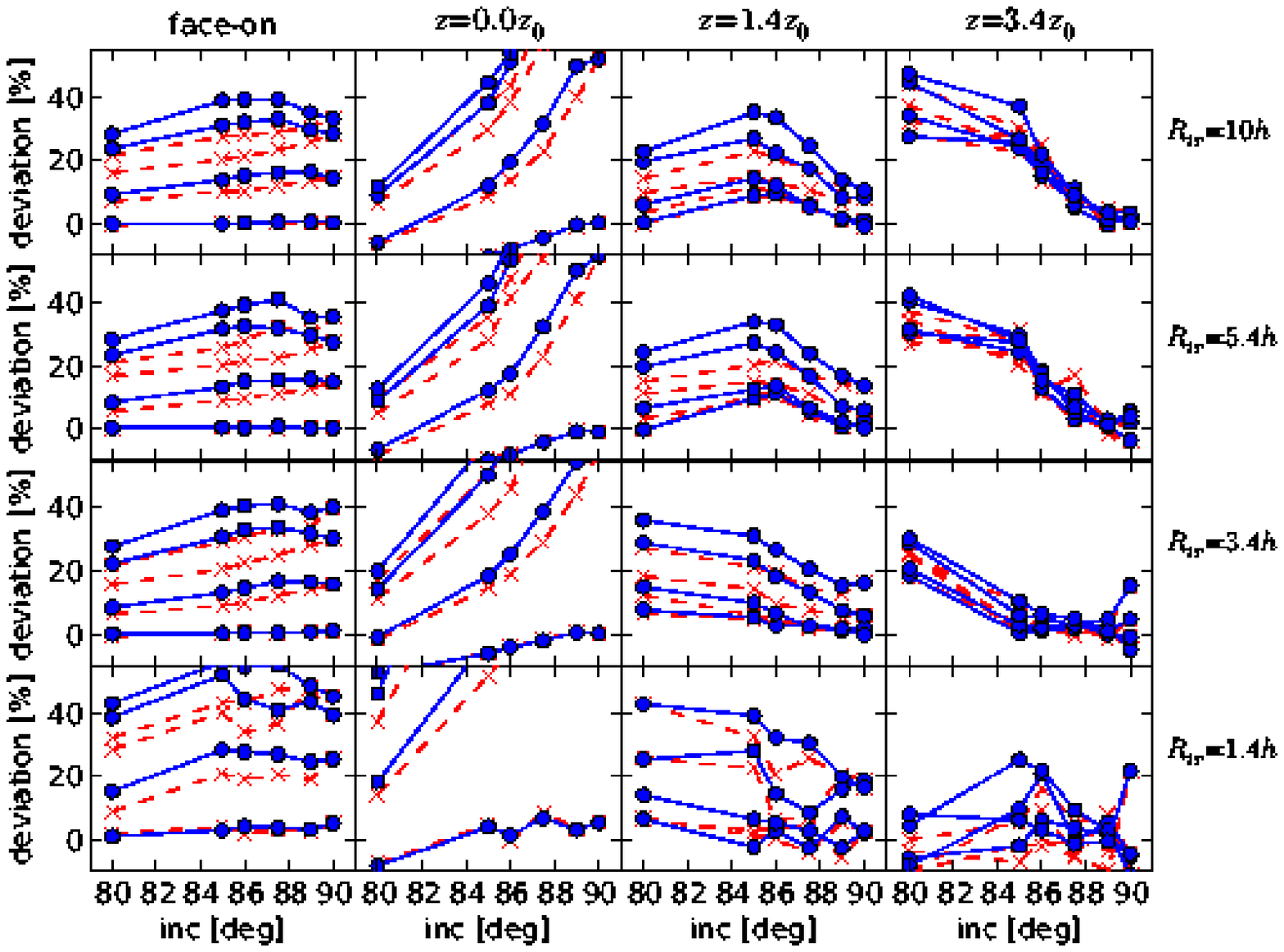}   
\caption{The systematic deviation of the measured scalelength with inclination
and dust influence. \newline Plotted is the deviation of the measured from the
known input scalelength (in percent) from fitting the 96 mock images with the
same setup compared to the real galaxies. From left to right the {\it four
columns} are the results for the face-on-equivalent profiles, for the major
axis, and for two cuts at $\!1.7$ and $3.4\!\cdot\!z_0$. The {\it four rows}
correspond to the different sharply truncated stellar input models, from top
to bottom: not-truncated (mimicked by truncating at $10 h$), long truncated
(truncated at $5.4 h$), intermediate truncated (truncated at $3.4 h$), and
finally short truncated (truncated at $1.4 h$).
The different symbols represent the two different ways to 
obtain an average quadrant: {\it filled circles} for the 
applied mean average out of four, {\it crosses} for the 
average out of two quadrants, excluding the dusty side. 
The different lines connecting the points in each panel
are associated to the different dust distributions (ordered
according to influence on the stellar disk). 
From bottom to top we have: the transparent case, the best 
case, an intermediate case, and finally the worst case.
See text for more details.
\label{dustfit}
}
\end{figure*}
\begin{figure*}
\includegraphics[width=5.0cm,angle=270]{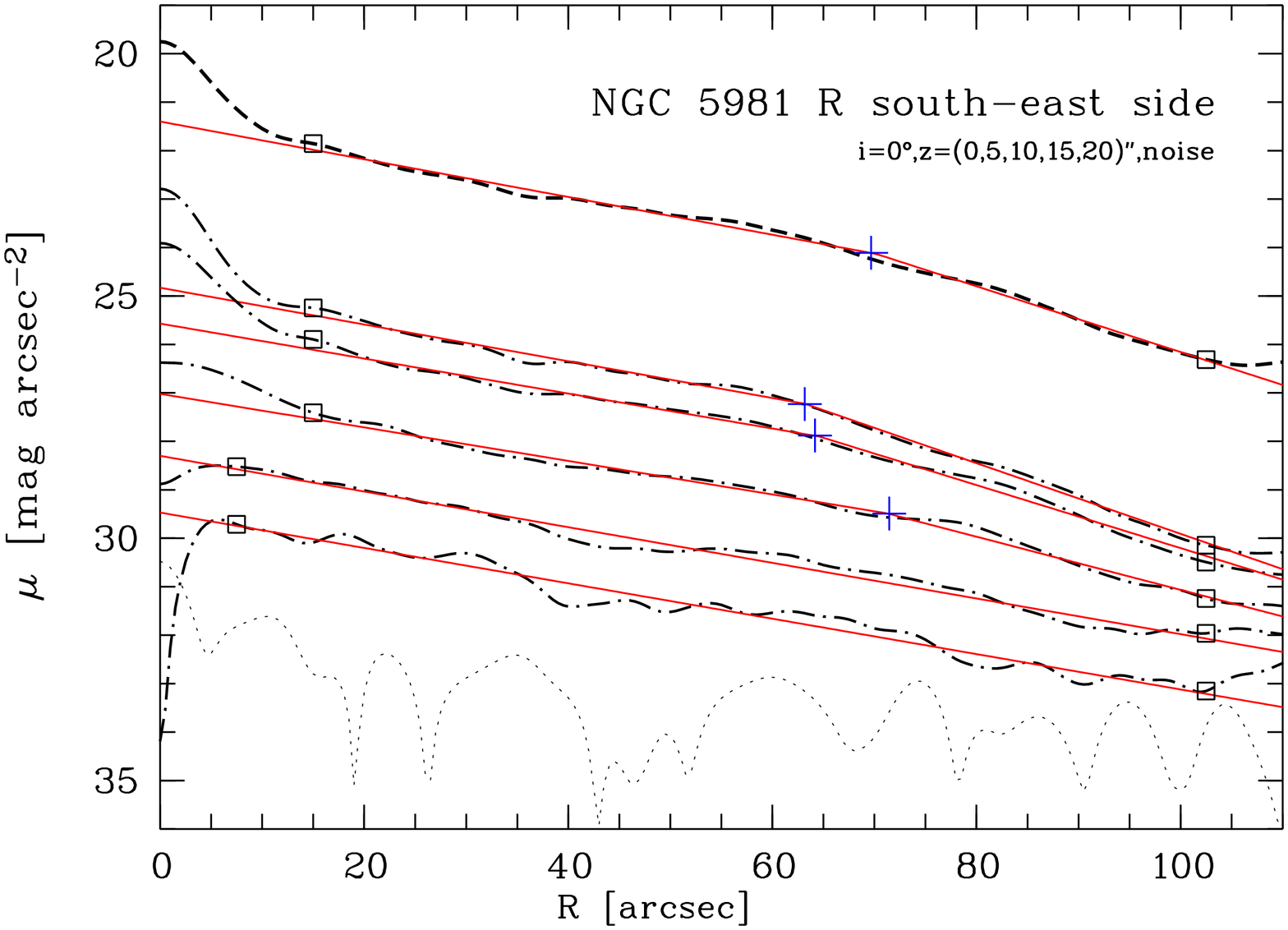}
\includegraphics[width=5.0cm,angle=270]{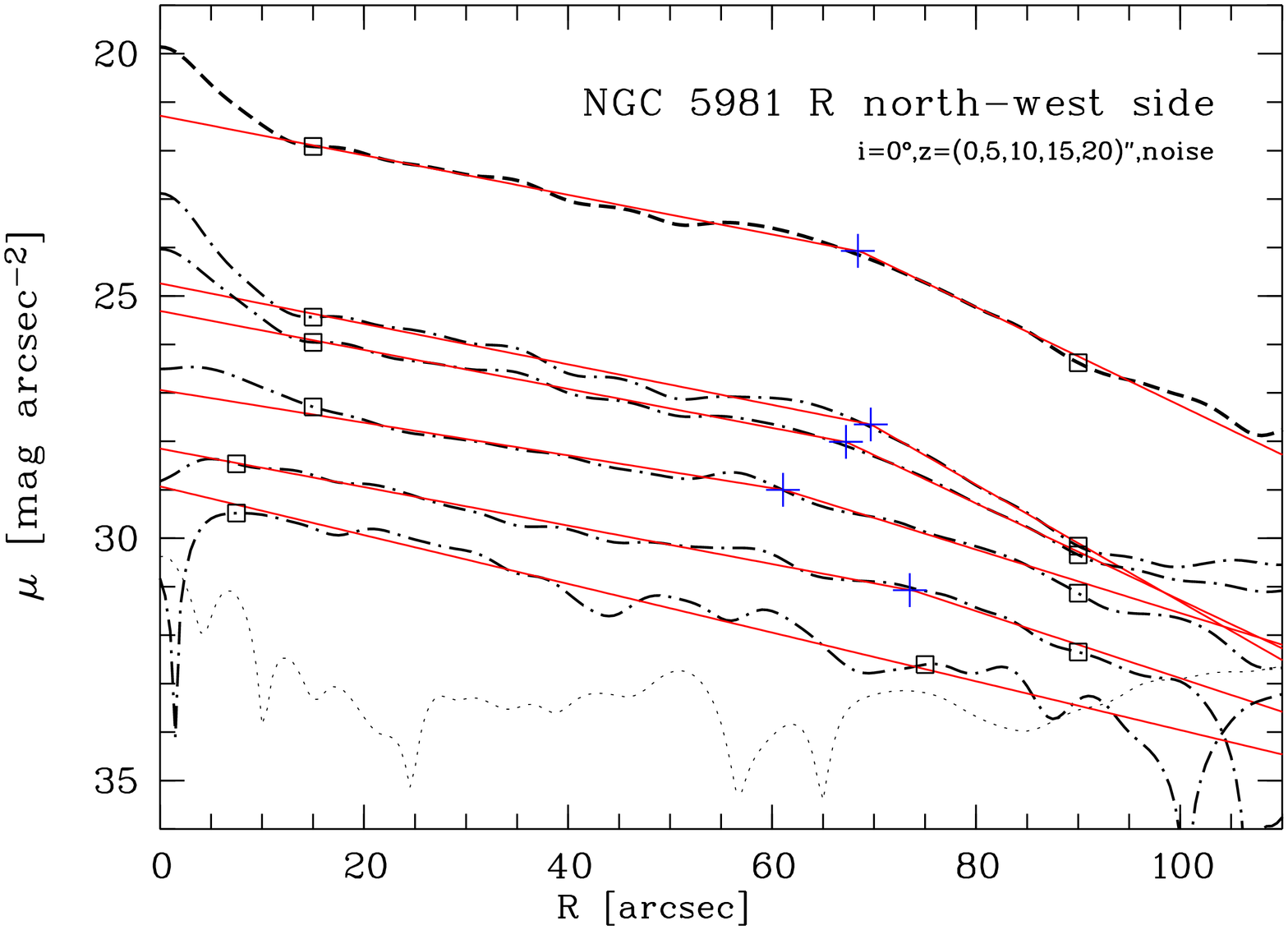}    
\includegraphics[width=5.0cm,angle=270]{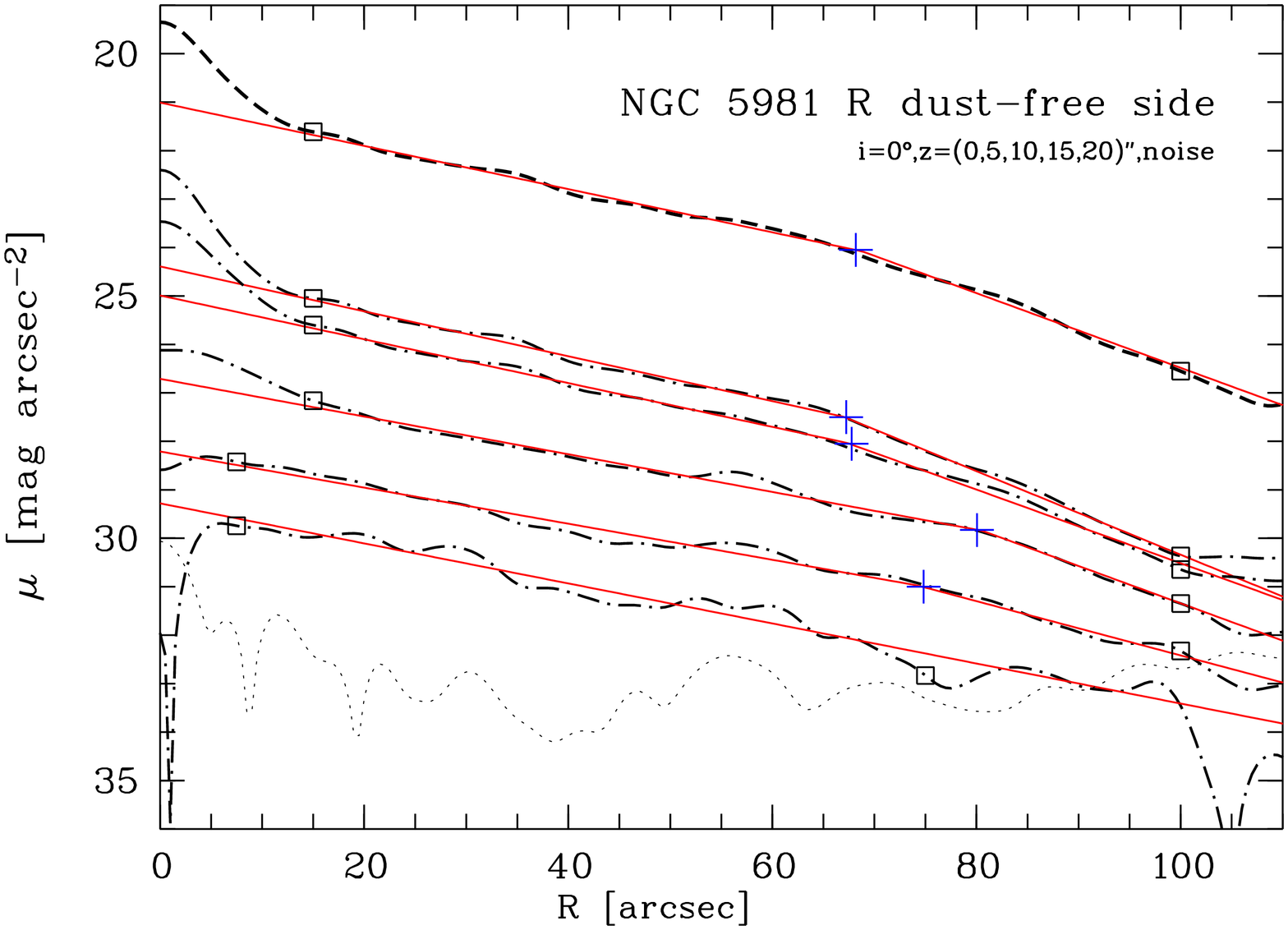}   
\includegraphics[width=5.0cm,angle=270]{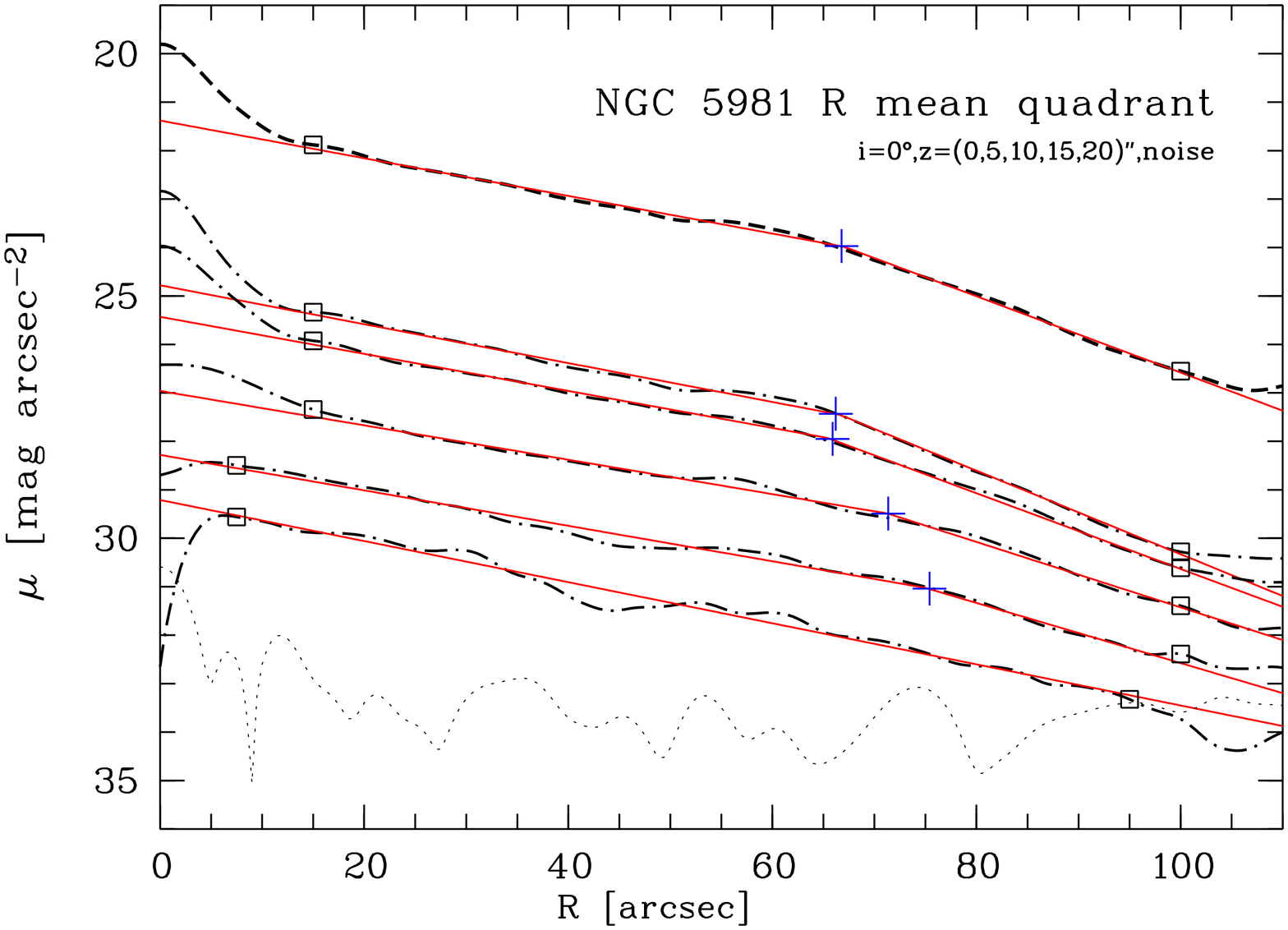} 
\caption{ 
Influence of asymmetries: Averaging quadrants \newline 
The deprojected surface brightness profiles of NGC\,5981 using four different
ways to combine the four quadrants. The {\it upper left} and {\it upper right
panel} show the result using only the two quadrants of the south-east and
north-west side respectively. The {\it lower left panel} shows the profile
using only a mean of the two dust-free quadrants. The {\it lower right panel} 
finally shows the resulting profiles after averaging all four quadrants.  
In each panel, the {\it dashed line} (the upper profile) is 
the face-on-equivalent, radial profile of
each galaxy. They are obtained by summing the light of the 
edge-on galaxy along the columns and deproject the result. 
The following {\it dotted-dashed lines} are the deprojected 
profiles of individual cuts from top to bottom starting from 
the major axis towards cuts at larger distance from the plane.
The proper distance $z$ from the plane is given in the caption
(together with name and filter) of each panel in the upper right 
corner. 
The {\it dotted line} (the lower profile) is the deprojected 
profile of a noise (or source free) cut outside of the 
galaxy. 
Overplotted are on each profile two {\it squares} marking the 
fitting region for the disk model. 
The result of the broken exponential fits are overplotted 
on each profile as {\it solid lines} where the {\it cross} marks 
the break radius.} 
\label{asym}  
\end{figure*}

%
The chosen inclination angles are $80\degc$, $85\degc$, $86\degc$,
$87.5\degc$, $89\degc$, and $90\degc$.
The stellar disk is a 3D single disk model with exponential profiles
both along the radial and vertical direction. The radial scalelength
is $h\!=\!33.4$ pix, and the vertical scaleheight is $z_0\!=\!0.125 h$.
The only parameter of the stellar distribution that is varied is the position
of an outer (sharp) truncation of the light distribution. The
truncation parameter has 4 values, $10 h$, $5.4 h$, $3.4 h$ and $1.4
h$.  The first value mimics an infinite exponential profile, and the
other three values represent a long truncated, an intermediate
truncated, and a short truncated stellar disk.

For the dust distribution a set of four different cases with
increasing influence on the measured R-band light distribution are
adopted. The parameters for these cases are within the range measured
by \cite{xil99}.  The first case is a dust free (transparent) model
which we use as our reference case.  The second case, which we call
the ``best case'', has the following dust parameters: $\tau_{\rm
R}\eq0.20$, $h_{\rm d}/h\eq1.0$, and $z_{\rm d}/z_0\eq0.30$. Here $\tau_{\rm
R}$ is the R-band optical depth, $h_{\rm d}$ is the dust scalelength,
$z_{\rm d}$ is the dust vertical scaleheight and $z_0$ is the stellar
scaleheight. The parameters of the third case, called the
``intermediate case'', are $\tau_{\rm R}\eq0.54$, $h_{\rm d}/h_{\rm
*}\eq1.4$, and $z_{\rm d}/z_{\rm *}\eq0.55$. Finally, the parameters of
the fourth case, which we call the ``worst case'' are $\tau_R\eq0.65$,
$h_{\rm d}/h_{\rm *}\eq1.5$, and $z_{\rm d}/z_{\rm *}\eq0.8$.
For the sake of simplicity the dust disk is truncated at the same
radius as that of the stars. In addition, we have tried several
other cases with variable dust truncations, however those did not show
any significant difference with respect to the ones presented here.
The number of images used to test the dust and inclination influence
on the reconstruction is then doubled (except in the 90$^\circ$ cases)
by using two different methods of symmetrising the far and near side
of the galaxy, insofar as the dustlanes are concerned. In the first
case we average all four quadrants, whereas in the second we only
average over the two far side quadrants which have less dust
influence and ignore the two near side quadrants with the dust
lane.
The artificial images are 3x3 pixels carbox smoothed.  The
deprojection is carried out on the vertically integrated profile,
hereafter, the face-on-equivalent view, as well as on the major
axis and two higher cuts at $z\eq1.7 z_0$ and $z\eq3.4 z_0$.
Since the input profile in this case is sharply truncated, we apply
only simple exponential radial fits ranging from the center to 3 times
the radial scalelength, except for the short truncated disk where we
go out to one scalelength.  From these fits we obtain the scalelength
and central surface brightness of the deprojected 3D distribution.
%
%
%
%
%
\subsubsection{Influence on the measured parameters}
To visualise the influence of the various dust distributions we
plotted the relative deviation in the recovered scalelength as a
function of the inclination angle (\fig\ref{dustfit}). Here the
relative deviation is defined explicitly as
$100\times(h_{\rm true}\!-\!h_{\rm recovered})/h_{\rm true}$.

From \fig\ref{dustfit}, it is obvious that using only the two
dust-free quadrants, shown as dashed-crossed lines, improves the
quality of the fit. However, above $88\deg$ the gain is only minimal
and countered by the advantages of an increased S/N -- especially
important to reach large vertical distances --  and by minimising the
residuals from masked objects.
Therefore, for the image preparation of the real data we average over all four
quadrants, even in the clear presence of dustlanes.

As expected our deprojected, face-on-equivalent, scalelengths are very
close to the true, 3D input scalelength in the absence of dust. There
is almost no influence of the non edge-on geometry (down to $80\degc$)
of the actual model galaxy which we treat in our deprojection to be
always at $90\degc$.
For the model galaxies with dust, we see that the error 
on the scalelength is almost always systematically 
smaller for galaxies with increasing deviation from the
perfect edge-on geometry.    

We also notice, from the dust-free case, that for the
short-truncated disk the deprojection improves the fits quite noticeably,
since it disentangles the truncated part of the profile and avoids
fitting for it, as is the case with the projected edge-on view.
For the latter we know that they result, slightly depending 
on the region used for fitting, in a systematic overestimation 
of the scalelength in the range of $20\%$ \cite[\eg][]{pohlen2001, 
kregel2002,kregel2003}. 
However, we note that our face-on-equivalent scalelengths are still
systematically overestimated by typically, $15\%\!-\!30\%$ due to neglecting
the existence of the dust component.
Naturally the effect is far the worse on the individually 
deprojected major axes (\hin overestimated by more than 50\%),
which probably make only sense for dust-deficient \s0 
galaxies or in the dust-reduced NIR bands. 
However, the narrowness of the dustlane works in our advantage, 
so that above $1.4$ times the vertical scaleheight the error 
in the measured scalelength already drops to less than 
$10\%$ (for $i\gtsim87\degc$). 
This means the scalelength measured on the individually 
deprojected higher vertical cuts are much better.

The discussion of the results that we obtain from the observed
galaxies takes the effects of the dust and inclination, 
that we learned from this section, into account. 
%
%

\subsection{Influence of intrinsic asymmetries}
\label{asymmetries}
\textit{Dustlanes:} In the last section we have estimated the
influence of the dust on the measured flattening of the inner
scalelength using our artificial images. Aside from a solely
quantitative correction to be made to the measured scalelength
in this case, one could argue that the dust in addition might be
responsible for the observed broken exponential structure. In this
picture the break would correspond to the point at which the dust lane
becomes less important. However, in almost all cases the location of
the break does not coincide with the, admittedly difficult to
estimate, location of the visible end of the dustlane on the image. An
obvious exception is the major axis profile ($z\!=\!0\arcsec$) of
NGC\,3390 where the dustlane extends out to $R\!\approx\!\pm55\arcsec$
and indeed causes an apparent break feature at this point (see
\fig\ref{dprofpfits}), hence it is excluded from the fit.
In addition, we still find profiles showing a broken exponential
structure at more than two times the stellar scaleheight above the
disk, where the effect of the dustlane is minimal. Also NGC\,4179 does
show a break, but has no dustlane at all.
Furthermore, if one assumes that the outer slope is the
intrinsic, dustfree slope, the extrapolation into the center will in almost
all cases come up too high (much brighter than the bulge peak). This would
require far too much dust in these galaxies.
Therefore, the dust quantitatively affects the measured inner
scalelengths, however, is not responsible for the break in the profiles
(\cf\fig\ref{asym}).
\noindent\textit{Intrinsic small scale asymmetries:} 
For real galaxies the residuals from masked objects will certainly affect the
recovered profile. The problem is ostensibly aggravated by the averaging
process of the four quadrants (\cf\sec\ref{prepa}), which will mirror any
residual on one, to all other sides.  However, one has to keep in mind that
our deprojection method cannot deal with `holes', so we have to somehow fill
these masked regions.  For example, in the case of NGC\,5981 the extended
companion superimposed on the upper left side of the bulge could be well
traced in the contour lines of the averaged galaxy image (see
\fig\ref{prepima}). Under these worst case circumstances we made sure that
this area is excluded from our fitting region. In the latter case we did
not fit beyond $z\!>\!20$\arcsec.

In addition, the final position of the break radius is strongly affected by
the small scale asymmetries (star-forming regions, dust concentrations,
residuals from masked objects) which vary in each of the vertical profiles.
Whereas this only slightly affects the measured scalelength, 
the meeting point of the two exponential fits (break point) 
is more susceptible to such variations.

\noindent\textit{Large scale asymmetries:} We do not expect significant large
scale asymmetries in our galaxies, since the input sample is morphologically
selected to be symmetric. NGC\,5981, which is asymmetric relative to the rest
of the sample, is a good test case to estimate the effect of averaging
quadrants from both sides. The asymmetry is clearly seen in the radial
profiles (\cf\fig\ref{mapspprofs}). The outer slope of the broken exponential
fit is significantly shallower on the NW(right)-side, where also the contour
map is more extended.
Fitting only a mean quadrant obtained from the NW(right)- or SE(left)-side of
the galaxy separately (\cf\fig\ref{asym}) we find differences of
$5\%\!-\!10$\% for the scalelengths of the inner region. In the outer disk
(beyond the break) the differences are clearly larger, \eg 40\% on the major
axis, due to the stronger asymmetry in the outer disk.  However, both sides
are equally well fit with the broken exponential function, so we do not
introduce a bias on the validity of the applied fitting function by averaging
the two sides.  
%
%
\section{Results} 
\label{results}
Now we move on to present the results obtained by applying the
deprojection method on our observed edge-on galaxies. Again here we
first apply the method on the vertically integrated image. This
results in a profile that is equivalent to the one obtained from
azimuthal averaging, for example by fits with concentric ellipses, of
face-on galaxies.  A direct comparison between the two views (see
\sec\ref{deprojmethod}) is carried out in \sec\ref{faceon}.

Furthermore, we deproject a number of vertical slices to explore the
existence and properties of, for example, a thick disk
component. Here the following questions are addressed: does the inner
slope flattens with increasing $z$, as expected for some
thick disk models? Does the break radius, if present, vary
significantly with $z$?  Finally, how does the vertical distribution
of the outer disk, beyond the break point, changes with $z$? These
question are addressed in \sec\ref{vertical}.

The radial profiles obtained from deprojecting all the galaxies in the
sample, together with the best fit broken exponential models are shown
in \fig\ref{dprofpfits}.
The results for the face-on-equivalent profile of each galaxy in the
sample are tabulated in \tab\ref{tabresults}.
\begin{table*}
\begin{center}
{\normalsize
\begin{tabular}{ l c c rrr c r cccc}
\hline
\rule[+0.4cm]{0mm}{0.0cm}
Galaxy
& Filter
& Profile
&$b_1$
&$b_2$
&\rbr
&\hin
&\hout
&\rbr
&\hin
&\mubr
&$A_{{\rm R,V,K}^{\prime}}$
\\[+0.1cm]
& 
& type
&[\arcsec]
&[\arcsec]
&[\arcsec]
&[\arcsec]
&[\arcsec]
&[\hin]
&[\hout]
&[\magsqarcsec]
&[mag]
\\
\rule[-3mm]{0mm}{5mm}{\scriptsize{\raisebox{-0.7ex}{\it (1)}}}
&{\scriptsize{\raisebox{-0.7ex}{\it (2)}}}
&{\scriptsize{\raisebox{-0.7ex}{\it (3)}}}
&{\scriptsize{\raisebox{-0.7ex}{\it (4)}}}
&{\scriptsize{\raisebox{-0.7ex}{\it (5)}}}
&{\scriptsize{\raisebox{-0.7ex}{\it (6)}}}
&{\scriptsize{\raisebox{-0.7ex}{\it (7)}}} 
&{\scriptsize{\raisebox{-0.7ex}{\it (8)}}}
&{\scriptsize{\raisebox{-0.7ex}{\it (9)}}}
&{\scriptsize{\raisebox{-0.7ex}{\it (10)}}} 
&{\scriptsize{\raisebox{-0.7ex}{\it (11)}}} 
&{\scriptsize{\raisebox{-0.7ex}{\it (12)}}} 
\\[-0.2cm]
\hline\hline \\[-0.4cm]
   NGC\,522  &R&II&22 & 87  & 63.1 & 52.7 &  8.7 & 1.2 & 6.1 & 22.8 & 0.23 \\
   NGC\,1596 &R&III&30 & 210 & 100.2& 27.1 & 53.8 & 3.7 & 0.5 & 24.4 & 0.03 \\
   NGC\,3390 &R&I &35 & 140 & $-$  & 26.9 & $-$  & $-$ & $-$ & $-$  & 0.21 \\
   NGC\,4179 &V&II &45 & 155 & 97.7 & 30.9 & 22.6 & 3.2 & 1.4 & 23.3 & 0.11 \\
ESO\,380-019 &V&II &12 & 117 & 87.6 & 27.9 & 11.5 & 3.1 & 2.4 & 22.9 & 0.26 \\
  NGC\,5290&\Km&II &30 & 110 & 82.5 & 31.2 & 13.6 & 2.7 & 2.3 & 21.0 & 0.00 \\
   NGC\,5290 &R&II &30 & 125 & 79.9 & 33.1 & 16.3 & 2.4 & 2.0 & 23.7 & 0.02 \\
   NGC\,5981 &R&II &15 & 100 & 66.8 & 27.9 & 13.8 & 2.4 & 2.0 & 24.0 & 0.07 \\
  UGC\,10459 &R&II &10 & 59  & 44.2 & 15.6 &  5.7 & 2.8 & 2.7 & 25.3 & 0.03 \\
    IC\,4871 &V&II &20 & 136 & 116.5& 40.4 &  9.7 & 2.9 & 4.2 & 24.4 & 0.27 \\
   FGC\,2339 &R&II &37 & 66  & 57.3 & 17.6 &  5.9 & 3.3 & 3.0 & 26.4 & 0.10 \\
ES0\,404-018 &V&II & 0 & 92  & 75.6 & 34.0 &  9.6 & 2.2 & 3.6 & 24.6 & 0.13 \\
\hline
\end{tabular}
}
\caption[]{Exponential disk parameters for the face-on-equivalent profiles. \newline
{\scriptsize{\it (1)}} Principal name, 
{\scriptsize{\it (2)}} filter,
{\scriptsize{\it (3)}} profile type (I: no break, II: downbending break, III: upbending break),
{\scriptsize{\it (4,5)}} inner and outer fitting boundaries,
{\scriptsize{\it (6,7,8)}} break radius, inner, and outer scalelength in units of arcsec, 
{\scriptsize{\it (9)}} break radius in relation to the inner scalelength,
{\scriptsize{\it (10)}} inner scalelength in relation to the outer scalelength,
{\scriptsize{\it (11)}} the surface brightness at the break radius (estimated 
at the crossing point of the two exponential fits),  
{\scriptsize{\it (12)}} galactic extinction for the appropriate band from NED according to 
\cite*{schlegel}.
\label{tabresults} 
}
\end{center}
\end{table*}
\begin{figure*}
\includegraphics[width=6cm,angle=270]{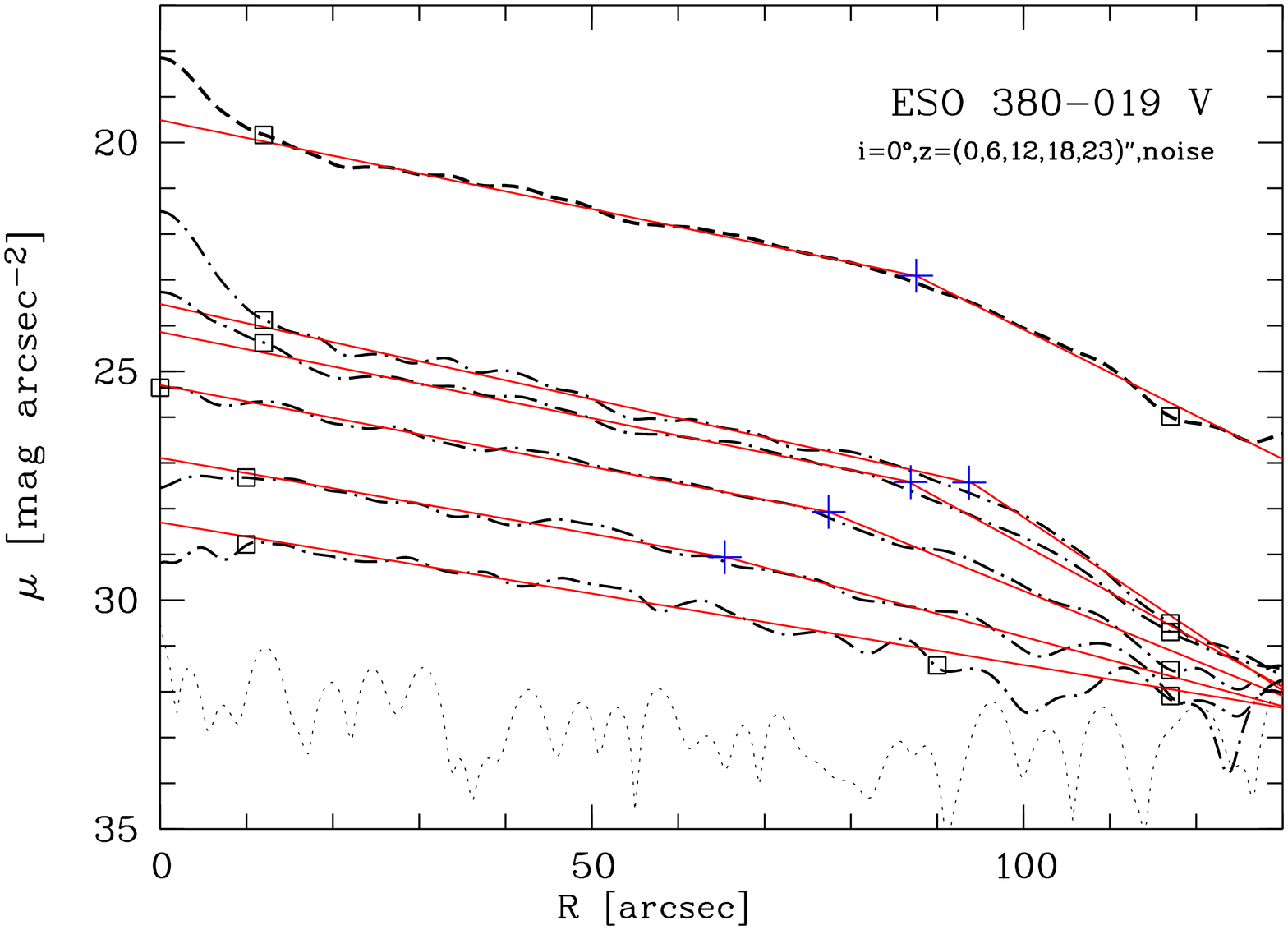}   
\includegraphics[width=6cm,angle=270]{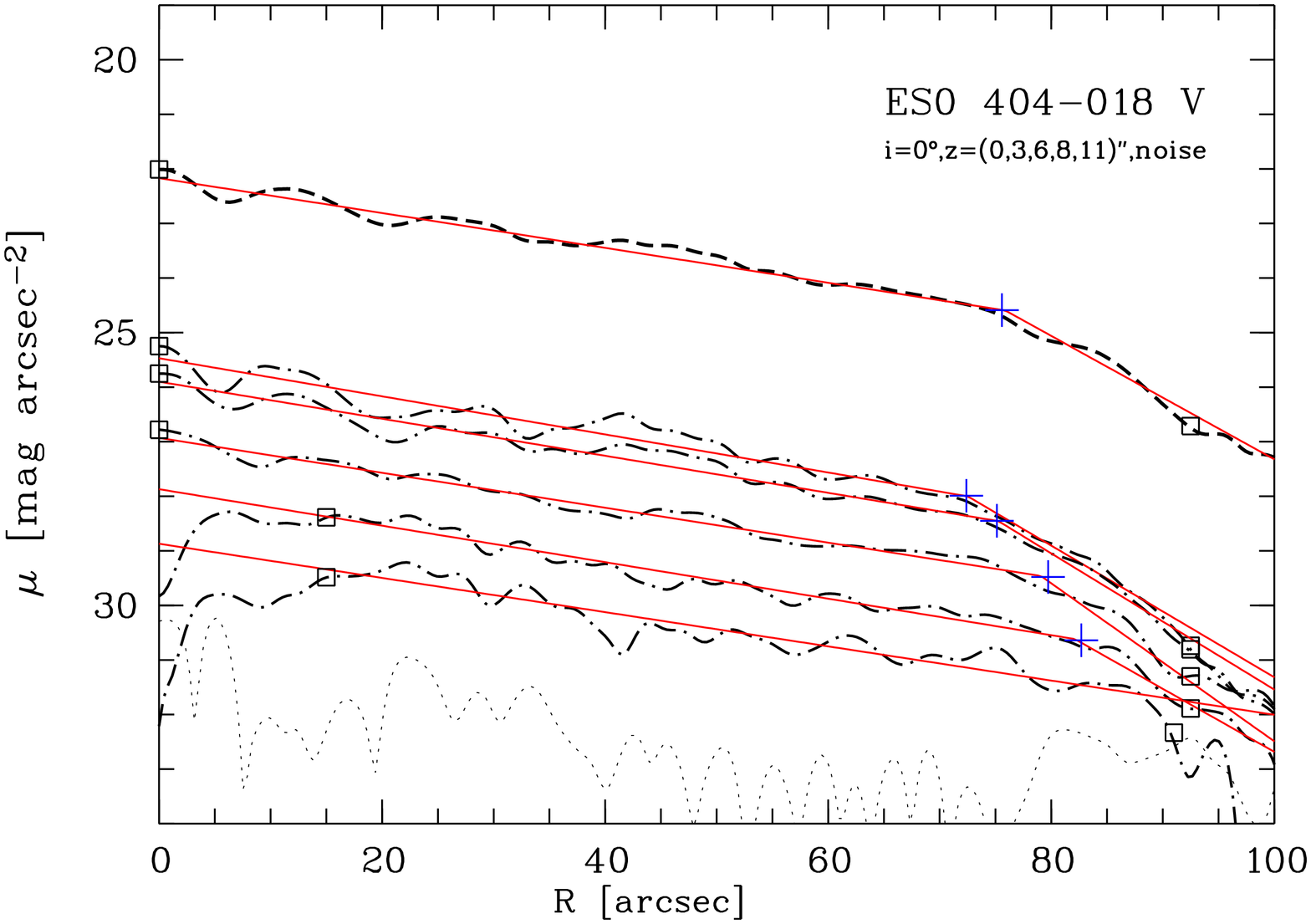}   
\includegraphics[width=6cm,angle=270]{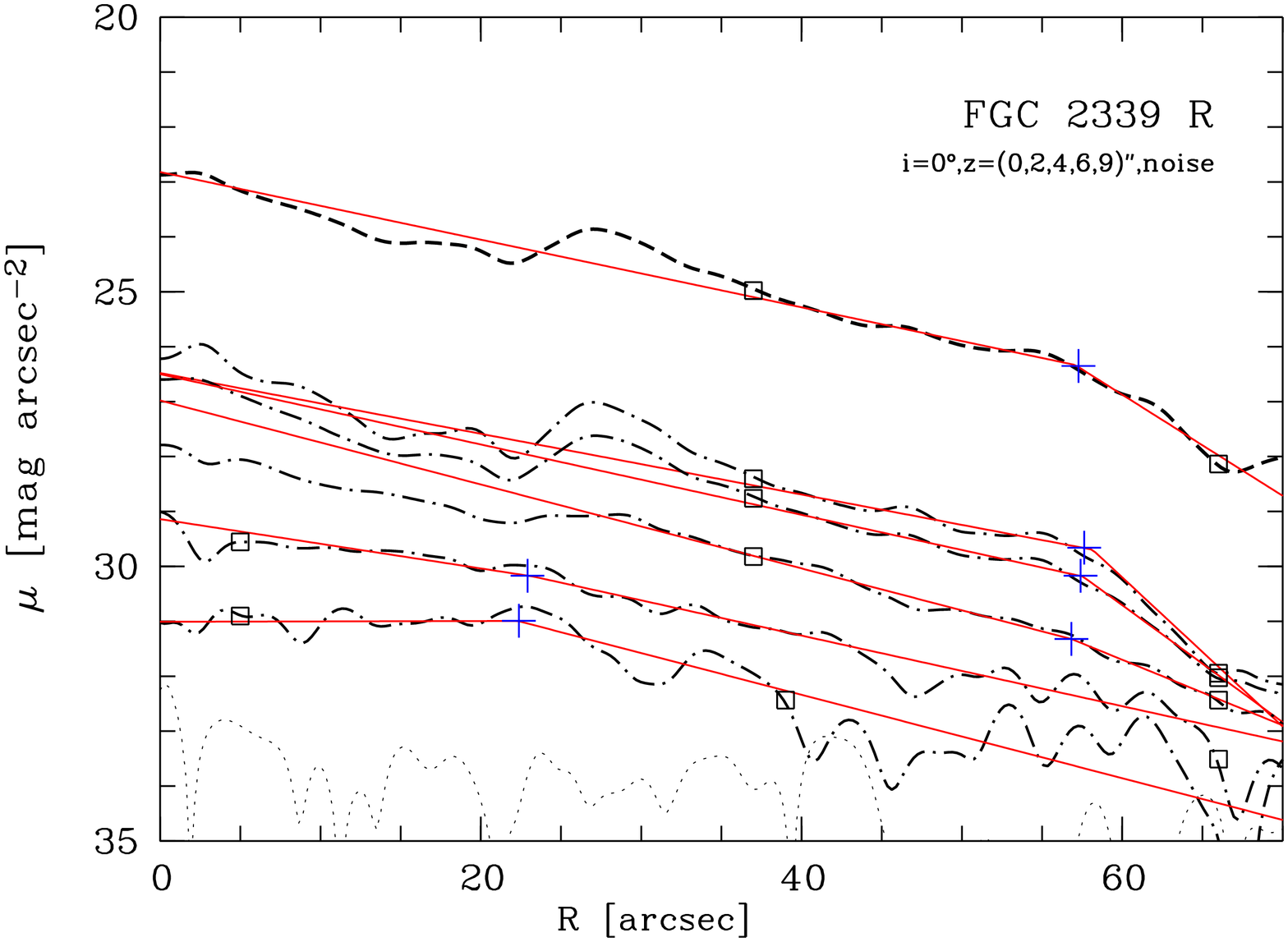}     
\includegraphics[width=6cm,angle=270]{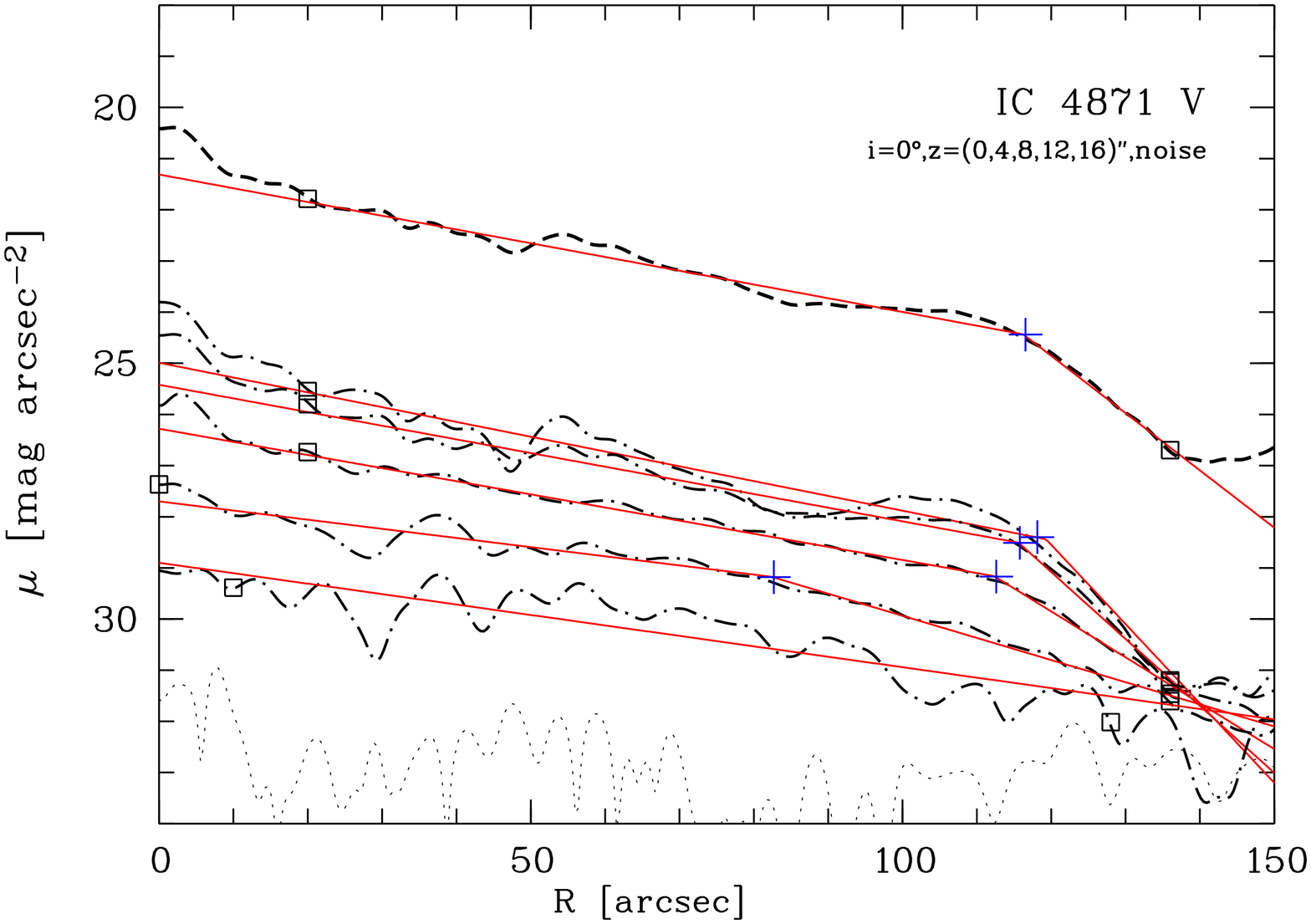}     
\includegraphics[width=6cm,angle=270]{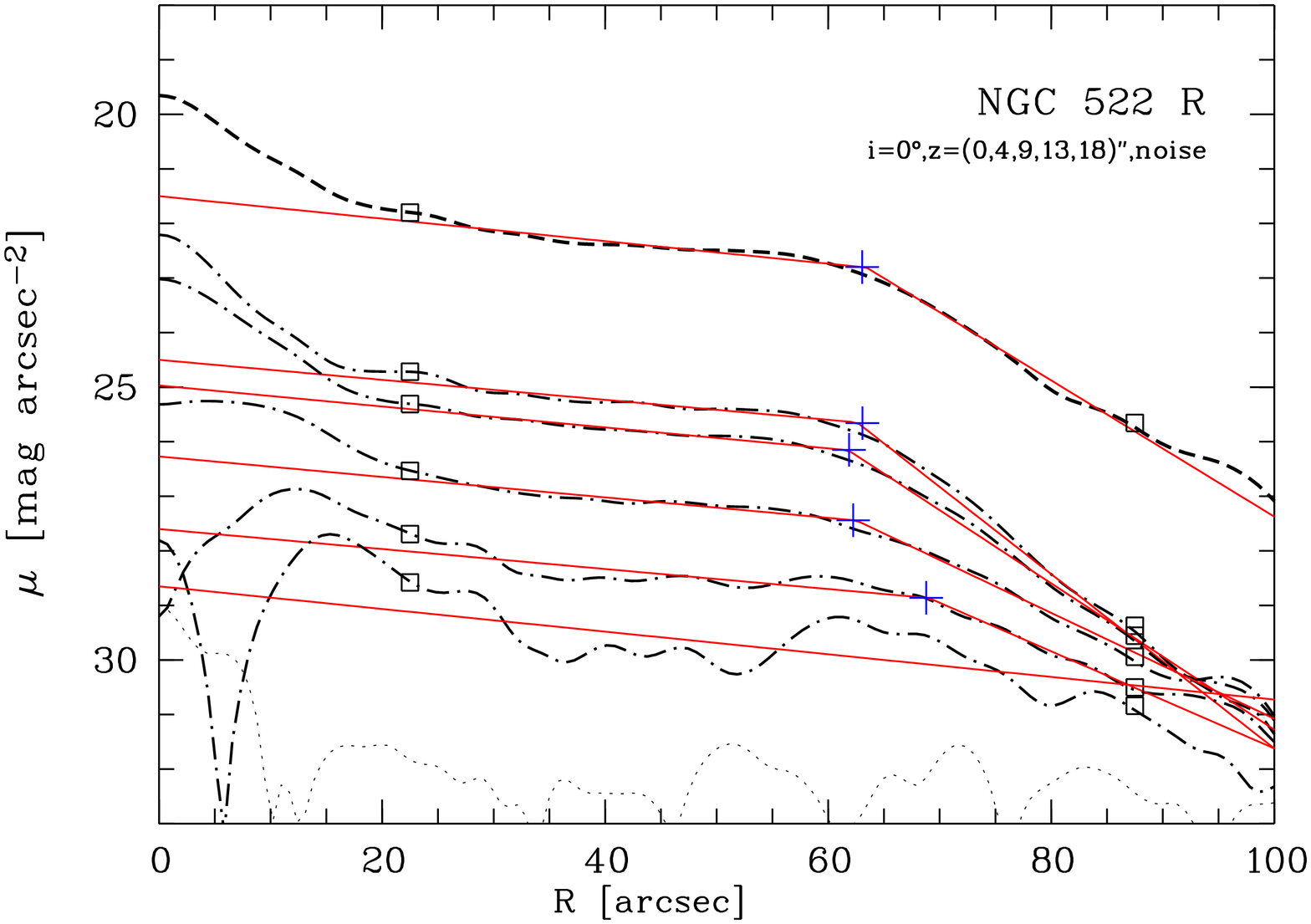}    
\includegraphics[width=6cm,angle=270]{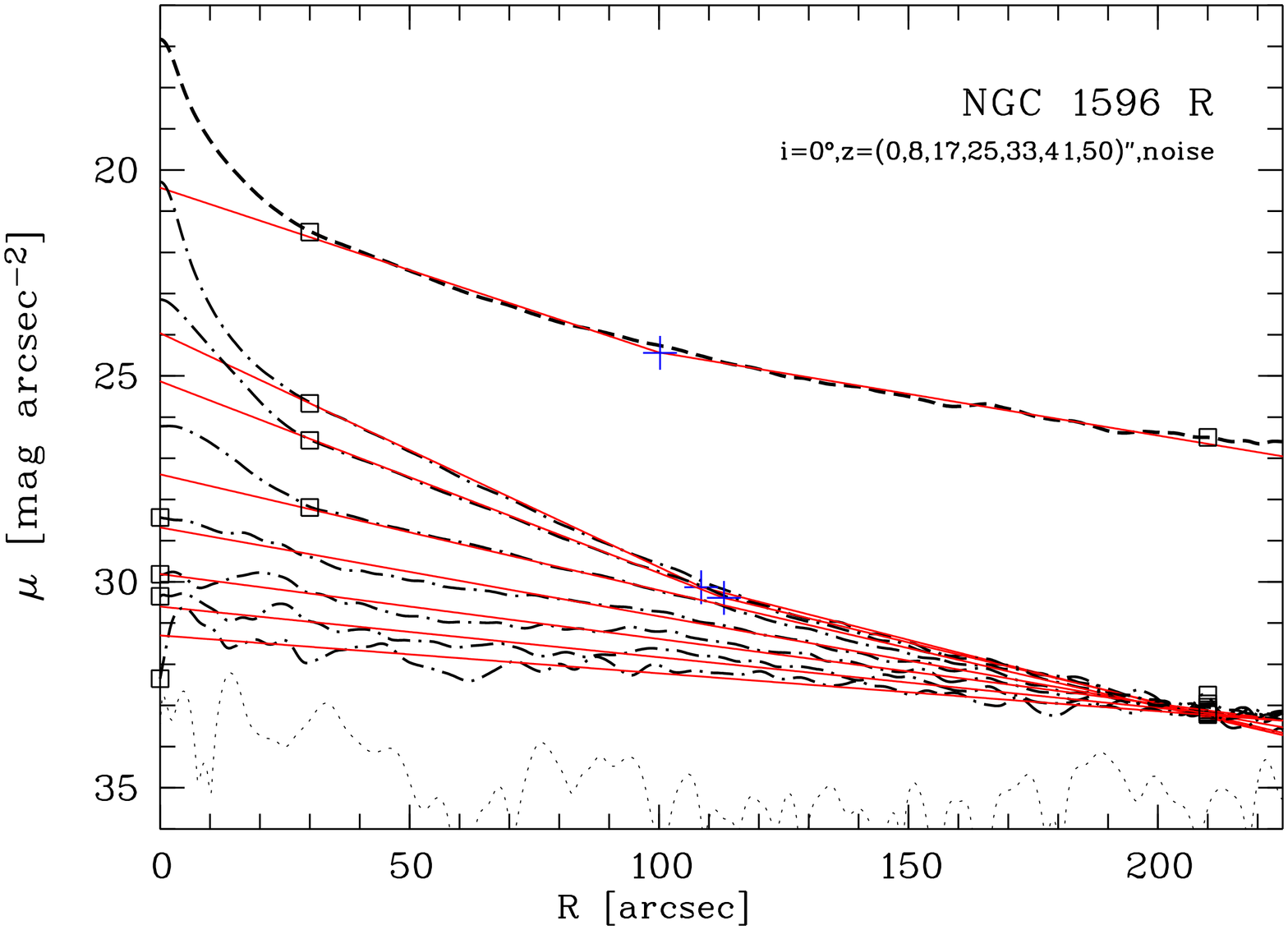}   
\caption{Deprojected radial profiles of the sample galaxies. \newline
The surface brightness profiles of each galaxy are shown in 
separate panels. 
The different line types are explained in \fig\ref{asym}.
\label{dprofpfits}
}
\end{figure*}
\addtocounter{figure}{-1}
\begin{figure*}
\includegraphics[width=6cm,angle=270]{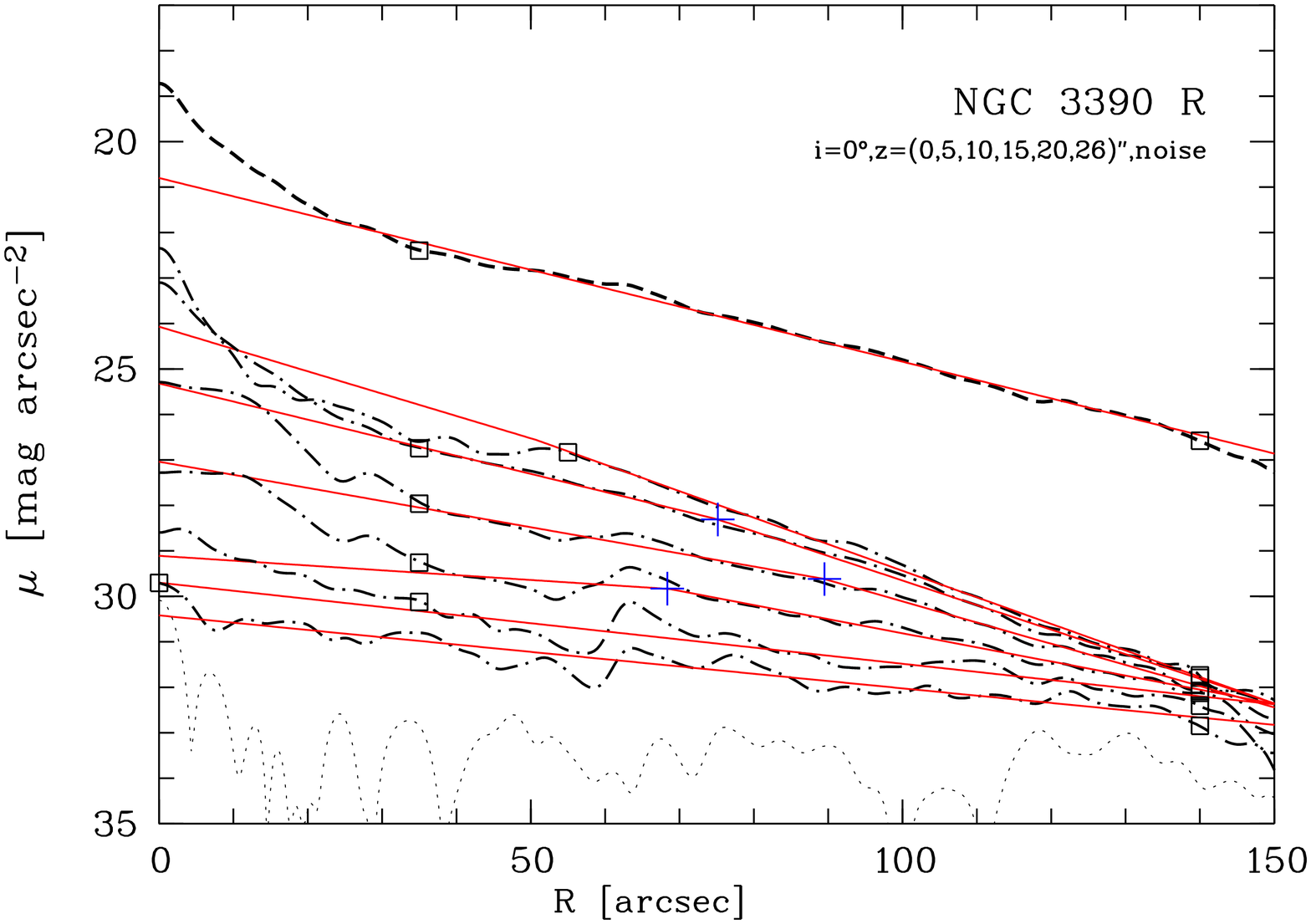} 
\includegraphics[width=6cm,angle=270]{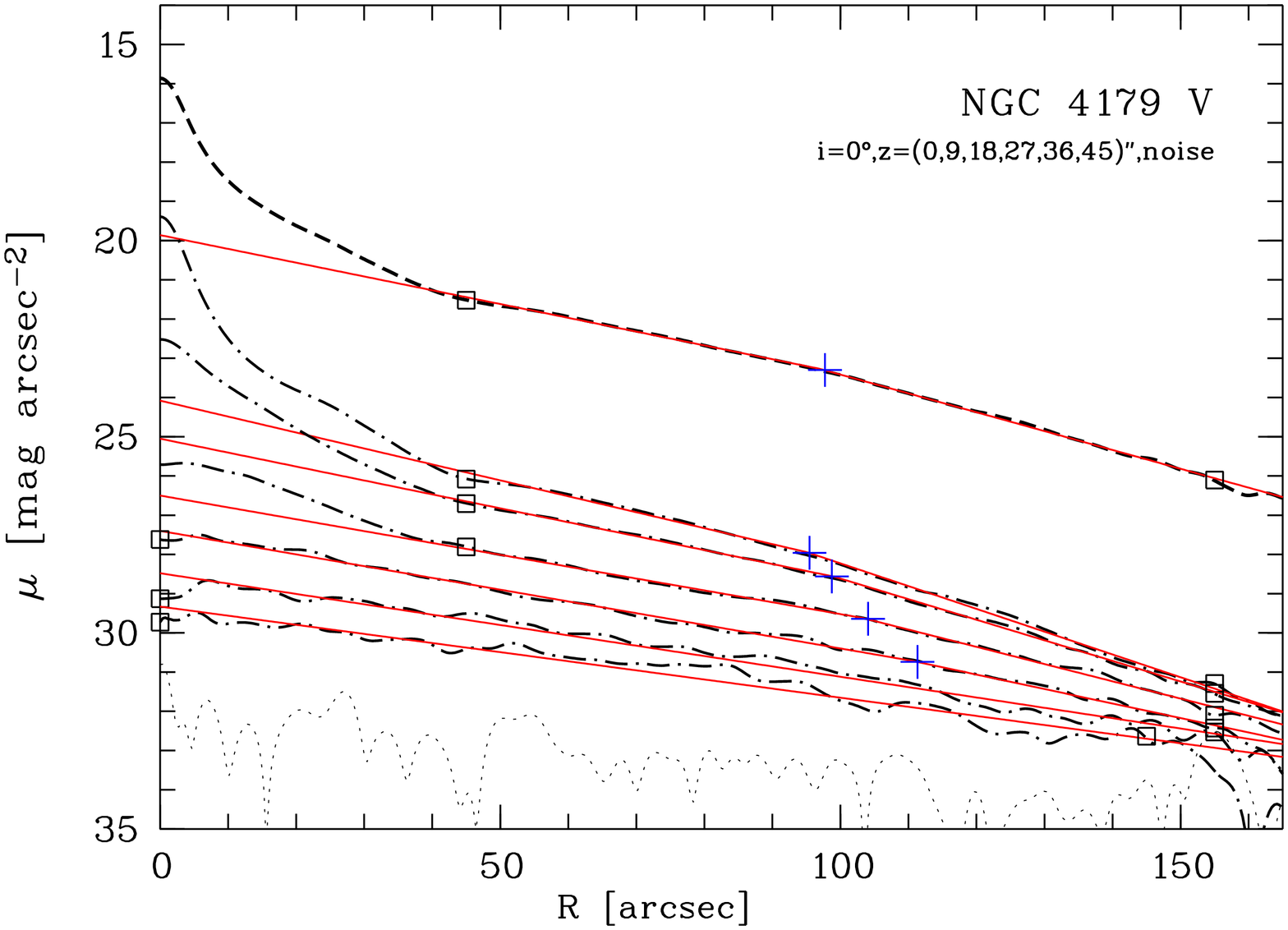} 
\includegraphics[width=6cm,angle=270]{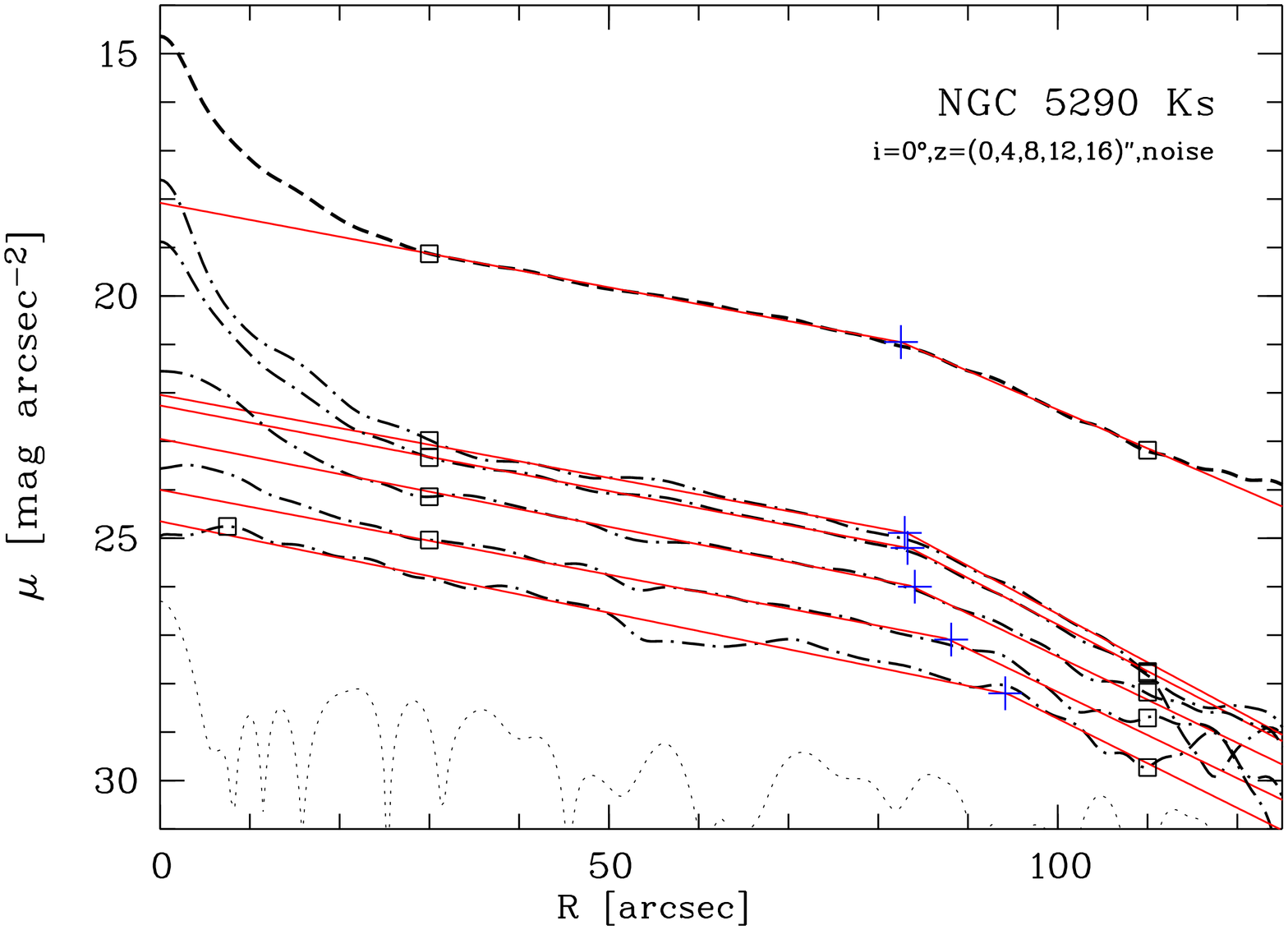}     
\includegraphics[width=6cm,angle=270]{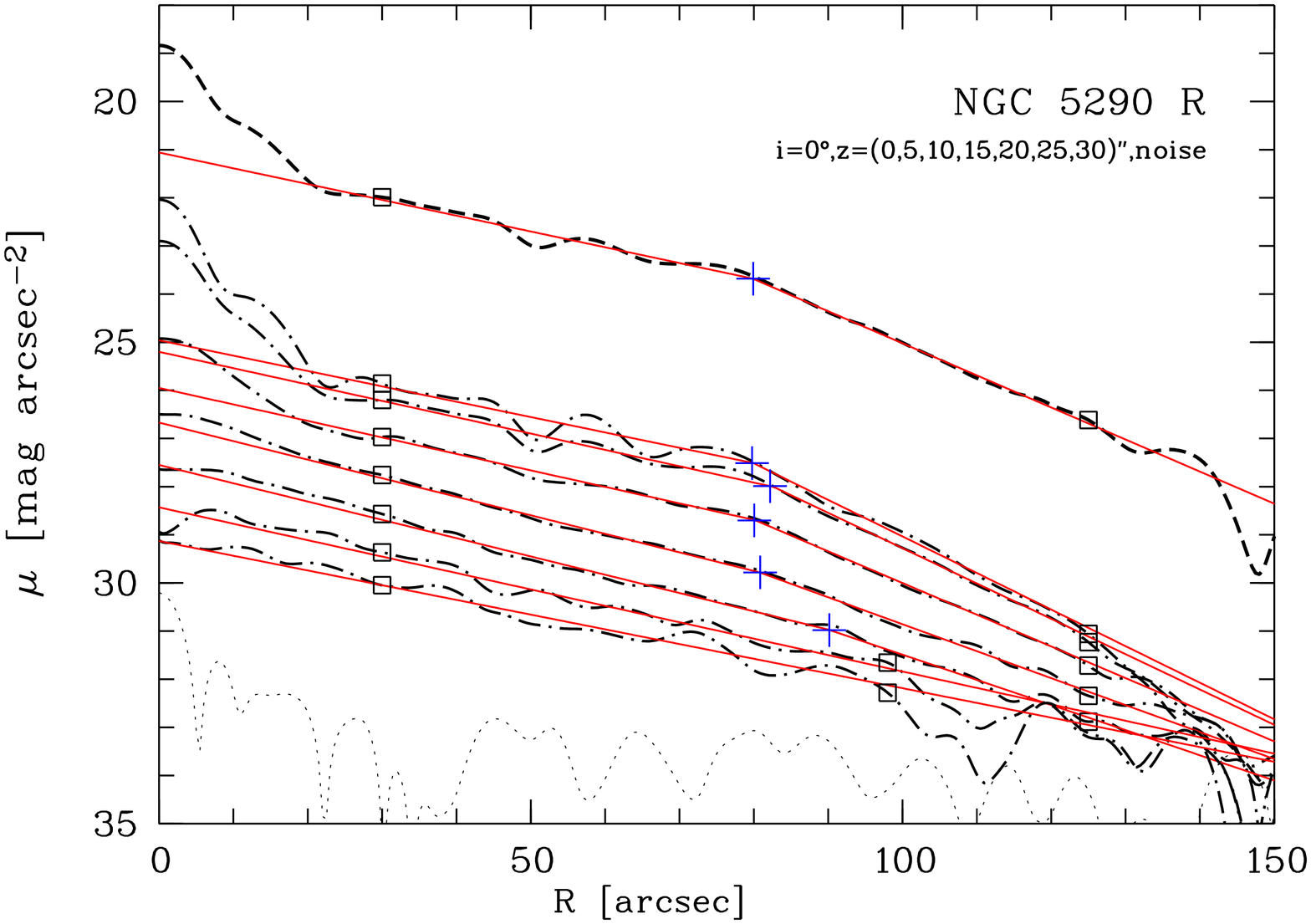}    
\includegraphics[width=6cm,angle=270]{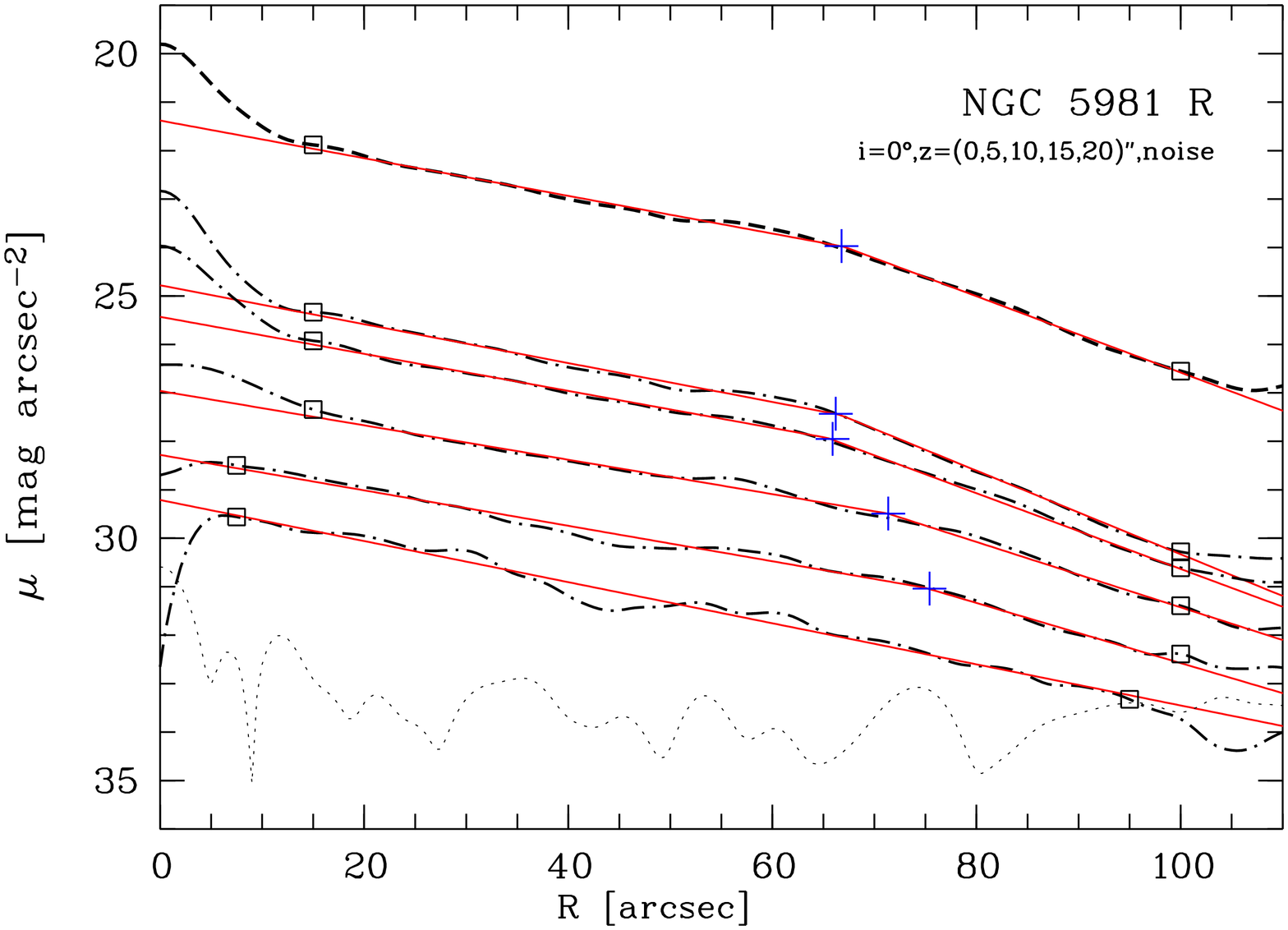}    
\includegraphics[width=6cm,angle=270]{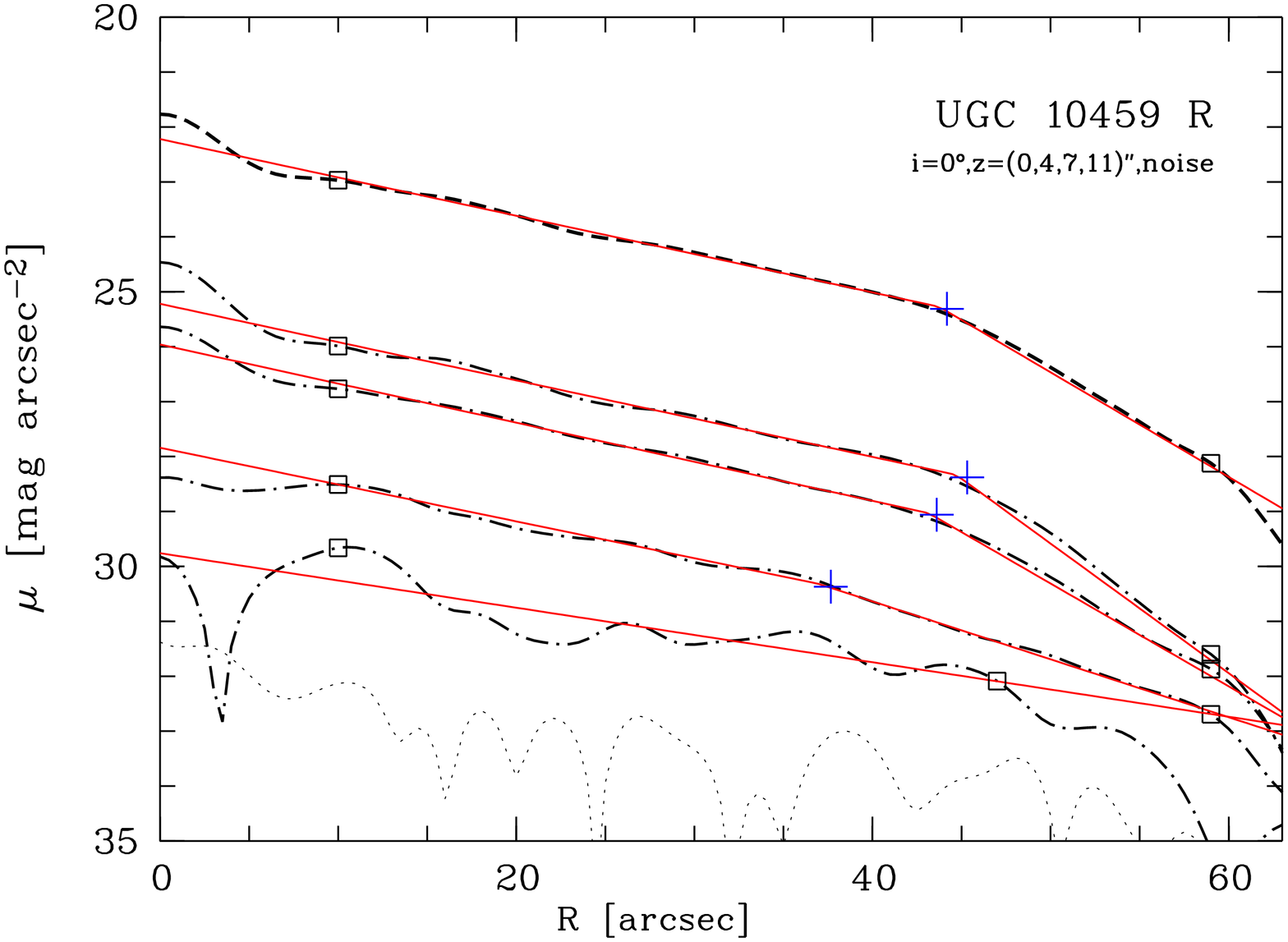}   
\caption{(continued): Deprojected profiles of the sample galaxies}
\end{figure*}
\subsection{The face-on-equivalent view} 
\label{faceon}
\subsubsection{Face-on {\it vs.} Edge-on: profile types.}
As mentioned earlier there are three basic classes of surface
brightness profiles that have been found for face-on galaxies,
\typeoc, \typetc, and \typeiiic~\citep{erwin2005,erwin2006,pohlen2006}.
Here, in the deprojected face-on-equivalent view, we also find at least 
one example for each of the three types of surface brightness profiles.
All eight Sbc and later-type galaxies (ESO\,380-019, 
ES0\,404-018, FGC\,2339, IC\,4871, NGC\,522, NGC\,5290, 
NGC\,5981 and UGC\,10459) show clear \typetc, downbending 
profiles. 
In comparison with \cite{erwin2006} and \cite{pohlen2006} we cannot increase
the level of classification (\ie assigning sub classes) in order to explore
the possible physical origin of the break, for example, star-formation
threshold or bar related break. This is mainly because inferring accurate bar
sizes in edge-on galaxies is almost impossible.
It is interesting to note that the NIR \Km-band profile of NGC\,5290,
typically assumed to be dust free, shows exactly the same behaviour as the
R-band. Therefore, the dust and recent star formation, clearly more prominent
in the R-band, do not play a role in the basic classification scheme applied
here.
The lenticular galaxy, NGC\,4179, also exhibits a \typet profile with
a ratio of inner to outer scalelength of $\hindhout\!=\!1.4$. This ratio is
somewhat low in comparison to the mean of
$\langle\hindhout\rangle\!=\!2.1$ reported by \cite{pohlen2006} for
late-types, or the median of $2.4$ reported by \cite{erwin2006} for
early-type galaxies.
One galaxy, NGC\,3390, although classified as an Sb galaxy, exhibits a
dominant outer spheroidal component closely resembling an \s0 galaxy
(see \fig\ref{mapspprofs}). Its face-on-equivalent profile is
consistent with being a classical unbroken profile, \typeoc.
Nevertheless, we can also fit to the recovered profile of this galaxy
a broken exponential, a \typetc. The scalelength ratio in this case is
only $\hindhout\!\ltsim\!1.2$, and so close to unity that in case of a
face-on galaxy this would not qualify as a \typetc.
The profile of the second \s0 galaxy, NGC\,1596, is very similar to the
antitruncations presented by \cite{erwin2005}. The outer profile is distinctly
shallower in slope than the main inner disk profile. Although it should be
classified as \typeiii in this sense, the profile is not well described by
our simple broken exponential fitting function. The whole disk profile (beyond
$R\!\gtsim\!\pm25$\arcsec) appears to be almost continuously upbending and one
may argue for an extended, exponential break region between breaks at
$R\!\approx\!70$\arcsec\ and $R\!\approx\!130$\arcsec. Nevertheless,
\cite{erwin2005, erwin2006} show some early-type, face-on counterparts 
with very similar profile shapes. 

\noindent
\subsubsection{Face-on {\it vs.} Edge-on: quantitative comparison.}
As discussed in \sec\ref{data} our sample is not unbiased. However, the
unusually high frequency of galaxies with \typet profiles suggests that we
indeed have selected against the two other types. 
For the nine \typet galaxies in our sample, we find that the break
radius is in the range between $1.2\!-\!3.3\hin$.
To compare this with the results of the face-on sample, we have to account for
the effect of dust on the inner scalelength. In \sec\ref{tests} we have shown
that in the case of dusty edge-on galaxies the resulting scalelengths are
typically 20\% overestimated (see \fig\ref{dustfit}).  Factoring this
correction\footnote{Without the small correction for the different wavelengths
used (R and V band here compared to r'-band for the face-on sample).} in one
obtains a break-radius in the range of $1.4-4.0\hin$, which matches the range
given by \cite{pohlen2006}.

The surface brightness at the break radius ranges from 21.9 to 26.3
R-\magsqarcsec, which also matches the full range for face-on galaxies
given by \cite{pohlen2006}. Here, the surface-brightness for each galaxy is
corrected for galactic extinction and, in the case of the V-band
images, transformed to the R-band assuming $(V\!-\!R)\!=\!0.7$.
%
\subsection{Vertical distribution}
\label{vertical}
\subsubsection{The inner vertical structure}
\begin{figure}
\includegraphics[width=5.9cm,angle=270]{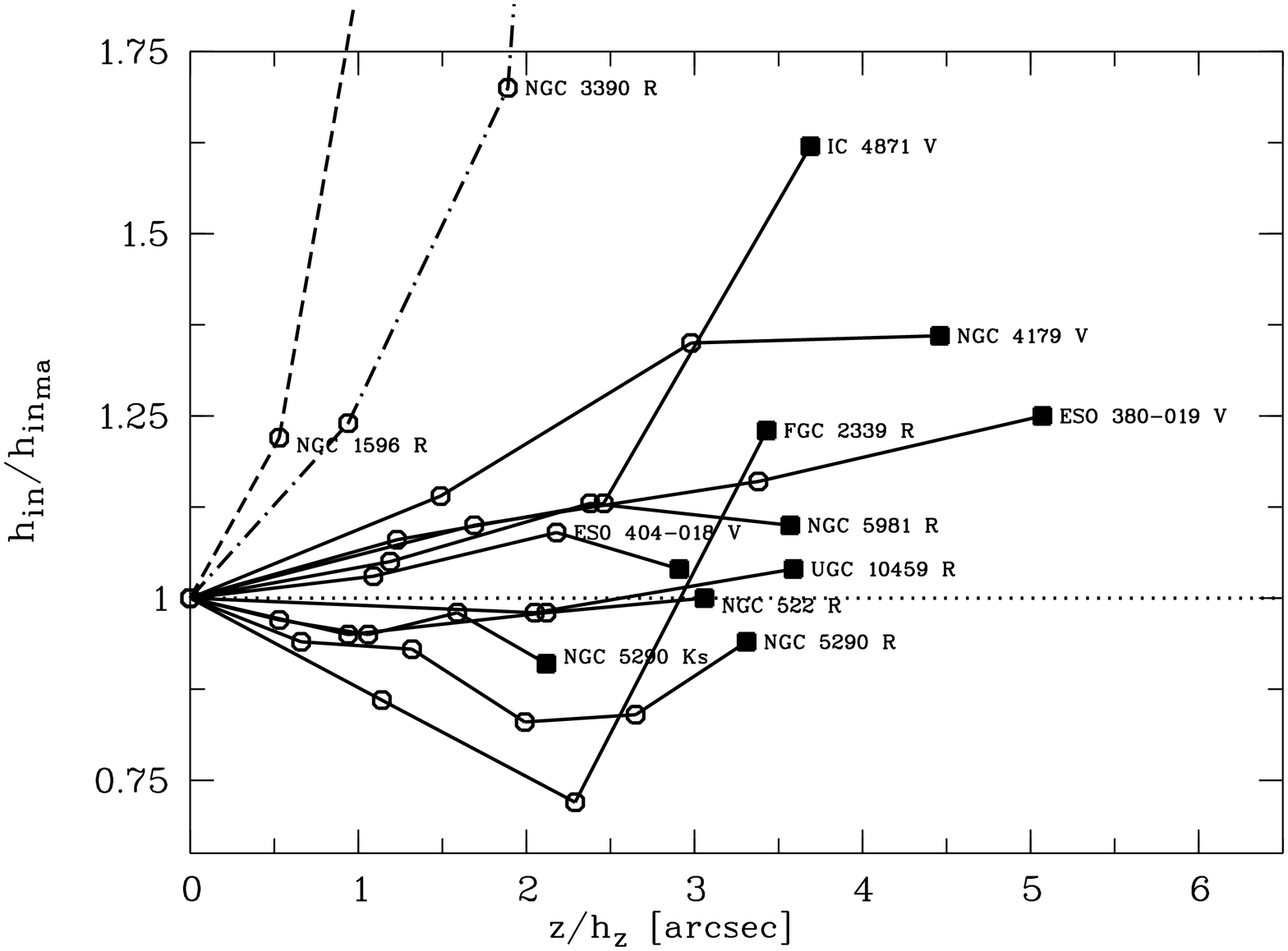}
\includegraphics[width=5.9cm,angle=270]{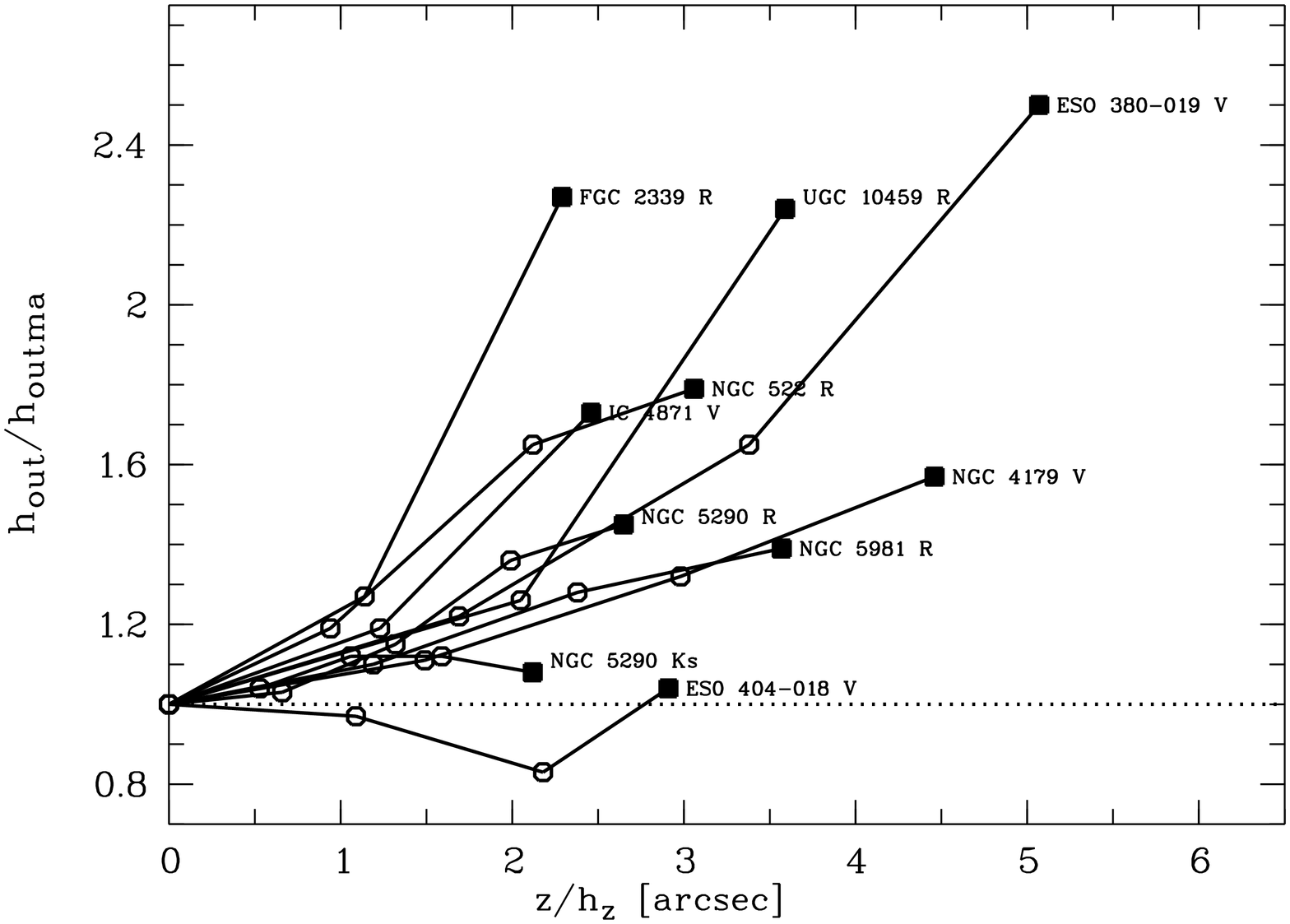}
\caption{Measured scalelength with increasing distance from the plane. \newline
{\it Upper panel:} The {\it open circles} are the measured inner scalelengths
($\hin$) normalised to the inner scalelengths measured on the major axis
$h_{\rm in_{\rm ma}}$ for each profile, versus the distance ($z$) from the
plane normalised to the scaleheight ($h_z$) given in \apx\ref{comments}. The
{\it filled squares} mark the values for the outermost profiles. All points
for an individual galaxy are connected ({\it solid lines} for \typetc, {\it
dotted-dashed line} for \typeoc, and {\it dashed line} for \typeiiic) and the
name of the galaxy is added. \newline
{\it Lower panel:} The {\it open circles} are the measured outer scalelengths
($\hout$) normalised to the outer scalelengths measured on the major axis
$h_{\rm out_{\rm ma}}$ for each profile versus the distance ($z$) from the
plane normalised to the scaleheight ($h_z$).  }
\label{hVSz}
\end{figure}

First we concentrate on the inner disk region, i.e., the region in
between the bulge/bar component and the location of the break.
For almost all of the late-type galaxies (Sbc and later, for details
see \apx\ref{comments}) we find an increase in the measured
scalelength with the distance from the plane.
The increase, at $3\!-\!5 z_0$, is typically of the order of about $20$\%,
relative to the major axis scalelength (see upper panel of \fig\ref{hVSz}).
Since we know that the omnipresence of dust extinction will mostly affect the
major axis profiles, as evident from the systematic flattening one sees for
our artificial galaxies (\fig\ref{dustfit}), our measurements of the increase
in scalelength in the optical, starting at the major axis, are only lower
limits. Consequently, the major axis slope for four galaxies is indeed
observed to be flatter than that obtained from adjacent $z$ cuts
(\cf\fig\ref{hVSz}).

  It is interesting to note that this type of flattening towards the
  midplane, which is for R-band images well explained by the dust effect, is
  also present at the level of approximately $15\%$ in our \Km-band image
  (NGC\,5290). If this is not an intrinsic feature of NGC\,5290, one could
  argue that the dust may still influence the scalelength determination 
  on the major axis even in the NIR band. On the other hand, this seems
  unlikely since the amount of dust inferred from the \Km-band flattening
  should have a more severe effect on the R-band profile (see
  \apx\ref{comments}).
In addition, tests with NGC\,5981 (see \apx\ref{comments}) show that
using the mean quadrant instead of the dustfree side, also results in
a systematic underestimation for the measured increase in scalelength
with distance from the plane.

\subsubsection{The outer vertical structure}
\begin{figure}
\includegraphics[width=5.9cm,angle=270]{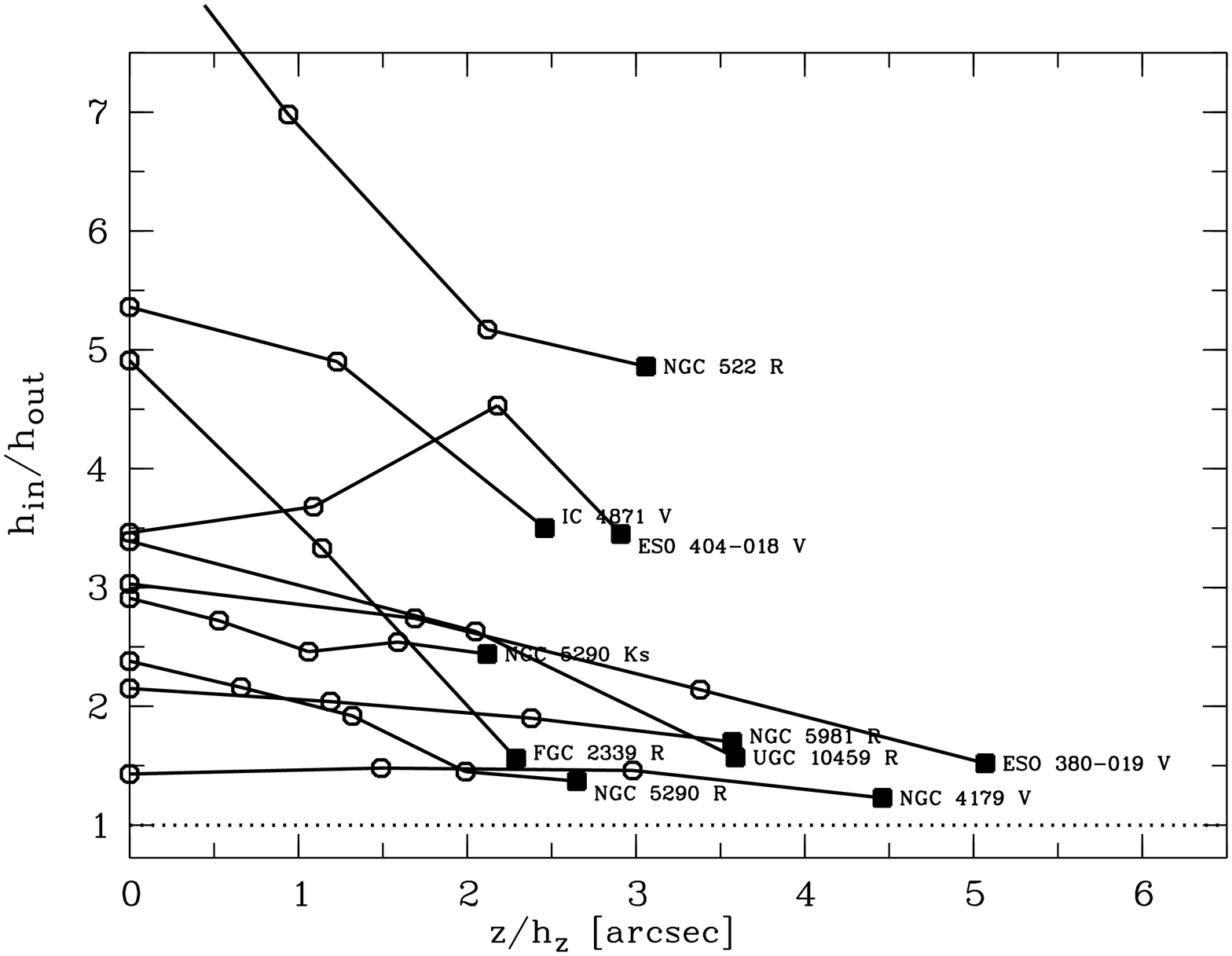}
\caption{Weakening of truncation with increasing distance from the
plane. \newline
The {\it open circles} are the measured ratios of inner to outer scalelength
(\hindhout) for each \typet profile with increasing distance ($z$) from the
plane normalised to the scaleheight ($h_z$). The {\it filled squares} mark the
values for the outermost profiles. All points for an individual galaxy are
connected ({\it solid lines}) and the name of the galaxy is added. \newline
\label{hinVShout}}
\end{figure}

Here we focus on the vertical structure of the outer disk, i.e., the region
beyond the break radius. For all of the late-type galaxies, namely, Sbc and
later, a significant increase of the measured scalelength with distance from
the plane is observed (see lower panel of \fig\ref{hVSz}).
Whereas for the inner disk the increase in the slope is by a
factor of 1.1-1.4, for the outer disk we find a factor of
1.4-2.4. This indicates an extreme flattening of the profile towards
larger distances from the plane.
Again, the dust, well known for flattening surface brightness
profiles, cannot be responsible here since it is concentrated in the
plane.
This could be regarded as a weakening of the truncation with increasing
distance from the plane (see \fig\ref{hinVShout}).
\subsubsection{The break radius as a function of $z$}
The position of the break radius as a function of vertical distance from the
plane ($z$) is another constraint that any theory that might explain the
truncation phenomenon should satisfy.
An inspection of the location of the break radius from the deprojected
profiles (\fig\ref{dprofpfits}) shows that this location does not
change significantly as a function of $z$ in the first two to three
$z$-slices. For the more distant $z$-slices the slopes of the inner
and outer regions become closer, due the more substantial flattening
in the outer disk, which makes it harder to confidently determine the
location of the break radius.  A clear demonstration of this is when
the inner and outer slopes are very close, in such a case the presence
of any genuine small scale structure in the image renders the location
of the crossing point very uncertain.
This is even more severe when the broken exponential model does not
provide a good description of the data, e.g., when the break region is
clearly extended (ESO\,380-019 or ESO\,404-018).

Resorting to the NIR image or concentrating on the smooth \s0 galaxies
in our sample, in which the substructure (star-forming region or dust
patches) are less prominent, does not improve the matter because for
these galaxies, as with the other, the flattening of the outer disk
dominates the uncertainty in the location of the crossing point.

To summarise, based on the first 2 to 3 vertical slices, the current data
suggest that the location of the break radius for all our galaxies is
consistent with being constant.
%

\section{Discussion} 
\label{discussion}
\noindent\textbf{Profile classification} 

\noindent
In our pilot sample of edge-on galaxies we do indeed find the same
three main types of surface brightness profiles as classified by
\cite{erwin2005, erwin2006} and \cite{pohlen2006} from their analysis
of face-on galaxy images.
The broken exponential parameters for the face-on-equivalent view
obtained from our edge-on galaxies are well consistent with those
found by the analysis of, the more complete, genuine face-on galaxy
samples. Only for FGC\,2339 the break seems to be at a slightly lower
than usual surface brightness, albeit at a typical distance from the
center.
However, this is also the galaxy with the value of the lowest central
surface brightness. Hence, similar to \cite{pohlen2006}, we still cannot
disentangle which of the two statements is correct: the surface
brightness at the break radius is constant, or it is coupled to the
central surface brightness.
We do not find a difference in the shape of the truncation comparing the
two profiles for NGC\,5920 obtained from the optical and the NIR image. Both
show clearly a broken exponential structure in the region between $30\ltsim
R\ltsim 110\arcsec$. This is consistent with the findings of
\cite{hunter2006b} for the profile shapes of several irregular galaxies using
$3.6\mu$ and $4.5\mu$ images from the Spitzer Space Telescope. Irregular
galaxies with double-exponential optical light profiles have the same in the
NIR. So baring an exception, NIR observation of truncations do
not seem to differ significantly from observation in the optical as proposed
by \cite{florido2006}.
\cite{pohlen2006} have shown that \typeo galaxies are rare in
late-type galaxies and, according to \cite{erwin2006}, more frequent
in \s0 galaxies.
Accordingly, our only \typeo candidate (NGC\,3390) exhibits an \s0-like
outer disk structure, despite being classified as Sb.
It is interesting to note that in this case the profiles of the
individual cuts, close to the major axis but uninfluenced by the dustlane,
show some indication of being better described by a broken exponential
structure. Therefore, the vertical integration in this case may obscure the
existence of a break close to the plane. This is clearly caused by the lack of
an apparent break in the vertical slices at larger vertical height above the
midplane. This might explain the high frequency of not-truncated disks seen in
the early-type face-on galaxy sample of \cite{erwin2006}.

Despite the discussion above about the general agreement between the
edge-on and face-on classifications, an edge-on counterpart of the
not-truncated, \typeo, late-type galaxies \cite[e.g., the face-on
NGC\,2776 by][]{pohlen2006}, is yet to be found. Given the expected
frequency of such galaxies ($\approx\!10\%$) and the size of the parent 
sample from which we select, we are not able to explain the absence 
of such galaxies.
If late-type \typeiii galaxies might be associated with interacting or merging
systems, as proposed by \cite{pohlen2006}, the detection of only one \typeiii
galaxy is in full agreement, as we excluded those systems from the original,
late-type dominated, sample.
In addition, the fact that our \typeiii galaxy is an early-type one 
is consistent with the observed increase in frequency of \typeiii
profiles for early-type galaxies \citep{erwin2006}.
To our knowledge, all edge-on \typeiii galaxies reported in the 
literature so far are early-types. These galaxies are NGC\,4762 
studied by several groups \citep{tsikoudi1980,wakamatsu1984, wozniak1994}, 
NGC\,4452 shown by \cite{hamabe1989} and NGC\,4696C\footnote{It is 
worth noticing, that NGC\,4696C is classified as an Sb galaxy 
\citep{rc3}. However, deep imaging, obtained by \cite{pohlen2001}, 
clearly shows an outer \s0-like envelope.} from \cite{pohlen2001}.
For face-on galaxies, \cite{erwin2005} differentiate between two
different types of \typeiii profiles. One where the excess light is
probably part of the disk which show a relatively sharp transition and
both parts are well described with exponentials (\typeiiidc). And
another one where the excess light is associated with the
halo/spheroid or bulge (\typeiiisc). The latter is characterised by a
small inner region fitted with an exponential and a continuously
upbending outer profile and a changing ellipticity with radius.
In the case of NGC\,1596 we would argue for a \typeiiis classification where
the outer structure is associated more with an envelope like structure. This
is supported by the difficulty one faces in fitting NGC\,1596 with a
consistent thin/thick disk combination \cite[cf.][]{pohlen2004b}. Any bulge
component would have to have an unlikely, extremely high, axis ratio (\cf
\apx\ref{comments}). In this case, the observed outer envelope is therefore
much better explained by interaction as recently supported by
\cite{chung2006}.

\medskip
\noindent\textbf{Inner Disk}

\noindent
Almost all of the analysed profiles exhibit an increase of the radial
scalelength with increasing distance from the plane. Notwithstanding the lack
of completeness in our sample, this finding could support the notion of {\it
ubiquitous thick disks} as proposed by \cite{dalcanton2002}, where these thick
disks have to have a systematically larger scalelength, compared to the thin
disk.
However, in the absence of a consistent two disk fit to the data at
all radii the thick disk model is not the only option. Instead one can
interpret the increase in the radial scalelength merely as a result
either of a radially flattened component above the inner disk or a
flattening of the inner disk itself.
For example, in the case of the late-type (Scd) galaxy FGC\,2339, 
the whole inner disk region is dominated by a ring- or bar-like 
structure in the midplane which also alters the shape of the disk
above/below the plane over the entire measured vertical extent.

\begin{figure*}
\includegraphics[width=5.0cm,angle=270]{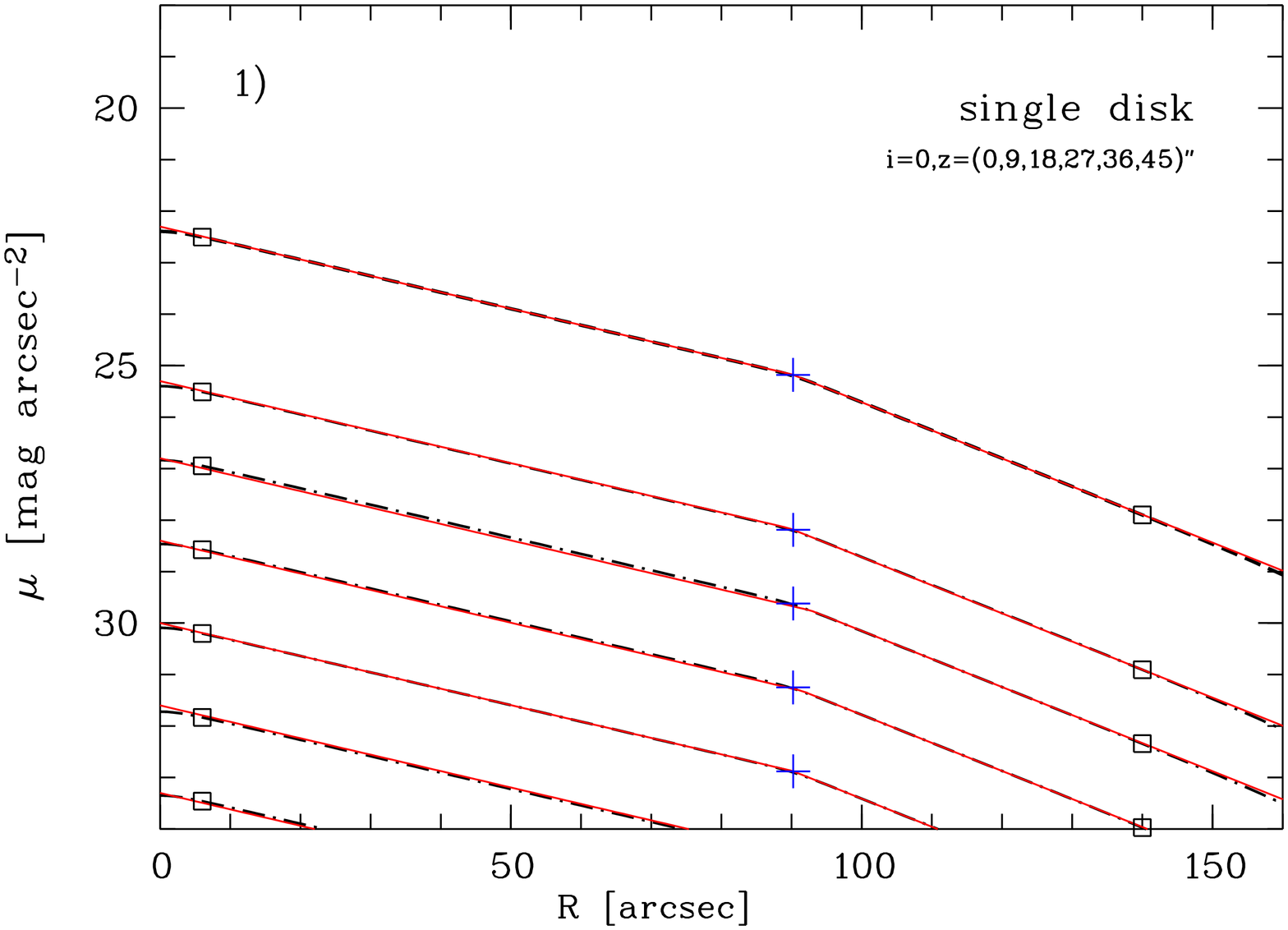}
\includegraphics[width=5.0cm,angle=270]{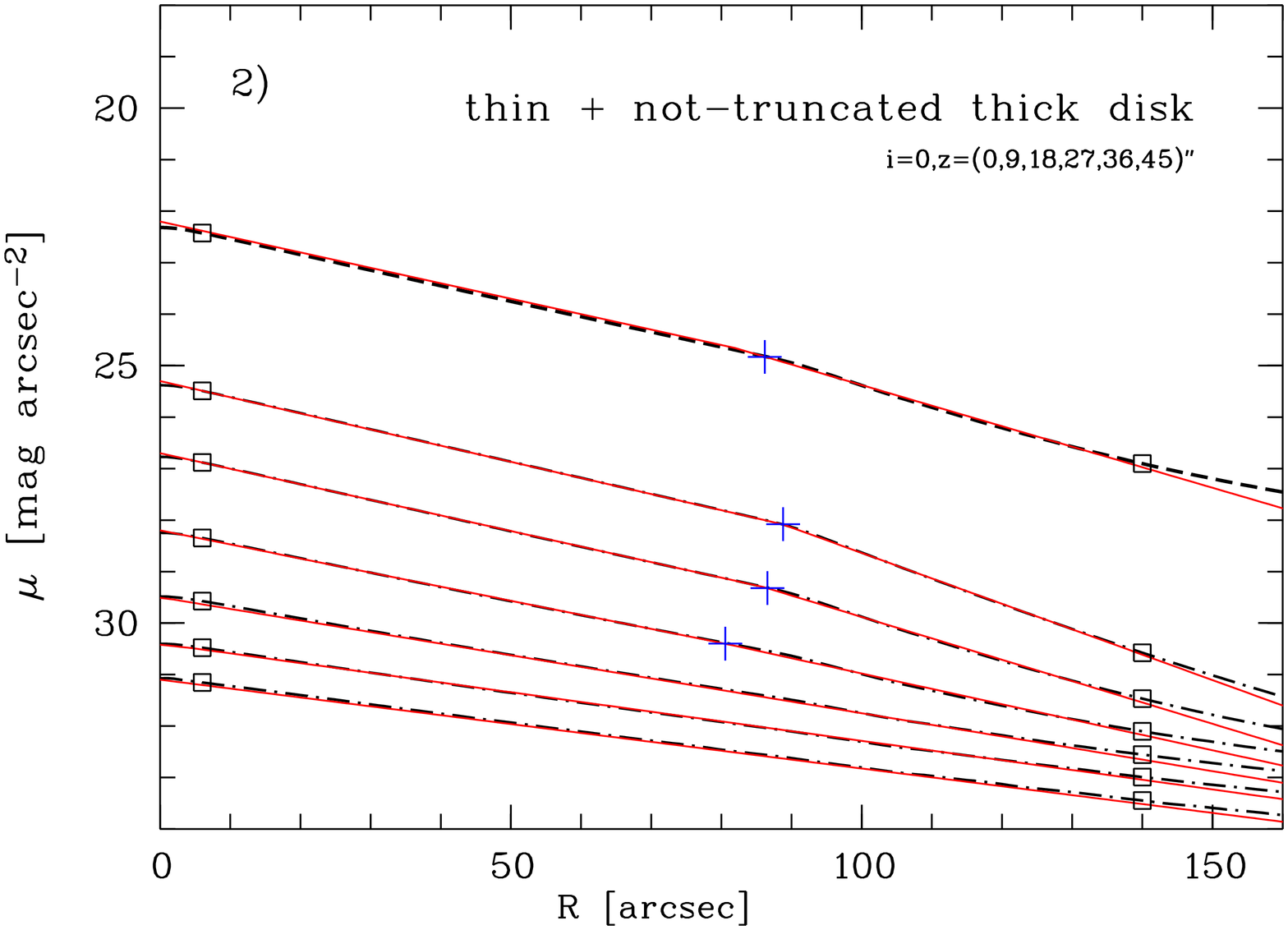}
\includegraphics[width=5.0cm,angle=270]{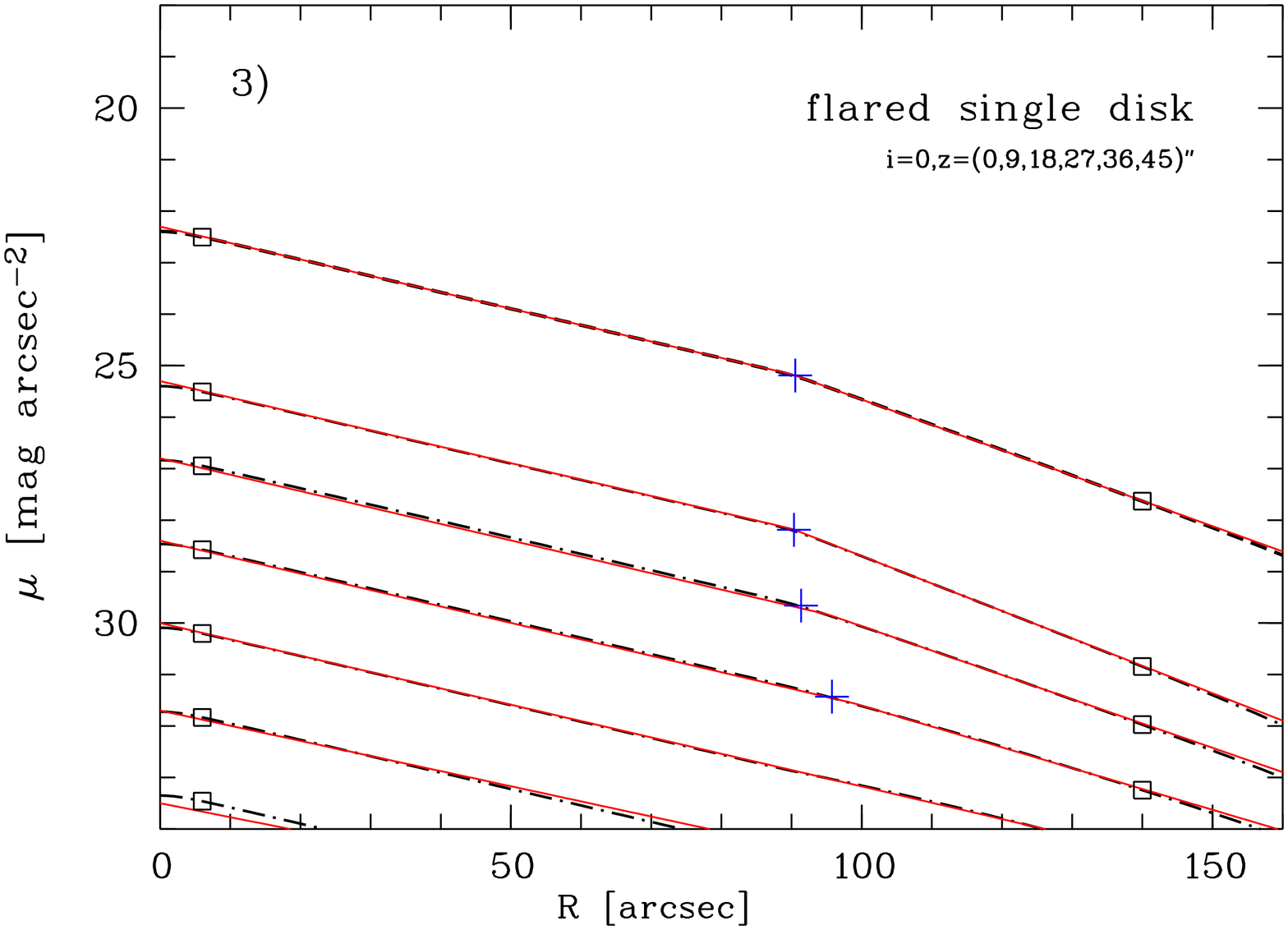}
\includegraphics[width=5.0cm,angle=270]{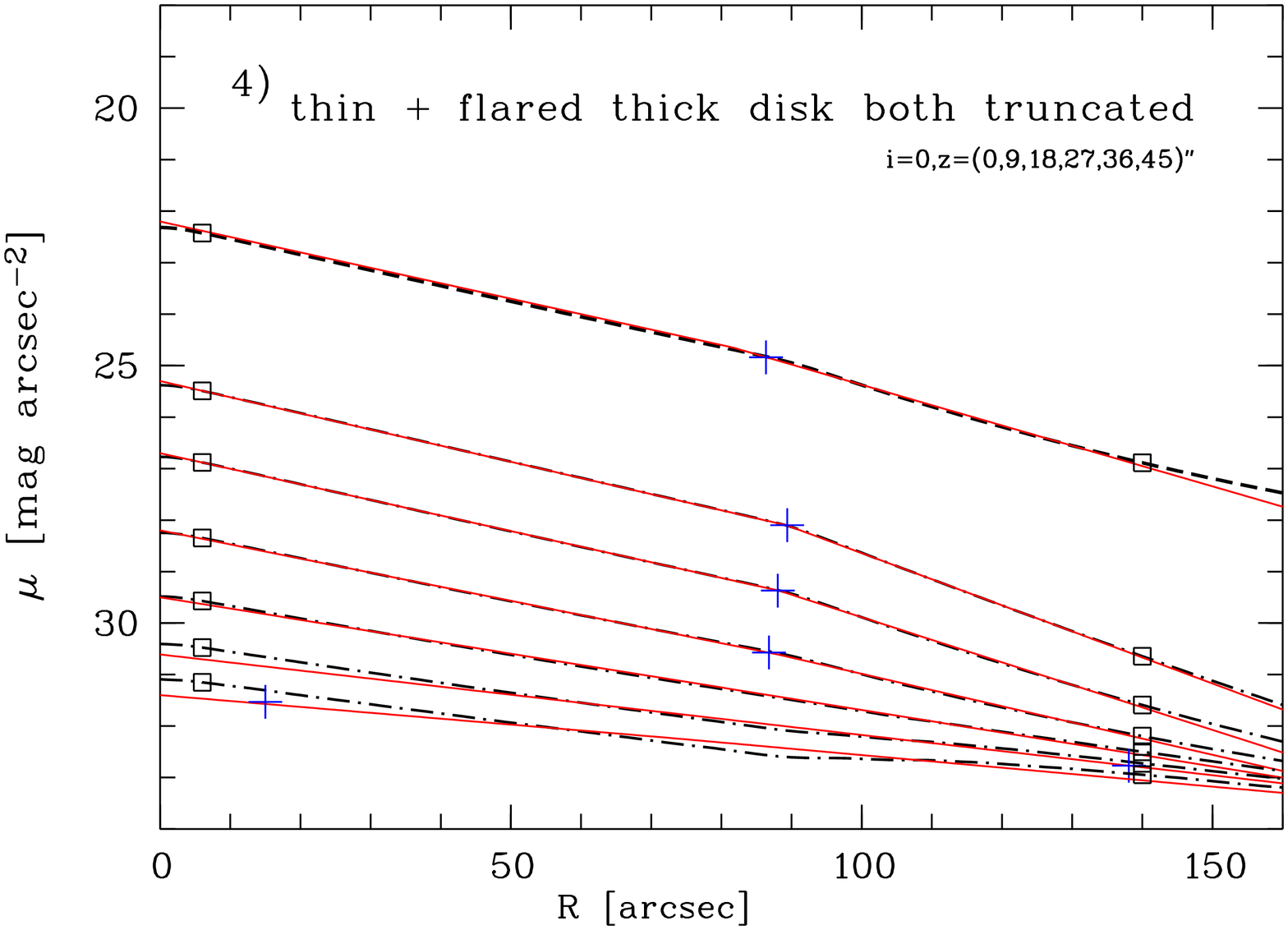}
\caption{Toy models to explain observed profile structure. \newline
Deprojected radial surface brightness profiles of six model galaxies: single
disk {\it (panel 1)}, truncated thin plus not-truncated thick disk {\it (panel
2)}, and two models with flaring {\it (panels 3-4)}. The figure has the same
setup as \fig\ref{dprofpfits}.}
\label{toymodels}
\end{figure*}

\medskip
\noindent
\textbf{Outer Disk} 
\noindent
In order to examine what could lead to the observed flattening of the
outer disk profile, we invoke several 3 dimensional (3D) toy
models. These 3D models are generated by combining separate
components, e.g., a truncated thin disk plus a truncated thick
disk. The resultant structures are qualitatively compared to those
reconstructed from the data. It is important to emphasise here that no
attempt is made to fit the data to these models, rather the intention
is to explore whether these toy models could generically reproduce the
increased flattening seen in the data. The 3D profiles of the models
are shown in \fig\ref{toymodels}.
To highlight the flattening in these models we fit broken exponential
profiles with the same fitting routine used for the real galaxies
(\cf\sec\ref{fitting}).
For example the upper-left panel (1) of \fig\ref{toymodels} shows a single
broken exponential disk model. This type of models does not show any
flattening in the reconstructed scalelengths and, therefore, is not
favoured by our data.

\medskip
\noindent
{\it Could we be seeing  de Vaucouleurs $R^{1/4}$  bulges?}
\noindent
\cite{bahcall1985} have shown that a thin disk plus a $R^{1/4}$ bulge can
mimic the shape of the isophotes of a thin disk plus thick disk model (without
bulge) down to a surface brightness level of about
$\mu_R\!\sim\!25.5$\magsqarcsec.
Furthermore, this combination can even produce some flattening of the radial
profiles with increasing vertical height beyond the break radius as seen in
our models.

\begin{figure}
\includegraphics[width=5.0cm,angle=270]{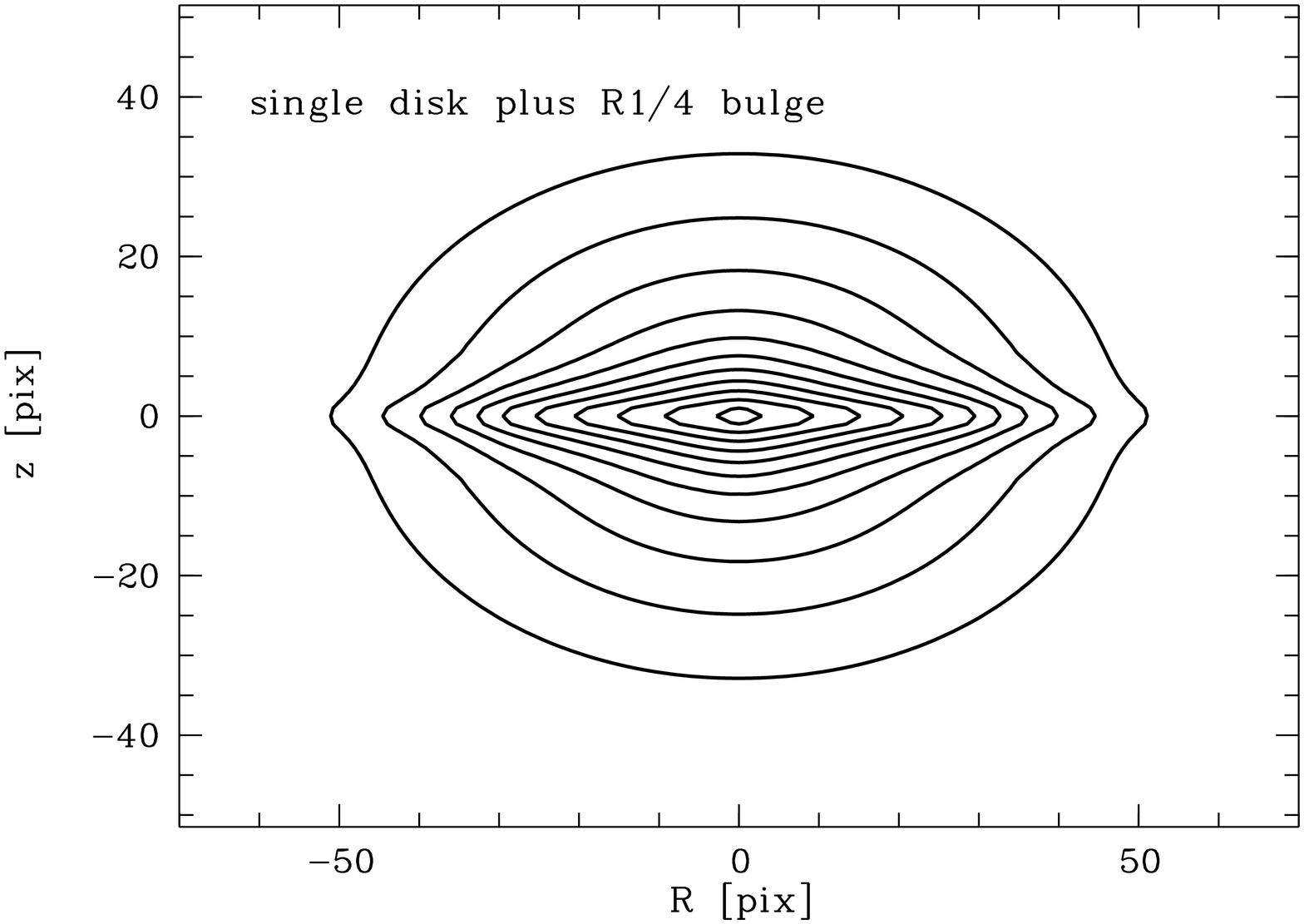}
\includegraphics[width=5.0cm,angle=270]{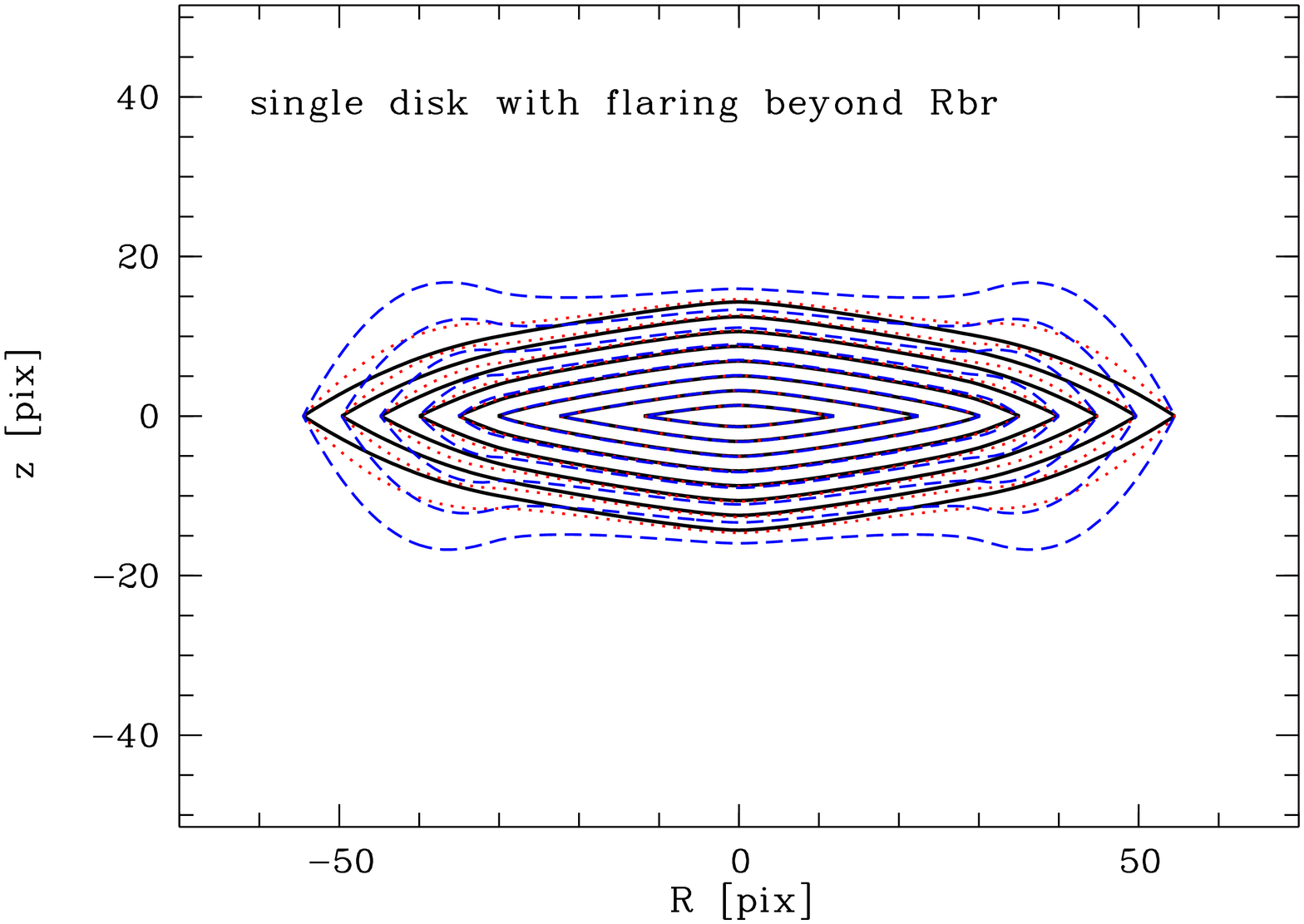}
\caption{{\it Upper panel:} Contour map of a radially broken
exponential, vertically exponential thin disk superimposed with a
standard de Vaucouleurs $R^{1/4}$ bulge. {\it Lower panel:} Three
combination of a the same thin disk and a thick disk, where the thick
disk scaleheight grows (flares) between the break radius (\rbr) and
three times \rbr with an increasing factor {\it (solid, dashed, dotted
lines)}.  }
\label{contour}
\end{figure}

Having said that however, there are strong arguments against
explaining the observed flattening with a spherical power law bulge
component. First, in a sample of 19 early-type galaxies (S0-Sbc)
\cite{balcells2003} find that none of the bulges have a $R^{1/4}$
behaviour.
Most of their bulges are better fit with exponentials which are not
expected to dominate the radial profiles at large
distances. Therefore, it is statistically unlikely to attribute the
flattening in the outer disk in most of our late-type edge-on
galaxies to a spherical bulge component.
Second, the observed shape of the outer contours in all the cases
studied here appear too much disk-like to be caused by a spheroid (see
\fig\ref{mapspprofs} and \fig\ref{contour}).
Third, it is almost impossible to create a $R^{1/4}$ bulge which will
leave the high S/N major axis profile purely exponential down to the
last measured point, but will dominate the radial profile at larger
vertical heights above the plane, and doing this all without being
clearly visible in a contour map.

In conclusion, since there is no clear sign of a de Vaucouleurs
$R^{1/4}$ in any of the observed images whereas the outer disk
flattening is found in the majority of galaxies in the sample, the
answer to the question posed here is, most probably: no.

\medskip
\noindent
{\it Does a thin plus thick disk combination produce the increased
flattening beyond the break?} 
\noindent
If one assumes that the thick disk scalelength is in general larger than
the thin disk scalelength, the increase in scalelength with $z$-height in the
inner disk will be always present. However, there are various possibilities
for producing the extended flattening in the outer disk beyond the break with
thin plus thick disk models. The exact shape of the flattening critically
depends on the, unknown, outer disk properties of the thick disk.

If, for example, we assume the thick disk to be not-truncated, we are
still able to produce some flattening beyond the break (see
panel (2) of \fig\ref{toymodels}). However, depending on the surface
brightness limit the not-truncated thick disk will become visible at large
radii.
This could be an explanation for some of the very few \typetpiii\ mixed
classifications found for late-type galaxies by \cite{pohlen2006}.
These galaxies show two breaks in their face-on surface brightness
profile. The first is downbending in the inner region, while the other
is upbending in the very outer region.
However, none of our edge-on sample galaxies shows this behaviour.
By adding a truncation to the thick disk one can produce a wide variety of
broken exponential structures depending, for example, on the ratio of inner to
outer disk scalelength (\hindhout). Models with unequal thin and thick disk
truncation radii provide even more choice to fit the data.

In conclusion, the answer to the question posed here is: yes. However,
the degree of flattening will depend on the details of the 3D model.

\medskip
\noindent
{\it Could the flattening be explained by intrinsic flaring\footnote{With {\sl
intrinsic flaring} we do not mean the sometimes observed fact of an increase
in scaleheight with radius \cite[e.g.][]{degrijs1997}, which is most probably
related to a combination of thin and thick disk with different scalelengths,
but the intrinsic change of the scaleheight for a single stellar disk
component.} of the stellar disk?}

\noindent
To produce an increased flattening in the outer-disk one can invoke models
with intrinsic flaring -- similar to those used for the Milky Way \cite[see
e.g.][]{gyuk1999} -- that starts beyond the break radius.
In the edge-on view, the contour map of the observed galaxy provides
a strong constraint on the degree of flaring.
The lower-panel of \fig\ref{contour} shows three different radially
broken exponential single disk models with flaring
factors\footnote{For the Milky Way \cite{gyuk1999} have models with
$F=5$ or $10$ between the solar radius and 25\,kpc.} of $F=2.5$, $5$,
and $10$.
If the flaring factor is chosen to be too large the resulting contour
map will blow up in the outer parts, a behaviour that is neither
observed in any of our sample galaxies nor in the much larger sample
of \cite{pohlen2001}.
In general, it seems that a strong flaring of the stellar disk is rarely
observed. In addition, in the few reported cases it appears to be rather
associated with \typeiii profiles, \eg NGC\,4762
\citep{wakamatsu1984,freeman2002}, and not with \typet as discussed here.
As shown in the panel (3) of \fig\ref{toymodels}, flaring of a
single disk beyond the break radius does not always produce the right
amount of flattening if compared to our observations.
A more promising attempt to apply flaring is obtained by allowing only the
thick disk component to flare beyond the break radius (see panel (4) of
\fig\ref{toymodels}).
This approach could help in cases where the inner disk indicates only
a very small value for the ratio of thin to thick disk
scalelength. The flaring is then needed to reproduce the observed
significant difference of the outer scalelength on the major axis
compared to high above the plane.
A flaring of the thick disk is also proposed by several authors
\cite[e.g.][]{siegel2002,du2006} for our Milky Way from star-count
studies. This flaring would account for the observed variable scaleheight 
with different direction of the analysed fields.
It is clear that the data can in principle discriminate between
possible 3D models, as each will produce different features in the
reconstructed profiles. However, we make no such attempt in this paper
except to point out the different possibilities.
\medskip
\noindent\textbf{Break Radius} 
\noindent
As we have shown earlier, the location of the break-radius in the
current sample is found to be consistent with being height
independent. In light of the enhanced flattening of the outer disk
relative to the inner disk, the determination of the break radius at
large vertical cuts is uncertain and more accurate modelling could
reveal significant variation in its location towards the larger
distance from the major axis.

Nevertheless, we would like to point out that if such a finding is
supported by more data, it could have far reaching implications on
theories that attempt to understand how the thin and thick disk are
assembled. As such a theory should explain how potentially distinct
components of the disk ``coordinate'' the location of their break
radii.

\section{Conclusions} 
\label{conclusion}
This paper has presented a new algorithm for reconstructing the intrinsic 3D
structure of galactic disks seen edge-on. The algorithm assumes axial-symmetry
and utilises the so-called Fourier Slice Theorem which in the case of edge-on
galaxies is equivalent to an inverse Abel-transform. The algorithm is applied
to a pilot set of 12 edge-on galaxy images, drawn from a parent sample of 72
galaxies, to obtain face-on-equivalent radial surface brightness profiles, and
reconstruct the radial profiles along several vertical cuts, including the
major axis.
Comparing the results obtained from the face-on-equivalent profiles with those
measured from ``real'' face-on samples we reach several conclusion. The
first, and probably most important, conclusion is that we find in the current
edge-on sample at least one example for each of the three basic profile types
found in face-on samples: no break, upbending break, and downbending break
\citep{erwin2005,erwin2006,pohlen2006}.
Despite this agreement, our morphologically selected parent sample
\citep{pohlen2001}, choosing undisturbed and prototypical disk galaxies,
clearly favours late-type galaxies with class \typet profiles, \ie downbending
breaks. This might be due to a connection between certain profile types and
distinct vertical structures. For example, if late-type \typeiii galaxies are
connected to interaction and/or merging processes, they would probably be
excluded from edge-on samples. In other words, they would exhibit distortions,
lopsidedness, or warping, which are all prominent in the edge-on view.
Although our parent sample is not unbiased in this sense, our results do
show that the parameters for the downbending breaks of the face-on-equivalent
profiles obtained from our edge-on galaxies are consistent with those seen in
face-on samples. 
This implies that truncated profiles of galaxies are independent of geometry
and there is no distinction in this regard between edge-on and face-on
studies.
In this paper we have been able to study, for the first time, the
deprojected radial profiles as a function of vertical height above the
plane, which is only possible to explore with edge-on galaxies.
Hence, this approach provides a unique opportunity to give
additional constraints on theories explaining the origin of the
various types of observed profiles in the context of disk formation
and evolution.
In this context, we find a generic flattening of the observed profiles in the
inner disk region with increasing distance from the plane, which cannot be
explained by dust extinction. This implies a more complex vertical structure
in the inner region than a single disk. It is possible to explain this
flattening by introducing a thick disk with systematically larger
scalelength compared to that of its host thin disk.
In the outer disk, beyond the break radius, we find an enhanced
flattening of the deprojected profiles, which again sets strong
constraints on theories that attempt to explain the observed broken
exponential structures.
Using some toy models we conclude that thin plus thick disk
combinations are, in principle, able to explain the observed
structures.  However, the details of this combination depend on the
exact shape of the outer region of the thick disk.  For example, a
thick disk without truncation will exhibit less flattening in the
outer disk than a thick disk with truncation. 
So far our data does not allow us to decide if thick disks are in general
truncated or not.
It remains unclear whether one can fine tune a simple model with broken
exponential thin plus thick disks in order to explain the observed
brightness distribution.

In addition, we find that the location of the break radius does not
change significantly as a function of $z$ in the first two to three
$z$-slices. Unfortunately the flattening of the radial profile beyond
the first few vertical cuts makes it hard to determine the location of
the break radius.
Although we do not know yet whether thick disks are indeed truncated,
we have shown that the vertical structure harbours clues to understand
the origin of the different radial profiles and should be pursued.
For example, luminous, not-truncated thick disk components
could be responsible for the high frequency of not-truncated 
lenticular galaxies in face-on surveys, while having
a broken exponential structure close to the midplane of
the galaxy. 
In conclusion, observations of the vertical structure set strong constraints
on modelling/simulating disk galaxies and the interpretation of their various
components. In the future, a larger, well selected sample of edge-on galaxies
will be analysed in order to shed light on some of the issues raised here.
 
%
\section*{acknowledgements}
The authors wish to thank Simone Bianchi, who kindly provided us with {\it
dusty-galaxy images} produced with his radiative transfer code. We thank Olof
van den Berg for using the reduced \Km-band image of NGC\, 5290. The authors
thank P.~van der Kruit, R.~Sancisi, M.~Verheijen, and E.~Valentijn for helpful
discussions, and the anonymous referee for his detailed and very helpful comments. 
Part of this work was supported by a Marie Curie Intra-European 
Fellowship within the 6th European Community Framework Programme.
This research has made use the Lyon/Meudon Extragalactic Database 
(LEDA, {\ttfamily http://leda.univ-lyon1.fr}) and the 
NASA/IPAC Extragalactic Database (NED) which is operated by 
the Jet Propulsion Laboratory, 
California Institute of Technology, under contract with the National 
Aeronautics and Space Administration. 
This research has made use of NASA's Astrophysics Data System
Bibliographic Services.  
{}

\appendix
\section{Notes on individual galaxies}
\label{comments}
{\bf ESO\,380-019:} \newline Owing to a not well defined thick patchy dust
lane the inclination is uncertain (clearly not 90\degc, but closer to
85\degc). The extent and size of the bulge component is also uncertain. The
major axis radial profile shows a broken exponential structure with a break at
$R\approx\pm(95\!-\!100)$\arcsec\ ($\mu_{\rm
V}\approx23.8$\,\magsqarcsec). The galaxy is moderately asymmetric. In the
contour map the SE(left)-side is obviously shorter.
The single disk fit by \cite{pohlen2001} yields a vertical scaleheight
of $\hz\eq3.5\arcsec$, a radial scalelength of $h\eq22.8\arcsec$, and an
inclination of $84.0\degc$.
\newline
The face-on-equivalent profile of ESO\,380-019 is well described as a
\typet with an inner shallow exponential (of
$\hin\eq27.9\arcsec\!\equiv\!2.4\hout$) and an outer steeper exponential
($\hout\eq11.5\arcsec$) with a break at a radial distance of
$\rbr\approx 88\arcsec\!\equiv\! 3.1\hin$ and a surface brightness of
$\mubr\eq22.9$V-\magsqarcsec.
\newline
We find an increase in scalelength of $\approx 25\%$ between the major
axis and about 5 times the vertical scaleheight (at 18\arcsec) in
the inner disk region. Since the galaxy is rather far from edge-on, 
the extended dustlane (with maximum extinction at $\approx 7\arcsec$)
could still play a role (\cf Fig. \ref{dustfit}).
The scalelength beyond the break in the outer disk increases
by a factor of 2.5 between $0$ and $5.1\!\cdot\!\hz$.
\medskip
\noindent
{\bf ESO\,404-018:} \newline This galaxy shows hardly any bulge component.
The disk structure is quite knotty and appears bended like a wave
\citep{weinberg2001} in addition to being slightly u-shaped at the edges. The
radial surface brightness profile shows quite clearly the deviation from a
simple not-truncated exponential model on both sides with a sharp break on the
NE(left)-side, and a smoother one on the SW(right)-side starting at
$R\!\approx\!\pm70$\arcsec\ and is well fitted by a broken exponential.
The single disk fit by \cite{pohlen2001} yields a vertical scaleheight
of $\hz\eq2.8\arcsec$, a radial scalelength of $h\eq31.3\arcsec$, and an
inclination of $90.0\degc$.
\newline
The face-on-equivalent profile of ESO\,404-018 is well described as a
\typet with an inner shallow exponential (of
$\hin\eq34.0\arcsec\!\equiv\!3.6\hout$) and an outer steeper exponential
($\hout\eq9.6\arcsec$) with a break at a radial distance of $\rbr\approx
76\arcsec\!\equiv\! 2.2\hin$ and a surface brightness of
$\mubr\eq24.6$V-\magsqarcsec.
\newline
We find an increase in scalelength of $\approx 12\%$ between the major
axis and about 4 times the mean vertical scaleheight (at 11\arcsec)
in the inner disk region. Although this galaxy is almost perfectly
edge-on we still find the steepest profile on the major axis, so the
dust is not playing a major role here.
The shape of the profile in the outer disk is not well described with
a single exponential. The break regions (starting at $\approx\!75$\arcsec)
seems to be extended and almost exponential itself.
This deviation from our simple broken exponential model strongly 
affects the measurement of the scalelengths beyond the break.
\medskip
\noindent
{\bf FGC 2339:} \newline This galaxy shows on both sides slightly
off-centered ring-, or bar-like knots at
$(R,z)\eq(+24\arcsec,+1\arcsec$) and $(R,z)\eq(-26\arcsec,+1\arcsec$). The
underlying disk is quite asymmetric on both sides. The major axis
profile on the N(left)-side is satisfactorily fitted with a broken
exponential showing a break at $R\approx\!\pm\!(55-60)$\arcsec\ and $\mu_{\rm
  R}\approx24.0$\,\magsqarcsec. On the S(right)-side the profile looks
more like a sharply-truncated profile.
The single disk fit by \cite{pohlen2001} yields a vertical scaleheight
of $\hz\eq1.8\arcsec$, a radial scalelength of $h\eq19.9\arcsec$, and an
inclination of $88.5\degc$.
\newline
The face-on-equivalent profile of FGC\,2339, beyond $35\arcsec$ can
be described as a \typet with an inner shallow exponential (of
$\hin\eq17.6\arcsec\!\equiv\!3.0\hout$) and an outer steeper exponential
($\hout\eq5.9\arcsec$) with a break at a radial distance of $\rbr\approx
57\arcsec\!\equiv\! 3.3\hin$ and a surface brightness of
$\mubr\eq26.4$R-\magsqarcsec.
\newline
For this galaxy the whole inner region $R\ltsim 35\arcsec$ is
dominated by a ring or barlike structure which introduces an extreme
flattening of the profiles at 2 vertical scaleheights and above 
\cite[cf.][]{olof}.
The small inner disk region in the range of $35\arcsec\ltsim R\ltsim
55\arcsec$ does not show a clear systematic flattening towards higher
$z$-cuts. The major axis profile is overall the flattest, so probably
heavily affected by the midplane dust. Nevertheless, there is a 19\%
increase in scalelength between $2.3$ and $3.4\!\cdot\!\hz$. 
The scalelength beyond the break in the outer disk increases
by a factor of 2.3 between $0-2.3\!\cdot\!\hz$.
\medskip
\noindent
{\bf IC\,4871:} (alternative name: IC\,4872) \newline 
This galaxy is overall very patchy with many distinct \htwo\  regions.
On the S(right)-side, along the major axis at $R\eq+100$\arcsec\ there
is an unusually warped patch of star formation visible. The major axis
radial profile can be well described with a broken exponential with a
very steep outer decline (dropping by $\!\approx\!2$\,mag within
$\approx 10$\arcsec) on both sides (with slightly different outer
slopes), starting at $R\approx\pm115$\arcsec
($\mu_{\rm V}\approx23.5$\,\magsqarcsec).
The single disk fit by \cite{pohlen2001} yields a vertical scaleheight
of $\hz\eq3.8\arcsec$, a radial scalelength of $h\eq42.3\arcsec$, and an
inclination of $87.5\degc$.
\newline
The face-on-equivalent profile of IC\,4871 is well described as a
\typet with an inner shallow exponential (of
$\hin\eq40.4\arcsec\!\equiv\!4.2\hout$) and an very steep outer exponential
($\hout\eq9.7\arcsec$) with a break at a radial distance of $\rbr\approx
117\arcsec\!\equiv\! 2.9\hin$ and a surface brightness of
$\mubr\eq24.4$V-\magsqarcsec.
\newline
We find an increase in scalelength of $\approx 13\%$ between the major axis
and about $2.5$ times the mean vertical scaleheight (at 8\arcsec) in the inner
region.
The scalelength beyond the break in the outer disk increases
by a factor of 1.7 between $0-2.5\!\cdot\!\hz$.
\medskip
\noindent
{\bf NGC\,522:} \newline Our image has some residual structure in the
flatfield which reaches the disk on the SW(right)-side at
$(R,z)\eq(+10\arcsec,-15\arcsec$)\ and at $(R,z)\eq(+60\arcsec,-30\arcsec$) 
causing the unusually shaped contours.
There is a galaxy cluster in the background with an edge-on 
spiral galaxy superimposed at $(R,z)\eq(-13\arcsec,-17\arcsec$) 
and another elliptical at $(R,z)\eq(-60\arcsec,-20\arcsec$). 
The major axis radial profile shows a clear broken exponential
behaviour with a very steep outer decline (dropping by
$\!\approx\!5$\,mag within $\approx\!25$\arcsec) on both sides,
starting at $R\approx\pm60$\arcsec\ ($\mu_{\rm
  R}\approx21.5$\,\magsqarcsec).
Even the edge-on profiles show clearly a decrease in the outer slope
with increasing $z$ (see below). The galaxy is slightly asymmetric with
the NE(left)-side shorter than the SW(right)-side.
The single disk fit by \cite{pohlen2001} yields a vertical scaleheight
of $\hz\eq4.3\arcsec$, a radial scalelength of $h\eq32.0\arcsec$, and an
inclination of $87.5\degc$.
\newline
The face-on-equivalent profile of NGC\,522 is well described as a
\typet with an inner shallow exponential (of
$\hin\eq52.7\arcsec\!\equiv\!6.1\hout$) and an very steep outer exponential
($\hout\eq8.7\arcsec$) with an `early' break at a radial distance of
$\rbr\approx 63\arcsec\!\equiv\! 1.2\hin$ and a surface brightness of
$\mubr\eq22.8$R-\magsqarcsec.
\newline
The major axis profile is overall the flattest, so probably heavily 
affected by the midplane dust. We find only a marginal increase in 
scalelength of $\approx 5\%$ between $0.9$ and $3.1$ times the mean 
vertical scaleheight (between $4-13$\arcsec) in the inner region.
The scalelength beyond the break in the outer disk increases
by a factor of 1.8 between $0-3.1\!\cdot\!\hz$.
\medskip
\noindent
{\bf NGC\,1596:} \newline 
See \cite{pohlen2004b} for additional notes. 
The face-on-equivalent profile of NGC\,1596 is classified as \typeiiic,
\ie a break with an upbending profile beyond. However, the profile is
not well described by a simple broken exponential. The whole disk
profile (beyond $R\!\gtsim\!\pm25$\arcsec) appears to be almost continuously
upbending. One may argue for an extended, exponential break region
between breaks at $R\!\approx\!70$\arcsec\ and $R\!\approx\!130$\arcsec.
Therefore, the exact position of the break is unclear.
Judging from the individually deprojected cuts this structure is
driven by the clear upbending breaks on the major axis and at
$z\!\eq\!8$\arcsec\ at a radial distance of $R\!\approx\!110$\arcsec.
All cuts at larger distance from the plane ($z\!\gtsim\!17$\arcsec) 
are consistent with having no break.
\newline
We find an increase in scalelength by a factor of 3.6 from the major
axis (steepest) to a vertical scaleheight of 33\arcsec\ (flatter) in
the inner region.
\medskip
\noindent
{\bf NGC\,3390:} \newline There is a nearby, low surface brightness
companion, of unknown distance, visible at
$(R,z)\eq(+140\arcsec,-125\arcsec$) and another unusually small and
round object at $R\eq+85$\arcsec\ on the major axis.
This galaxy exhibits a quite unusual appearance. The inner high surface
brightness part looks like a normal edge-on Sb galaxy with a well
defined dust-lane. However, the outer low surface brightness part
shows that the galaxy is embedded in a huge spheroidal envelope much more
like an \s0. The galaxy was already classified as an \s0-a by
\cite{lauberts} and \s0? by \cite{sa}, whereas \cite{rc3} and
\cite{pgc} quote Sb or Sbc.
The radial profiles close to the major axis, similar to other 
\s0 galaxies, decline almost straight into the noise with only 
a mild break at $R\!\approx\!\pm90$\arcsec\ in cuts higher 
above the plane. This structure makes it impossible to fit 
a sharply truncated or broken exponential model. 
Even the edge-on profiles show clearly a decrease in the outer slope
with increasing z (see below). 
\newline
The face-on-equivalent profile of IC\,3390 is reasonably well
described as a \typeo (see discussion in main text, \sec\ref{faceon}).
For NGC\,3390 we find an increase in scalelength by a factor of 3.1 between
the major axis (steepest) and a vertical distance of 26\arcsec\ (flatter) in
the inner disk region as expected from the \s0 envelope.
A scaleheight of $5.3\arcsec$, measured in the I-band by \cite{degrijs1998},
is used in \fig\ref{hVSz} and \fig\ref{hinVShout}.

\medskip
\noindent
{\bf NGC\,4179:}\\
See \cite{pohlen2004b} for additional notes. 
We find that the face-on-equivalent profile of NGC\,4179 is well
described as a \typet with an inner shallow exponential (of
$\hin\eq30.9\arcsec\!\equiv\!1.4\hout$) and an outer slightly steeper
exponential ($\hout\eq22.6\arcsec$) with a break at a radial distance of
$\rbr\approx 98\arcsec\!\equiv\! 3.2\hin$ and at a surface brightness of
$\mubr\eq23.3$V-\magsqarcsec.
\newline
In the inner disk we find an increase in scalelength of 36\% 
between the major axis and $4.5$ times the thin disk scaleheight 
($6.1\arcsec$) reported by \cite{pohlen2004b}. Overall the 
scalelength increases by a factor of 1.8 to the highest measured
cut at $z\eq45\arcsec$.
This is very similar to the factor of 1.9 reported by
\cite{pohlen2004b} for the ratio of thick to thick disk scalelength,
obtained with a completely different method (3D model fitting).
Their thin disk scalelength of $26.2\arcsec$ is almost identical 
to our scalelength of $26.7\arcsec$ at the major axis. Our scalelength of 
$46.8\arcsec$ for the vertical cut with the largest distance from 
the plane (at $z\!\eq\!45$\arcsec) is $\approx 9\%$ smaller 
compared to their thick disk scalelength of $51.0$\arcsec, which
we may reach for cuts even higher above.  
\medskip
\noindent 
{\bf NGC\,5290:} \newline The inner disk is embedded in an outer,
clearly thicker disk component or in a low surface brightness
extension of the bulge. The galaxy exhibits a thick and patchy
dustlane, and the two sides are rather asymmetric. In the contour map
the E(left)-side seems to be thicker and shorter, whereas the
W(right)-side is longer and tapers off.
This asymmetry is reflected in the major axis radial profile. When
fitted with a broken exponential structure, the slope of the outer
region on the E(left)-side is much steeper compared to the
W(right)-side. The latter seems to be almost $\!\approx\!15$\arcsec\ 
`longer'.
The single disk fit to the R-band image by \cite{pohlen2001} yields a
vertical scaleheight of $\hz\eq7.5\arcsec$, a radial scalelength of
$h\eq25.4\arcsec$, and an inclination of $88.5\degc$.
\newline
The face-on-equivalent profile of NGC\,5290 is in both bands (R and
\Km) well described as a \typet with an inner shallow exponential (of
$\hin\eq33.1\arcsec\!\equiv\!2.0\hout$ in R and
$\hin\eq31.2\arcsec\!\equiv\!2.3\hout$ in \Km) and an outer steeper
exponential ($\hout\eq16.3\arcsec$ in R and $\hout\eq13.6\arcsec$ in \Km).
The break radius is at a similar radial distance ($\rbr\approx
80\arcsec\!\equiv\! 2.4\hin$ in R, and $\rbr\approx 83\arcsec\!\equiv\!
2.7\hin$ in \Km) and at a surface brightness of
$\mubr\eq23.7$R-\magsqarcsec and $\mubr\eq21.0$\Km-\magsqarcsec
respectively.
\newline
In the \Km-band, the scalelength of the major axis in the inner disk region is
the flattest compared to all other vertical cuts above (see
\fig\ref{dprofpfits}). For the R-band it is almost as flat as the scalelength
at a distance of 4 times the vertical scaleheight.
We observe a systematic decrease in scalelength in both bands up to a
vertical distance of $z\approx15\arcsec$, after which the scalelength
starts to rise again for the deeper R-band image out to the highest
vertical cut at $30\arcsec$.
If this is not an intrinsic feature of NGC\,5290 one could argue that the dust
may still influence the scalelength determination even in the NIR
band. However, quantitatively this seems very unlikely since the amount of
dust inferred from the \Km-band flattening should have a more severe effect on
the R-band profile. 
In the deeper optical image the scalelength increases by 26\%
between $2$ and $4$ vertical scaleheight.
In the R-band the scalelength beyond the break in the outer disk 
increases by a factor of 1.5 between $0-2.6\!\cdot\!\hz$.
\medskip
\noindent
{\bf NGC\,5981:} \newline There is a possible dwarf elliptical
companion visible at $(R,z)\eq(-10\arcsec,+30\arcsec$), and another
companion superimposed on the disk at $(R,z)\eq(+77\arcsec,-5\arcsec$).
For both we have no velocity information, but there is another
large elliptical nearby (NGC\,5982, $6.2$\arcmin\  away with $\Delta
v\!\approx\!1300$\,\kms).
NGC\,5981 is slightly disturbed and asymmetric. The inner disk is
$\approx\!1.5$\deg tilted against the outer disk and the bulge component
is furthermore slightly tilted against the inner disk.
This asymmetry is reflected in the radial profiles. The outer slope 
of a broken exponential fit is significantly shallower on the 
NW(right)-side, where also the contour map is more extended. 
The single disk fit by \cite{pohlen2001} yields a vertical scaleheight
of $\hz\eq4.2\arcsec$, a radial scalelength of $h\eq24.1\arcsec$, and an
inclination of $86.5\degc$.
\newline
The face-on-equivalent profile of NGC\,5981 is well described as a
\typet with an inner shallow exponential (of
$\hin\eq27.9\arcsec\!\equiv\!2.0\hout$) and an outer steeper exponential
($\hout\eq13.8\arcsec$) with a break at a radial distance of
$\rbr\approx 67\arcsec\!\equiv\! 2.4\hin$ and at a surface brightness of
$\mubr\eq24.0$R-\magsqarcsec.
\newline
We find an increase in scalelength of $\approx 10\%$ between the major 
axis and about 3.6 times the mean vertical scaleheight 
(at 15\arcsec) in the inner region. 
Using only the two quadrants from the dust free side for 
averaging (in opposite to all four) this increase in scalelength 
for the inner disk grows to $\approx 25\%$. 
In this case -- as expected from the model fits (\cf
Fig\ref{dustfit}) -- the scalelength on the major axis is less affected
by the dust, so smaller, whereas the scalelength at 3.6 vertical
scaleheights is almost identical ($29.7\arcsec$ for the mean quadrant
versus $29.2\arcsec$ for the dust free part).
The scalelength beyond the break in the outer disk increases by a
factor of 1.4 between $0-3.6\!\cdot\!\hz$ in our mean quadrant.
Owing to the described asymmetries in the disk on both sides this
galaxy is well suited to study their effects on the deprojected
profiles.
Fitting only a mean quadrant obtained from the NW(right)- or
SE(left)-side of the galaxy separately we find differences of 
$5\%\!-\!10$\% for the scalelengths of the inner region. In the outer disk
(beyond the break) the differences are clearly larger, \eg 40\% on the
major axis, due to the obvious asymmetry.
\medskip
\noindent 
{\bf UGC\,10459:} \newline This galaxy has only a very small bulge
component. There is a possible background galaxy, with unknown
distance, visible at $(R,z)\eq(+33\arcsec,-9\arcsec$).
The major axis profile shows an obvious broken exponential structure
with a break radius at $R\!\approx\!\pm45$\arcsec\ ($\mu_{\rm
  R}\!\approx\!23.0$\,\magsqarcsec) separating a shallow inner and a
steeper outer slope.
The single disk fit by \cite{pohlen2001} yields a vertical scaleheight
of $\hz\eq2.0\arcsec$, a radial scalelength of $h\eq13.4\arcsec$, and an
inclination of $87.0\degc$.
\newline
The face-on-equivalent profile of UGC\,10459 is well described as a
\typet with an inner shallow exponential (of
$\hin\eq15.6\arcsec\!\equiv\!2.7\hout$) and an outer rather steep
exponential ($\hout\eq5.7\arcsec$) with a break at a radial distance of
$\rbr\approx 44\arcsec\!\equiv\! 2.8\hin$ and at a surface brightness of
$\mubr\eq25.3$R-\magsqarcsec.
\newline
We find an increase in scalelength of $\approx 40\%$ between the major
axis and about $5.6$ times the vertical scaleheight (at 11\arcsec) in
the inner disk region.
The scalelength on the major axis is slightly larger compared to
scalelength of the cut at $2.1\!\cdot\! \hz$, which is well explained by
the influence of the dust.
The scalelength beyond the break in the outer disk increases
by a factor of 2.2 between $0-3.6\!\cdot\!\hz$.
\begin{table*}
\begin{center}
{\normalsize
\begin{tabular}{ l c c c rrr c r ccc}
\hline
\rule[+0.4cm]{0mm}{0.0cm}
Galaxy
& Filter
& Profile
& $z$ cut
&$b_1$
&$b_2$
&\rbr
&\hin
&\hout
&\rbr
&\hin
&\mubr
\\[+0.1cm]
& 
& type
&[\arcsec]
&[\arcsec]
&[\arcsec]
&[\arcsec]
&[\arcsec]
&[\arcsec]
&[\hin]
&[\hout]
&[\magsqarcsec]
\\
\rule[-3mm]{0mm}{5mm}{\scriptsize{\raisebox{-0.7ex}{\it (1)}}}
&{\scriptsize{\raisebox{-0.7ex}{\it (2)}}}
&{\scriptsize{\raisebox{-0.7ex}{\it (3)}}}
&{\scriptsize{\raisebox{-0.7ex}{\it (4)}}}
&{\scriptsize{\raisebox{-0.7ex}{\it (5)}}}
&{\scriptsize{\raisebox{-0.7ex}{\it (6)}}}
&{\scriptsize{\raisebox{-0.7ex}{\it (7)}}} 
&{\scriptsize{\raisebox{-0.7ex}{\it (8)}}}
&{\scriptsize{\raisebox{-0.7ex}{\it (9)}}}
&{\scriptsize{\raisebox{-0.7ex}{\it (10)}}} 
&{\scriptsize{\raisebox{-0.7ex}{\it (11)}}} 
&{\scriptsize{\raisebox{-0.7ex}{\it (12)}}}
\\[-0.2cm]
\hline\hline \\[-0.4cm]
ESO 380-019 &V & II & foe & 12 & 117 & 87.6 & 27.9 & 11.5 & 3.1 & 2.4 & 22.9 \\
ESO 380-019 &V & II & 0 & 12 & 117 & 93.7 & 26.1 & 8.6 & 3.6 & 3.0 & 27.4 \\
ESO 380-019 &V & II & 6 & 12 & 117 & 86.9 & 28.8 & 10.5 & 3.0 & 2.7 & 27.4 \\
ESO 380-019 &V & II & 12 & 0 & 117 & 77.4 & 30.4 & 14.2 & 2.6 & 2.1 & 28.1 \\
ESO 380-019 &V & II & 18 & 10 & 117 & 65.4 & 32.7 & 21.5 & 2.0 & 1.5 & 29.1 \\
ESO 380-019 &V & I & 23 & 10 & 90 & $-$ & 34.9 & $-$ & $-$ & $-$ & $-$ \\
ES0 404-018 &V & II & foe & 0 & 92 & 75.6 & 34.0 & 9.6 & 2.2 & 3.6 & 24.6 \\
ES0 404-018 &V & II & 0 & 0 & 92 & 72.4 & 31.1 & 9.0 & 2.3 & 3.5 & 28.0 \\
\hline
\end{tabular}
}
\caption[]{Exponential disk parameters for all fitted profiles. \newline
{\scriptsize{\it (1)}} Principal name, 
{\scriptsize{\it (2)}} filter,
{\scriptsize{\it (3)}} profile type (I: no break, II: downbending break, III: upbending break),
{\scriptsize{\it (4)}} vertical position $z$ of the deprojected slices (`feo' means the face-on-equivalent profile), {\scriptsize{\it (5,6)}} inner and outer fitting boundaries, {\scriptsize{\it (7,8,9)}} break radius, inner, and outer scalelength in units of arcsec, {\scriptsize{\it (10)}} break radius in relation to the inner scalelength,
{\scriptsize{\it (11)}} inner scalelength in relation to the outer scalelength,
{\scriptsize{\it (12)}} the surface brightness at the break radius (estimated 
at the crossing point of the two exponential fits. The full version of this table is available online.
\label{detailedresults} 
}
\end{center}
\end{table*}
\section{Figures}
\label{figures}
The following figures show radial surface brightness profiles 
and isophote maps for four galaxies per page (in alphabetical 
order).
For the isophote map {\sl (lower panels)} each galaxy is rotated 
to the major-axis. The magnitude of the outer contour 
($\mu_{\rm lim}$, defined by a 3$\sigma$ criterion of the 
background) is indicated in each plot. The consecutive contours 
are equidistant spaced by 0.5\,mag. 
In the plot, the contour lines are drawn with increasing
smoothing towards the outer parts. Note however that for the actual
fitting a constant scale smoothing is used. For the inner contours
out to where the noise begins to increase no smoothing was
applied. The following 2-3 contours are smoothed by replacing each
pixel by the mean of $3^2$-pixels ``around'' the central pixel. For
the outer two contours this smoothing is increased to $5^2$-pixels.
The  {\sl upper panels} displays the major-axis surface brightness 
profile {\it (top, solid line)} and two parallel radial profiles 
{\it (lower, solid lines)} each above and below the major axis. 
The exact vertical positions ($z$) for the plotted profiles are 
indicated in the upper right corner of the plot.  
In some cases the best fit, sharply truncated, single disk model 
obtained by \cite{pohlen2001} is overplotted {\it (dashed line)}. 
\newpage
\begin{figure*}
\includegraphics[width=8cm]{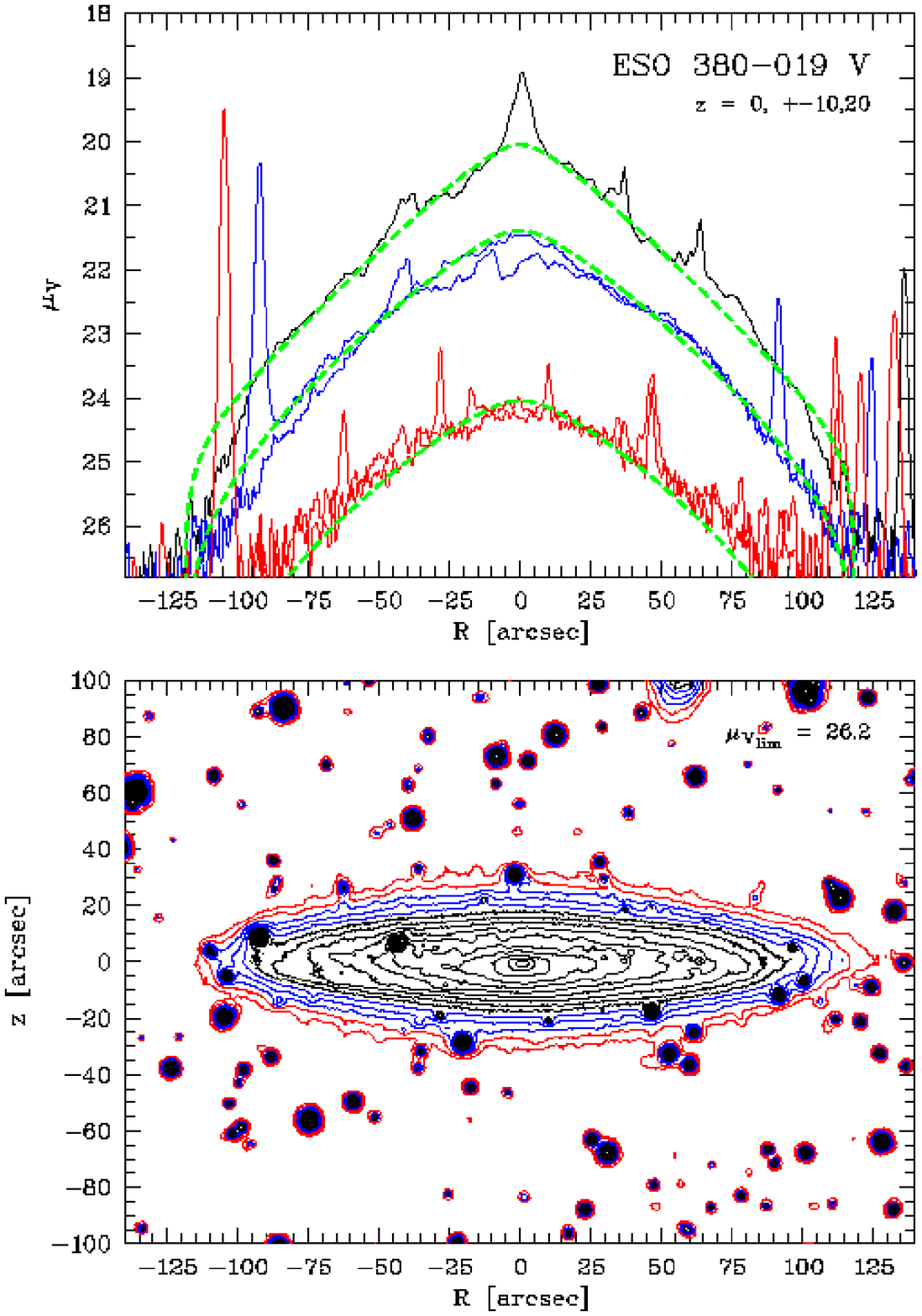}
\includegraphics[width=8cm]{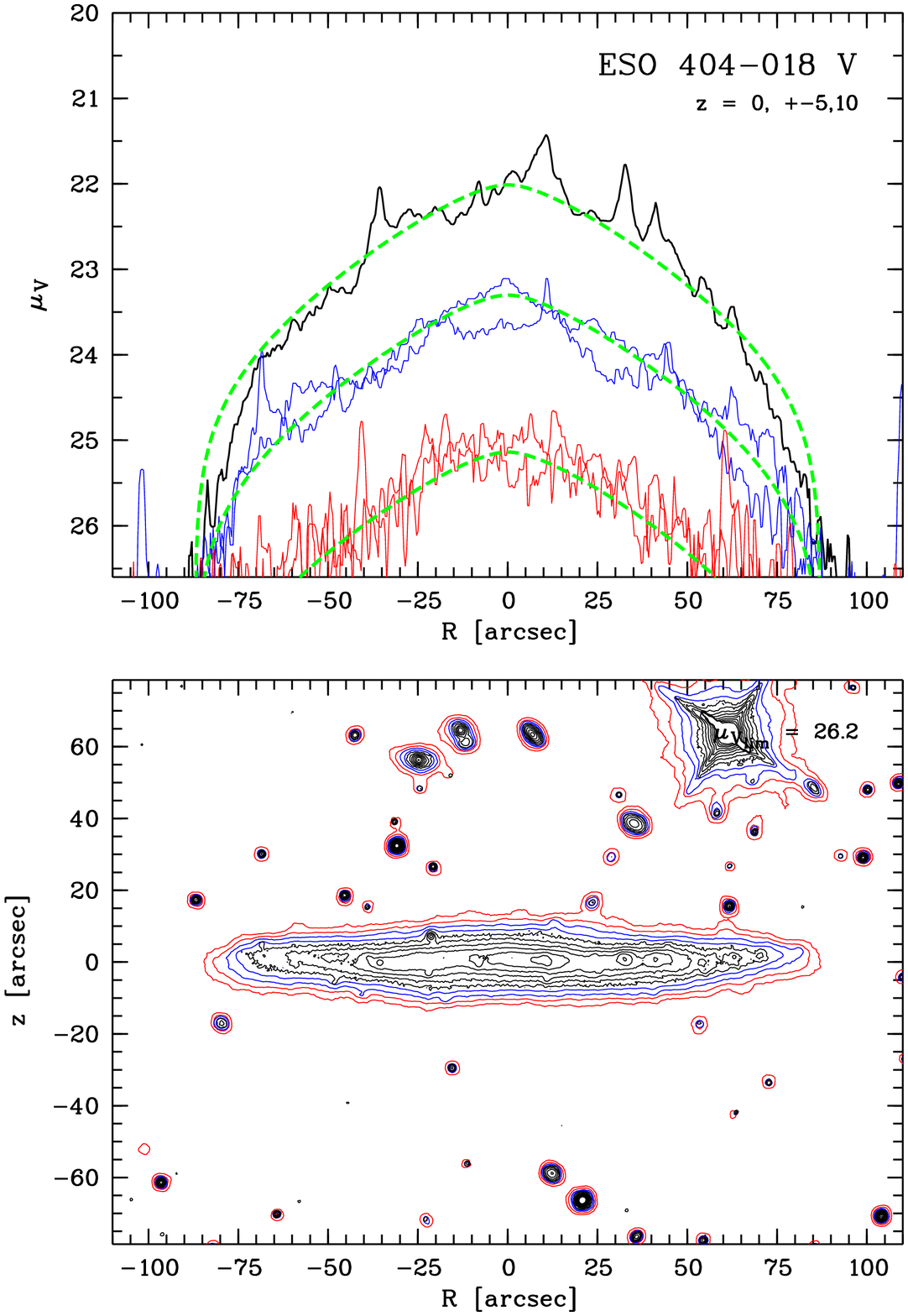} \\
\includegraphics[width=8cm]{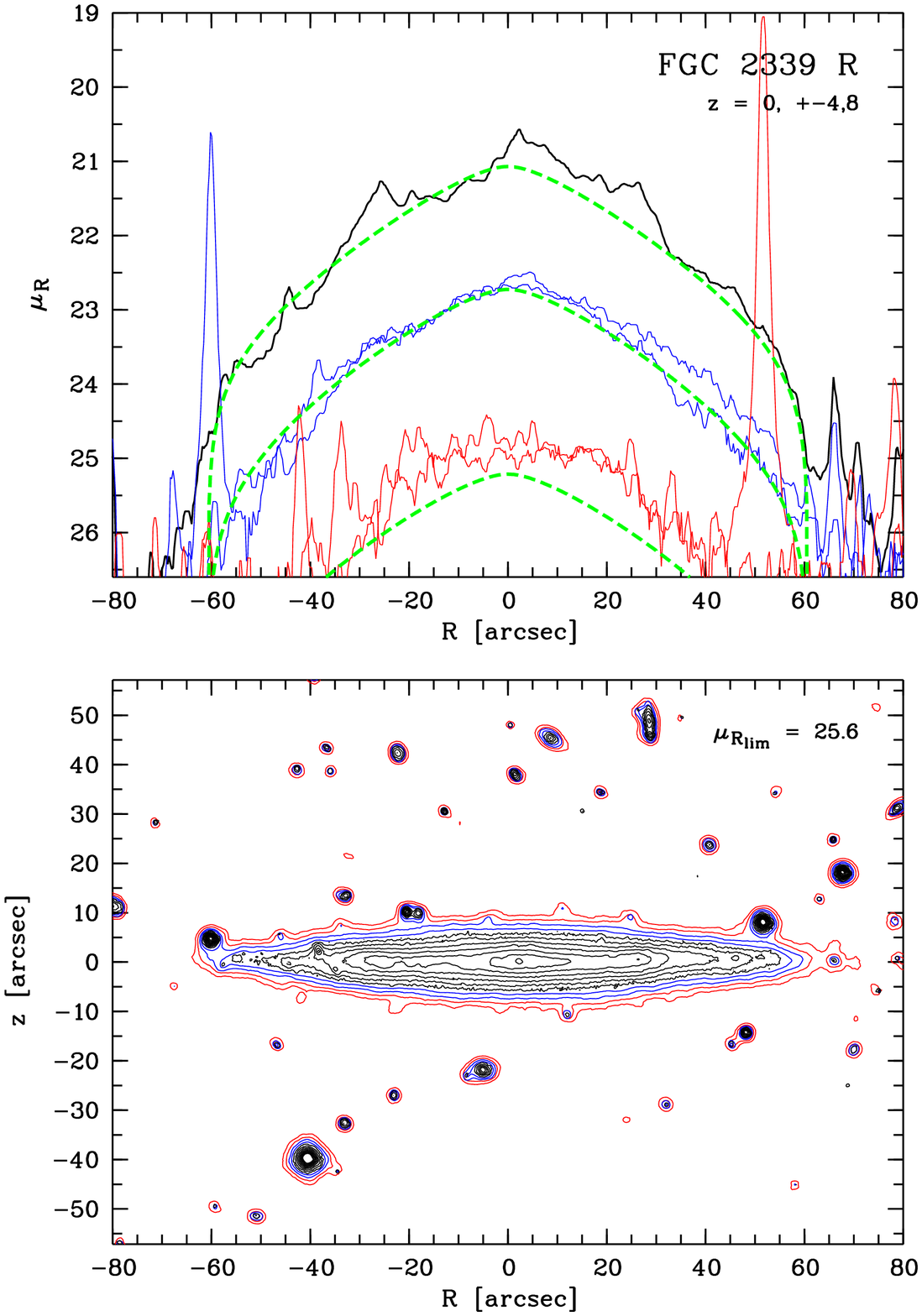}
\includegraphics[width=8cm]{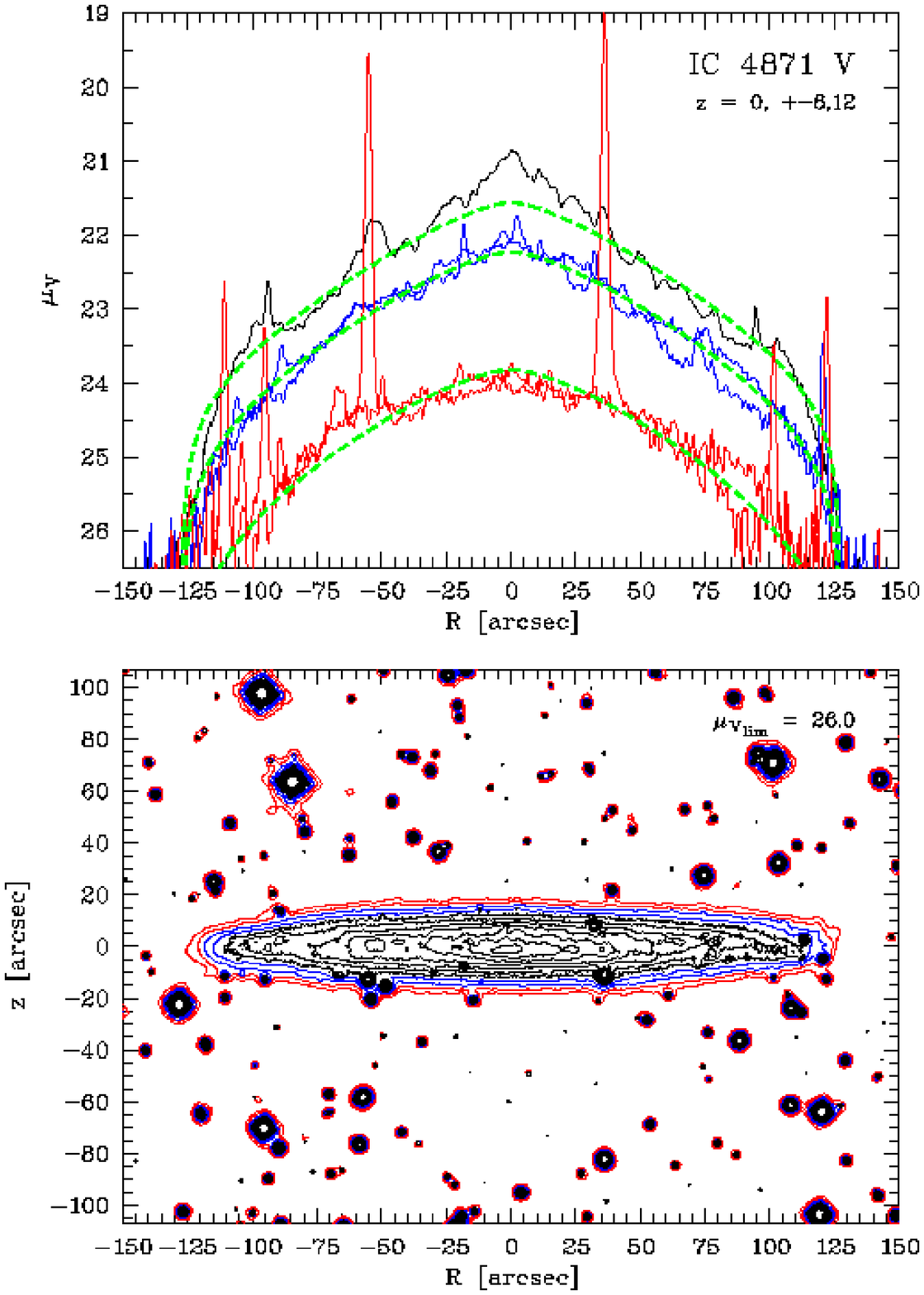}
\caption{Radial surface brightness profiles and isophote maps for 
ESO\,380-019 V-band, ESO\,404-018 V-band, FGC\,2339 R-band, and 
IC\,4871 V-band {\it (from top left to bottom right)}.
See text for further explanations. 
}
\label{mapspprofs}
\end{figure*}
\addtocounter{figure}{-1}
\begin{figure*}
\includegraphics[width=8cm]{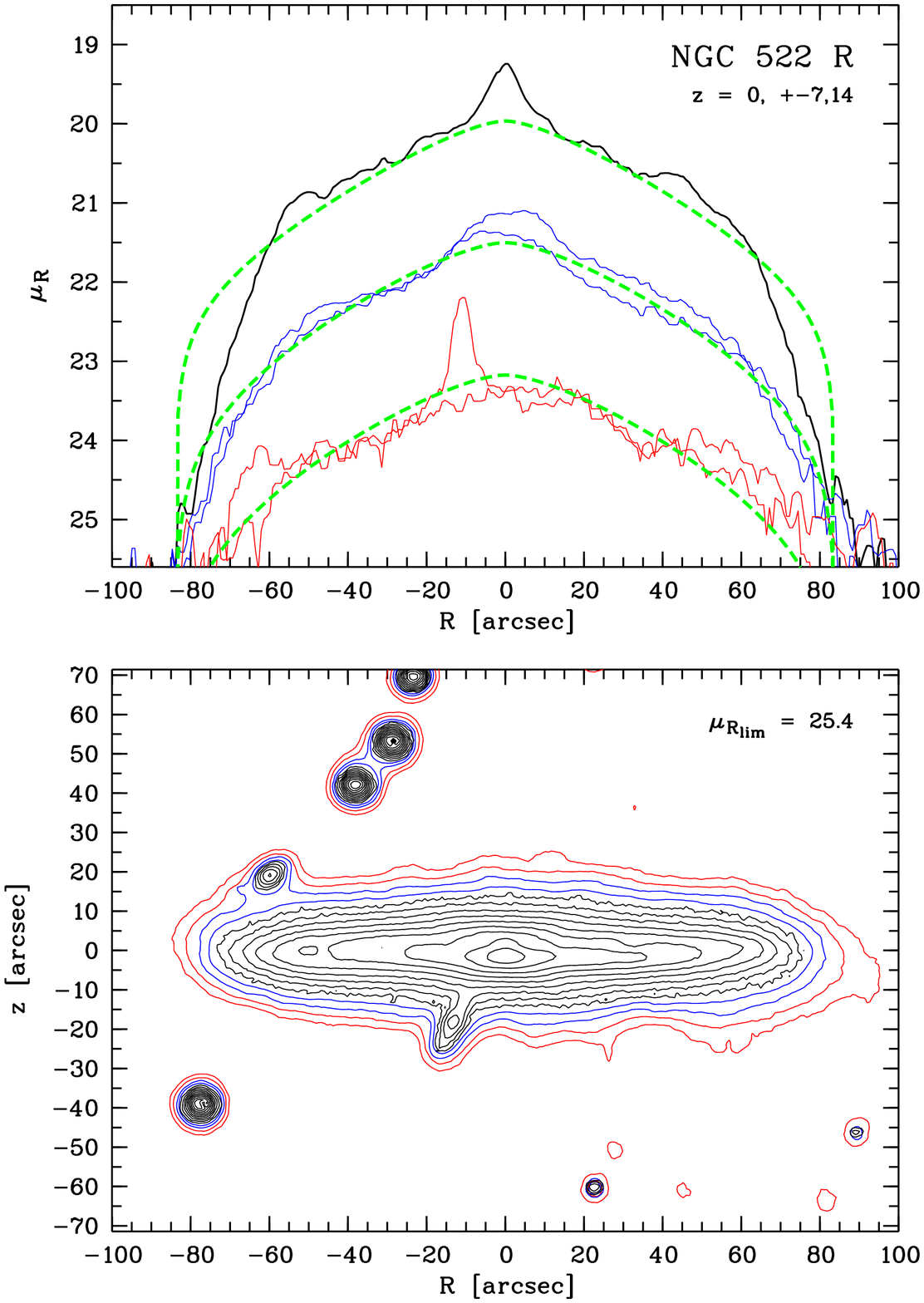}
\includegraphics[width=8cm]{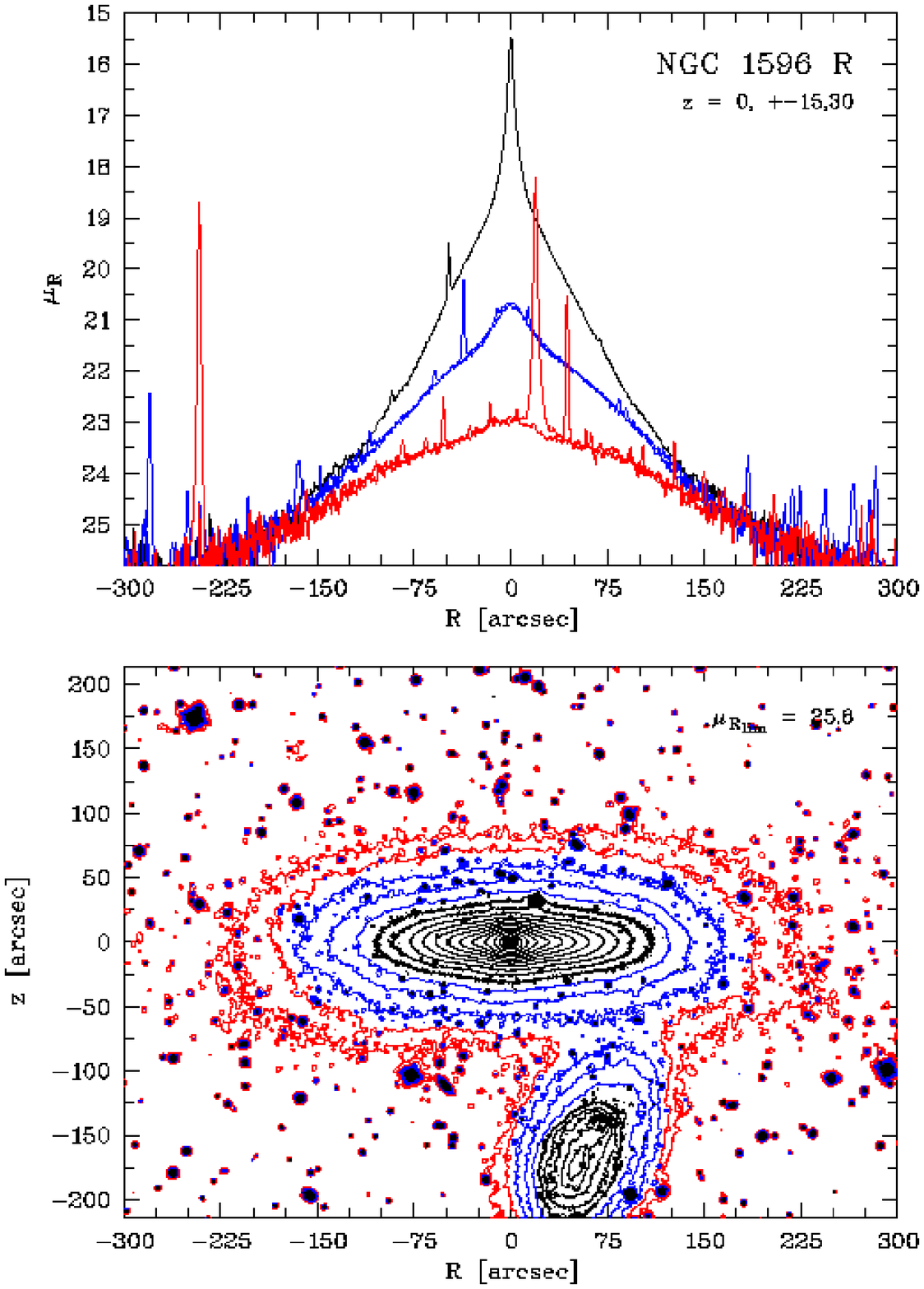} \\
\includegraphics[width=8cm]{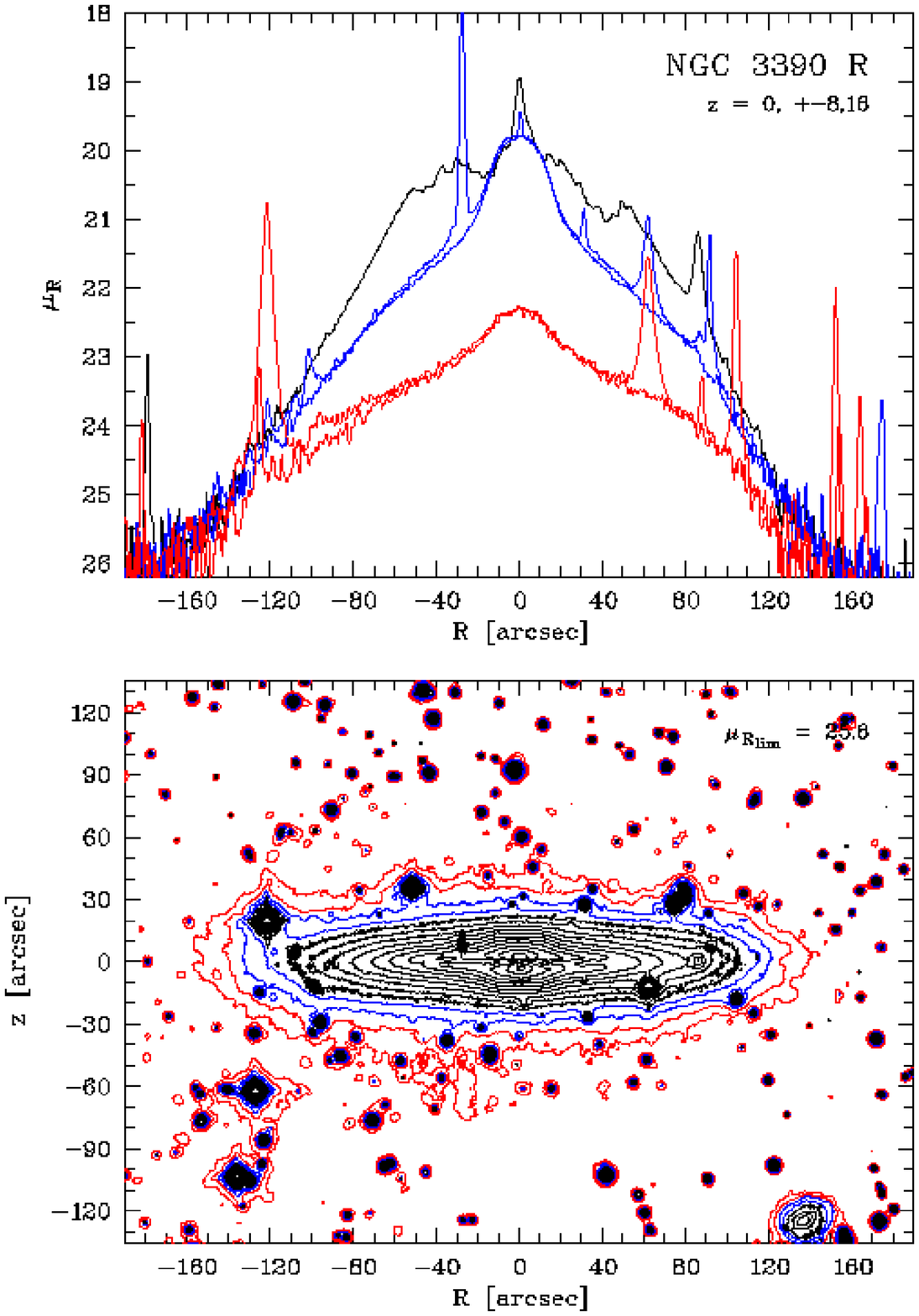}
\includegraphics[width=8cm]{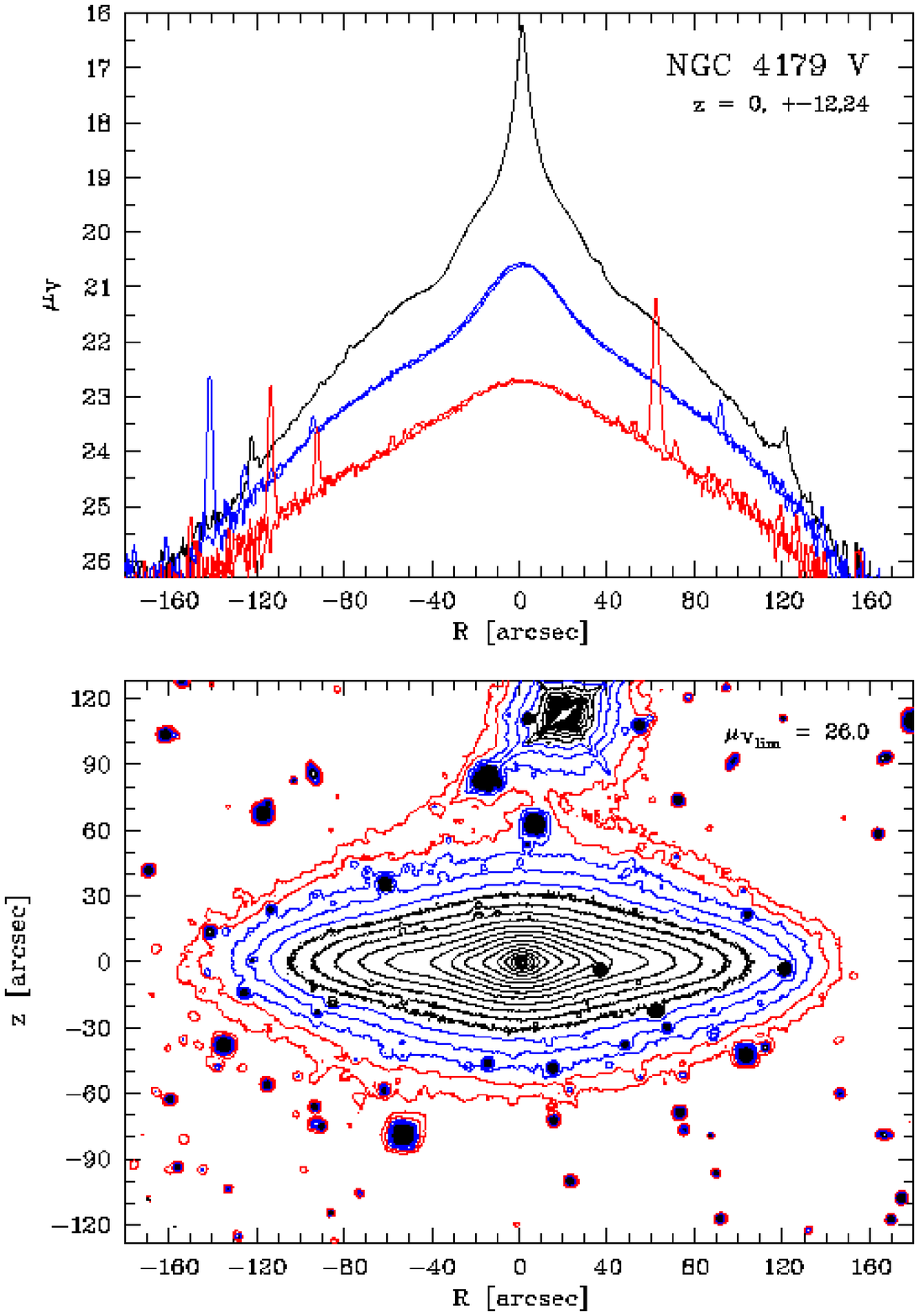}
\caption{Radial surface brightness profiles and isophote maps for 
NGC\,522 R-band, NGC\,1596 R-band, NGC\,3390 R-band, and 
NGC\,4179 V-band {\it (from top left to bottom right)}.}
\end{figure*}
\addtocounter{figure}{-1}
\begin{figure*}
\includegraphics[width=8cm]{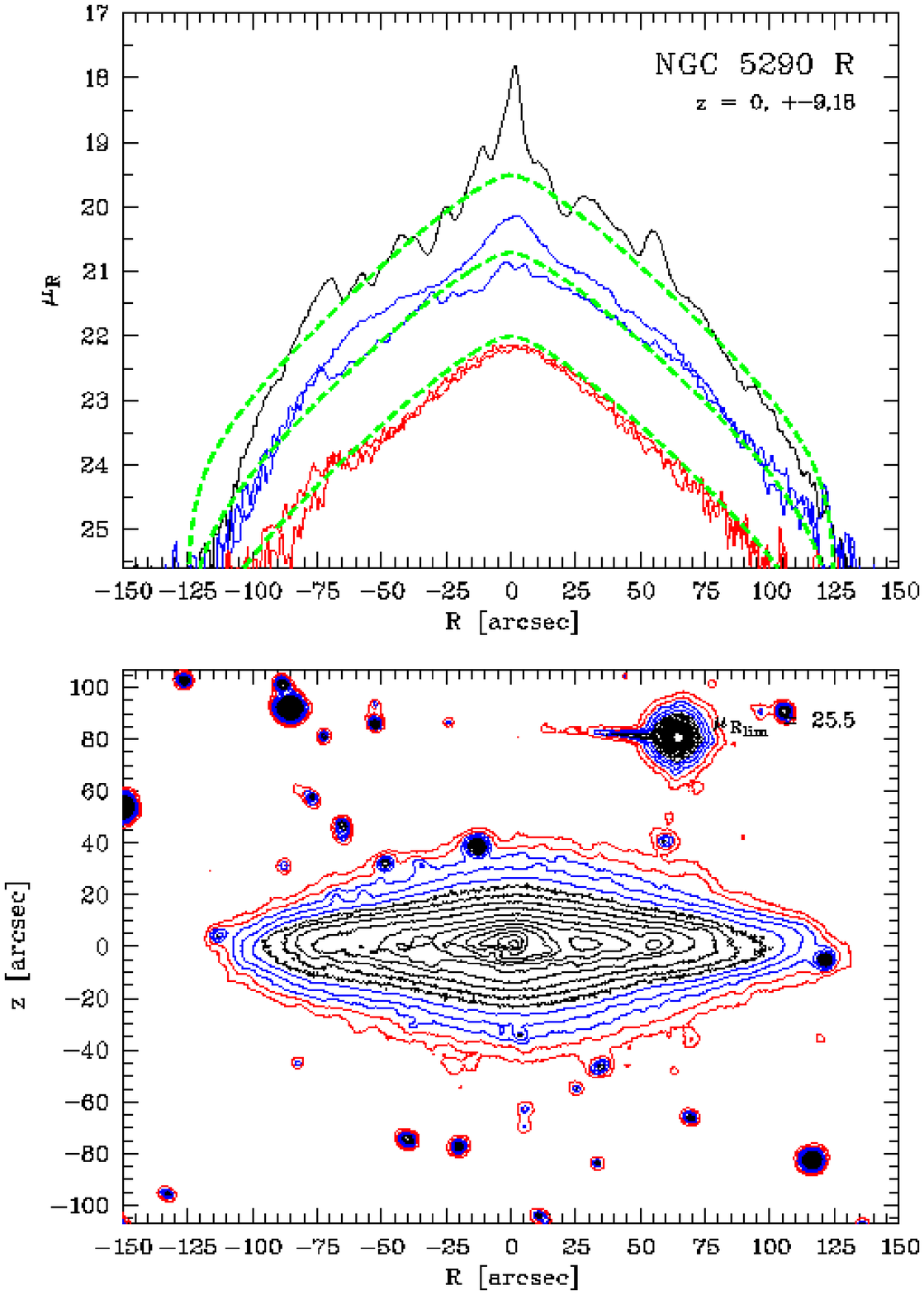}
\includegraphics[width=8cm]{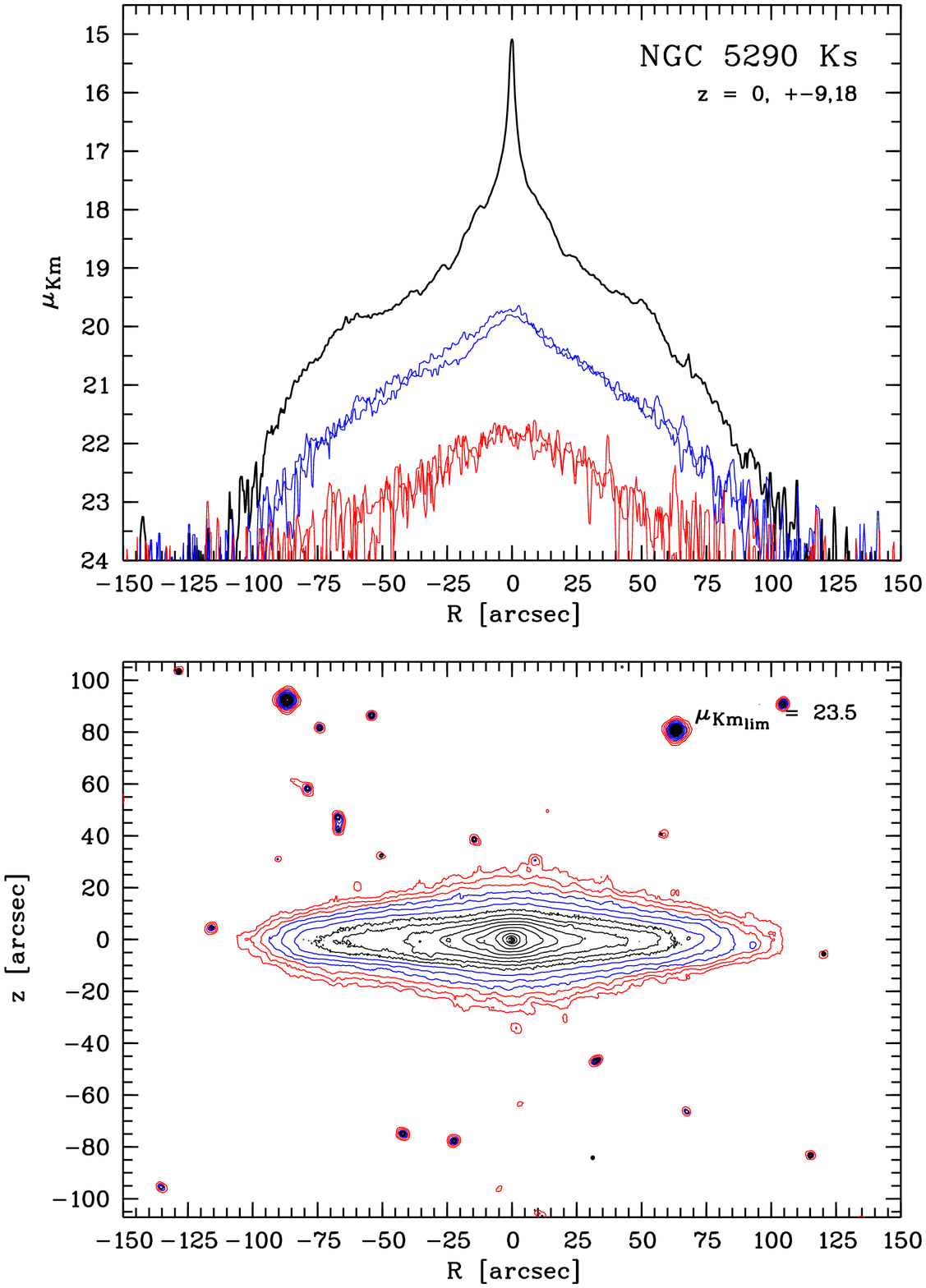}
\includegraphics[width=8cm]{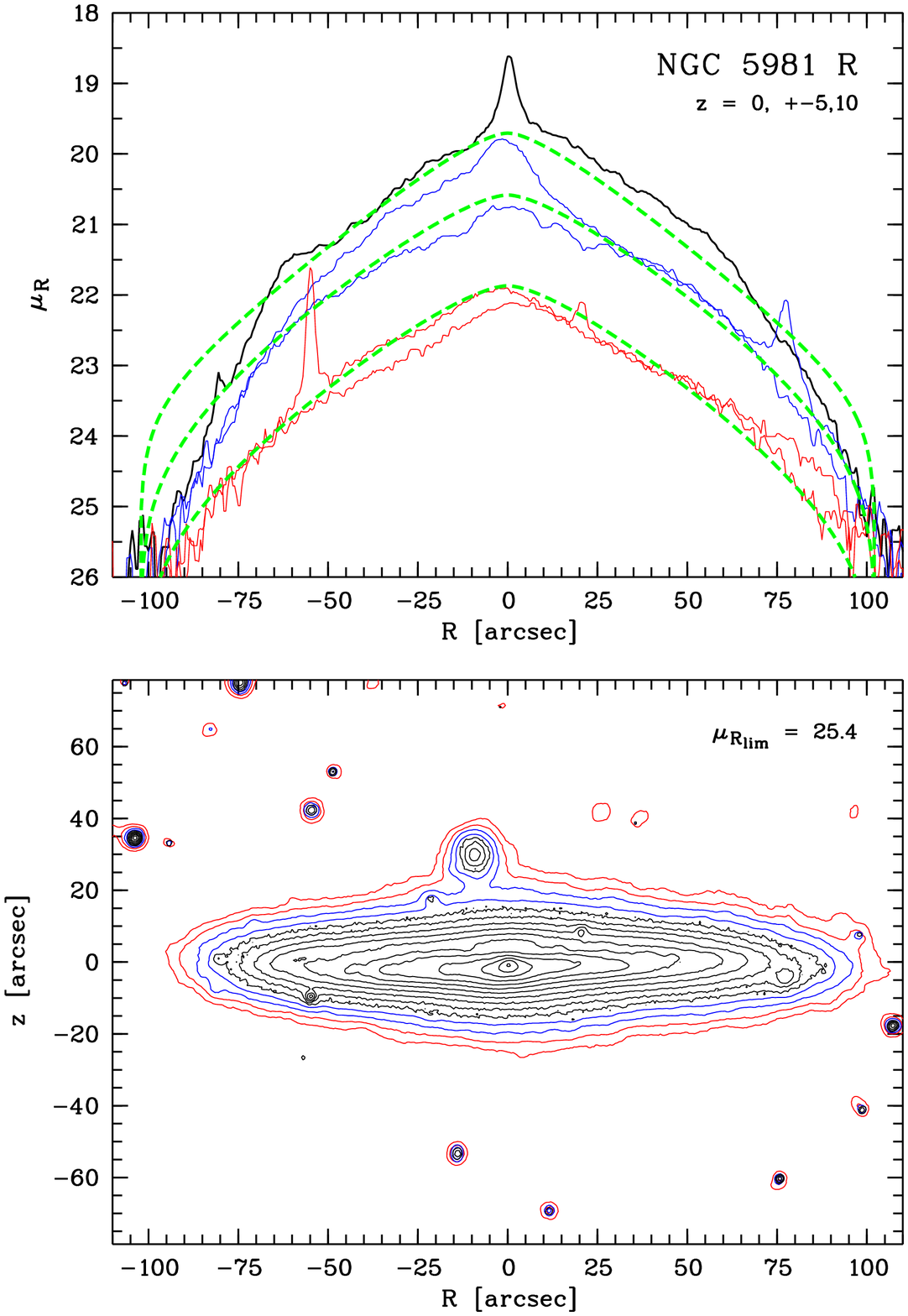}
\includegraphics[width=8cm]{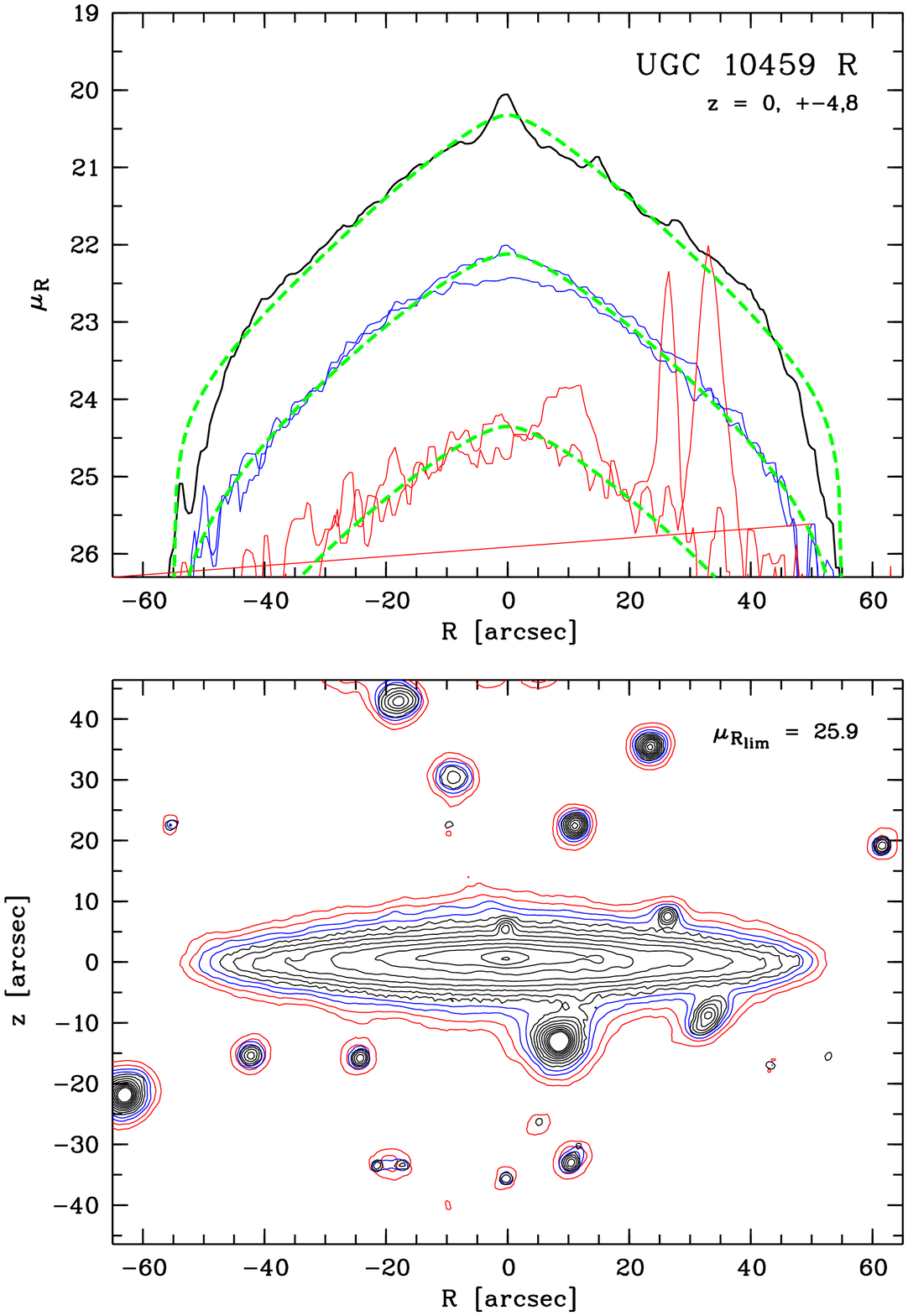}
\caption{Radial surface brightness profiles and isophote maps for 
NGC\,5290 R-band, NGC\,5290 \Km-band, NGC\,5981 R-band, and 
UGC\,10459 R-band {\it (from top left to bottom right)}.}
\end{figure*}


\begin{thebibliography}{}
%
\bibitem[Bertin \& Arnouts(1996)]{bertin1996} Bertin, E., \& 
Arnouts, S.\ 1996, A\&AS, 117, 393 

\bibitem[Bahcall \& Kylafis(1985)]{bahcall1985} Bahcall, J.~N., 
\& Kylafis, N.~D.\ 1985, ApJ, 288, 252 

\bibitem[Balcells et al.(2003)]{balcells2003} Balcells, M., 
Graham, A.~W., Dom{\'{\i}}nguez-Palmero, L., \& Peletier, 
R.~F.\ 2003, ApJL, 582, L79 

\bibitem[Barteldrees \& Dettmar(1994)]{barteldrees1994} 
Barteldrees, A., \& Dettmar, R.-J. 1994, A\&AS, 103, 475

\bibitem[Bianchi, Ferrara, \& Giovanardi(1996)]{bianchi96} 
Bianchi S., Ferrara A., Giovanardi C., 1996, ApJ 465, 127

\bibitem[Binney, Davies \& Illingworth(1990)]{bdi90} Binney, 
J. J.,Davies, R. L., \& Illingworth, G. D. 1990,  ApJ, 361, 78

\bibitem[Binney \& Tremaine (1987)]{binney87} Binney, J. J., Tremaine,
S., Galactic dynamics. Princeton, NJ, Princeton University Press


\bibitem[Byun(1998)]{byun} Byun, Y.~-I. 1998, Chinese Journal 
of Physics, 36, 677

\bibitem[Chung et al.(2006)]{chung2006} Chung, A., Koribalski, 
B., Bureau, M., \& van Gorkom, J.~H.\ 2006, MNRAS, 370, 1565	

\bibitem[Dalcanton \& Bernstein(2002)]{dalcanton2002} Dalcanton, 
J.~J., \& Bernstein, R.~A.\ 2002, AJ, 124, 1328 

\bibitem[Debattista et al.(2006)]{debattista2006} Debattista, V.~P., 
Mayer, L., Carollo, C.~M., Moore, B., Wadsley, J., \& Quinn, T.\ 2006, 
ApJ, 645, 209 

\bibitem[Daubechies(1988)]{daubechies1988} Daubechies I., 1988, Comms.
Pure Appl. Math., 41, 909

\bibitem[Dehnen \& Gerhard(1993)]{dehnen93} Dehnen, W., \& 
Gerhard, O. E. 1993, MNRAS, 261, 311

\bibitem[Dehnen \& Gerhard(1994)]{dehnen94} Dehnen, W., \& 
Gerhard, O. E. 1993, MNRAS, 268, 1019

\bibitem[de Grijs(1998)]{degrijs1998} de Grijs, R.\ 1998, MNRAS, 
299, 595 

\bibitem[de Grijs \& Peletier(1997)]{degrijs1997} de Grijs, R., \& 
Peletier, R.~F.\ 1997, A\&A, 320, L21 

\bibitem[de Grijs et al.(2001)]{degrijs2001} de Grijs, R., 
Kregel, M., \& Wesson, K.~H.\ 2001, MNRAS, 324, 1074

\bibitem[de Vaucouleurs(1959)]{devaucouleurs1959} de 
Vaucouleurs, G.\ 1959, Handbuch der Physik, 53, 311 

\bibitem[de Vaucouleurs et al.(1991)]{rc3} de Vaucouleurs, G., 
de Vaucouleurs, A., Corwin, H.G., Buta, R.J., Paturel, G., 
\& Fouque, J.B. 1991, Third reference catalogue of bright 
galaxies, Springer-Verlag New York

\bibitem[de Zeeuw(2007)]{dezeeuw2007} de Zeeuw, T., 2007, Astrophysics and
  Space Science Proceedings, Springer (Dordrecht), Vol.\ 3, 571, ``Island
  universes: structure and evolution of disk galaxies'', R.S. de Jong (ed.)

\bibitem[Donoho(1992)]{donoho1992} Donoho D.L., Stanford University
Dept. of Statistics, 1992

\bibitem[Donoho \& Johnstone(1994)]{donoho1994} Donoho D.L., Johnstone I.M.,w
Biometrika 1994,81,425

\bibitem[Du et al.(2006)]{du2006} Du, C., Ma, J., Wu, Z., \& 
Zhou, X.\ 2006, MNRAS, 372, 1304	

\bibitem[Elmegreen \& Hunter(2006)]{elmegreen2006} Elmegreen, B.~G., 
\& Hunter, D.~A.\ 2006, ApJ, 636, 712 

\bibitem[Erwin et al.(2005)]{erwin2005} Erwin, P., 
Beckman, J.E., \& Pohlen, M.\ 2005, ApJL, 626, 81 

\bibitem[Erwin et al.(2006)]{erwin2006} Erwin, P., 
Pohlen, M., \& Beckman, J.E.\ 2006, in prep.

\bibitem[Fabricant et al.(1984)]{fabricant84} Fabricant, D., 
Rybicki, G., \& Gorenstein, P. 1984, ApJ, 286, 186

\bibitem[Ferguson et al.(2006)]{ferguson2006} Ferguson, A., 
Irwin, M., Chapman, S., Ibata, R., Lewis, G., \& Tanvir, 
N.\ 2006, astro-ph/0601121 

\bibitem[Florido et al.(2001)]{florido2001} Florido, E., 
Battaner, E., Guijarro, A., Garz{\'o}n, F., \& 
Jim{\'e}nez-Vicente, J.\ 2001, A\&A, 378, 82 

\bibitem[Florido et al.(2006)]{florido2006} Florido, E., Battaner, E.,
Guijarro, A., Garz{\'o}n, F., \& Castillo-Morales, A.\ 2006, A\&A, 455, 467

\bibitem[Freedman et al.(2001)]{hst_h0} Freedman, W.L.~et al.~
(some 14 other authors) 2001, ApJ, 553, 47

\bibitem[Freeman(1970)]{free70} Freeman K.C. 1970, ApJ, 160, 811

\bibitem[Freeman \& Bland-Hawthorn(2002)]{freeman2002} Freeman, 
K., \& Bland-Hawthorn, J.\ 2002, ARA\&A, 40, 487 

\bibitem[Gerhard \& Binney(1996)]{gerhard96} Gerhard, O., 
\& Binney, J. 1996, MNRAS, 279, 993

\bibitem[Governato et al.(2007)]{governato2007} Governato, F., 
Willman, B., Mayer, L., Brooks, A., Stinson, G., Valenzuela, O., Wadsley, 
J., \& Quinn, T.\ 2007, MNRAS, 374, 1479 

\bibitem[Gyuk et al.(1999)]{gyuk1999} Gyuk, G., Flynn, C., 
\& Evans, N.~W.\ 1999, ApJ, 521, 190 

\bibitem[Hamabe \& Wakamatsu(1989)]{hamabe1989} Hamabe, M., \& 
Wakamatsu, K.-I.\ 1989, ApJ, 339, 783 

\bibitem[Holley-Bockelmann \& Mihos(2001)]{holley2001} 
Holley-Bockelmann, J.~K., \& Mihos, J.~C.\ 2001, 
Bulletin of the American Astronomical Society, 33, 798 

\bibitem[Hunter \& Elmegreen(2006)]{hunter2006a} Hunter, D.~A., \&
Elmegreen, B.~G.\ 2006, ApJS, 162, 49 

\bibitem[Hunter et al.(2006)]{hunter2006b} Hunter, D.~A., 
Elmegreen, B.~G., \& Martin, E.\ 2006, AJ, 132, 801 

\bibitem[Jensen \& Thuan(1982)]{jensen} Jensen, E.B., \& Thuan, T.X. 
1982, ApJS, 50, 421 

\bibitem[Kregel(2003)]{kregel2003} Kregel, M. 2003, PhD Thesis, 
University Groningen

\bibitem[Kregel et al.(2002)]{kregel2002} Kregel, M., van der 
Kruit, P.~C., \& de Grijs, R.\ 2002, MNRAS, 334, 646 

\bibitem[Lauberts(1982)]{lauberts} Lauberts, A.\ 1982, 
ESO/Uppsala Survey of the ESO(B) atlas, Garching: European 
Southern Observatory (ESO) 

\bibitem[Li et al.(2006)]{li2006} Li, Y., Mac Low, M.-M., \& 
Klessen, R.~S.\ 2006, ApJ, 639, 879 

\bibitem[Lucy(1974)]{lucy74} Lucy, L. 1974, AJ, 79, 745

\bibitem[Matthews(2000)]{matthews2000} Matthews, L.~D.\ 2000, 
AJ, 120, 1764 

\bibitem[M{\"o}llenhoff et al.(2006)]{moellenhoff2006} 
M{\"o}llenhoff, C., Popescu, C.~C., \& Tuffs, R.~J.\ 2006, 
A\&A, 456, 941

\bibitem[N\"aslund \& J\"ors\"ater(1997)]{naeslund} N\"aslund, 
M., \& J\"ors\"ater, S. 1997, A\&A, 325, 915

\bibitem[Nelder \& Mead(1965)]{nelder65} Nelder, J.A., and 
Mead, R.\ 1965, Computer Journal, vol. 7, 308

\bibitem[Patterson(1940)]{patterson1940} Patterson, 
F.~S.\ 1940, Harvard College Observatory Bulletin, 914, 9 

\bibitem[P{\'e}rez(2004)]{perez2004} P{\'e}rez, I.\ 2004, 
A\&A, 427, L17 

\bibitem[Pohlen(2001)]{pohlen2001} Pohlen, M., 2001, 
Ph.~D.~ Thesis, Ruhr-University Bochum, Germany

\bibitem[Pohlen \& Trujillo(2006)]{pohlen2006} Pohlen, M., 
\& Trujillo, I.\ 2006, A\&A, 454, 759  

\bibitem[Pohlen et al.(2000a)]{pohlen2000a} Pohlen, M., Dettmar, 
R.-J., L\"utticke, R.\ 2000a, A\&A, 357, L1 

\bibitem[Pohlen et al.(2000b)]{pohlen2000b} Pohlen, M., 
Dettmar, R.-J., L\"utticke, R., \& Schwarzkopf, U. 2000b, 
A\&AS, 144, 405 

\bibitem[Pohlen et al.(2002a)]{pohlen2002a} Pohlen, M., Dettmar, R.-J., 
L\"utticke, R., \& Aronica, G.\ 2002, A\&A, 392, 807

\bibitem[Pohlen et al.(2002b)]{pohlen2002b} Pohlen, M., Dettmar, 
R.~J., L{\"u}tticke, R., \& Aronica, G.\ 2002, ASP Conf.~Ser.~275: Disks of 
Galaxies: Kinematics, Dynamics and Peturbations, 275, 15 

\bibitem[Pohlen et al.(2003)]{pohlen2003} Pohlen, M., Balcells, 
M., L{\"u}tticke, R., \& Dettmar, R.-J.\ 2003, A\&A, 409, 485 

\bibitem[Pohlen et al.(2004a)]{pohlen2004a} Pohlen, M., Beckman, 
J.~E., H{\"u}ttemeister, S., Knapen, J.~H., Erwin, P., 
\& Dettmar, R.-J.\ 2004a, Penetrating Bars Through Masks 
of Cosmic Dust, 713 

\bibitem[Pohlen et al.(2004b)]{pohlen2004b} Pohlen, M., 
Balcells, M., L{\"u}tticke, R., \& Dettmar, R.-J.\ 2004b, 
A\&A, 422, 465 

\bibitem[Prugniel \& Heraudeau(1998)]{pgc} Prugniel, P., 
\& Heraudeau, P.\ 1998,  A\&AS, 128, 299 

\bibitem[Rybicki(1987)]{rybicki87} Rybicki, G. B. 1987 in 
Structure and Dynamics of Elliptical Galaxies, IAU Symp. 127, 
ed. de Zeeuw, P. T., Kluwer, Dordrecht, 397

\bibitem[Sandage \& Tammann(1980)]{sa} Sandage, A., \& Tammann, 
G.A. 1980, A Revised Shapley-Ames Catalog of Bright Galaxies, 
Washington: Carnegie Institution

\bibitem[Sasaki(1987)]{sasaki} Sasaki, T. 1987, PASJ, 39, 849

\bibitem[Schlegel et al.(1998)]{schlegel} Schlegel, D.~J., 
Finkbeiner, D.~P., \& Davis, M.\ 1998, ApJ, 500, 525 

\bibitem[Siegel et al.(2002)]{siegel2002} Siegel, M.~H., Majewski, 
S.~R., Reid, I.~N., \& Thompson, I.~B.\ 2002, ApJ, 578, 151 

\bibitem[Tamm \& Tenjes(2006)]{tamm2006} Tamm, A., \& Tenjes, 
P.\ 2006, A\&A, 449, 67 

\bibitem[Trujillo \& Pohlen(2005)]{trujillo2005} Trujillo, I., 
\& Pohlen, M.\ 2005, ApJL, 630, L17 

\bibitem[Tsikoudi(1980)]{tsikoudi1980} Tsikoudi, V.\ 1980, 
ApJS, 43, 365 

\bibitem[van den Berg et al.(2006)]{olof} van den Berg, O. et
al.\ 2006, in prep.

\bibitem[van der Kruit(1979)]{vdk1979} van der Kruit, 
P.~C., 1979, A\&AS, 38, 15

\bibitem[van der Kruit \& Searle(1981)]{vdk1981} van der Kruit, 
P.~C., \& Searle, L.\ 1981, A\&A, 95, 105 

\bibitem[Wakamatsu \& Hamabe(1984)]{wakamatsu1984} Wakamatsu, 
K.-I., \& Hamabe, M.\ 1984, ApJS, 56, 283 

\bibitem[Weinberg(2001)]{weinberg2001} Weinberg, M.~D. 2001, 
ASP Conf.~Ser.~231: Tetons 4: Galactic Structure, Stars and 
the Interstellar Medium, p.\,53

\bibitem[Wozniak(1994)]{wozniak1994} Wozniak, H.\ 1994, 
A\&A, 286, L5 

\bibitem[Xilouris et al.(1999)]{xil99} Xilouris E.M., Byun Y.I., 
Kylafis N.D., Paleologou E.V., Papamastorakis P., 1999, A\&A, 344, 868

\bibitem[Yoachim \& Dalcanton(2006)]{yoachim2006} Yoachim, P., 
\& Dalcanton, J.~J.\ 2006, AJ, 131, 226 

\bibitem[Zaroubi \& Goelman(2000)]{zaroubi2000} Zaroubi S., Goelman G., 
2000, Magnetic Resonance Imaging, 18, 59

\bibitem[Zaroubi et al.(1998)]{zaroubi1998} Zaroubi S., 
Squires G., Hoffman Y., \& Silk J., 1998, ApJ, 500, L87 

\bibitem[Zaroubi et al.(2001)]{zaroubi2001} Zaroubi, S., 
Squires, G., de Gasperis G., Evrard A.~E., Hoffman Y., \& Silk J., 
2001, ApJ, 561, 600


\end{thebibliography}
\end{document}